\documentclass{article}
\usepackage[utf8]{inputenc}
\usepackage{amsmath}
\usepackage{amssymb}
\usepackage{graphicx}
\usepackage{float}
\usepackage{bm}
\usepackage{hyperref}
\usepackage{physics}
\usepackage[style=numeric-comp, sorting=none]{biblatex}
\usepackage{arxiv}
\usepackage{physics}
\usepackage{amsfonts}
\usepackage{mathtools}
\usepackage[ruled,linesnumbered]{algorithm2e}
\usepackage[shortlabels]{enumitem}

\SetNlSty{textbf}{\normalsize}{}

\newcommand{\add}[2]{\the\numexpr #1+#2}
\newcommand{\MA}[1]{
    \begin{bmatrix}
    | & | &  & |\\
    #1_{0} & #1_{1} & \cdots  & #1_{n}\\
    | & | &  & |
    \end{bmatrix}
}
\newcommand{\MB}[1]{
    \begin{bmatrix}
    | & | &  & |\\
    #1_{0} & #1_{1} & \cdots  & #1_{n-1}\\
    | & | &  & |
    \end{bmatrix}
}
\newcommand{\MC}[1]{
    \begin{bmatrix}
    | & | &  & |\\
    #1_{1} & #1_{2} & \cdots  & #1_{n}\\
    | & | &  & |
    \end{bmatrix}
}
\newcommand{\Mi}[1]{
    \begin{bmatrix}
    | & | &  & |\\
    #1_{i} & #1_{1+i} & \cdots  & #1_{n+i}\\
    | & | &  & |
    \end{bmatrix}
}

\title{Dynamic mode decomposition of magnetohydrodynamic bubble chain flow \\ in a rectangular vessel}

\author{
   Martins Klevs\\
  Institute of Numerical Modelling\\
  University of Latvia (UL)\\
  Riga, Latvia, Jelgavas 3, 1004 \\
  \texttt{martins.klevs@lu.lv} \\
   \And
  Mihails Birjukovs\\
  Institute of Numerical Modelling\\
  University of Latvia (UL)\\
  Riga, Latvia, Jelgavas 3, 1004 \\
  \texttt{mihails.birjukovs@lu.lv} \\
   \And
  Peteris Zvejnieks\\
  Institute of Numerical Modelling\\
  University of Latvia (UL)\\
  Riga, Latvia, Jelgavas 3, 1004 \\
   \And
  Andris Jakovics\\
  Institute of Numerical Modelling\\
  University of Latvia (UL)\\
  Riga, Latvia, Jelgavas 3, 1004
  }

\begin{document}
\maketitle

\begin{abstract}
We showcase the dynamic mode decomposition (DMD) code developed for applications in two-phase flow analysis. Vertical bubble chain flow in a rectangular vessel filled with liquid gallium is studied without and with applied static horizontal magnetic field (MF) and DMD is applied to the velocity fields computed via volume of fluid simulations. Flow patterns are investigated in the vessel and bubble reference frames. We demonstrate the effect of applied MF and gas flow rate on bubble wake flow and larger scale flow structures within the liquid metal vessel by examining velocity field mode statistics over trajectory time and total flow time, as well as the computed mode velocity fields.
\end{abstract}

\keywords{Dynamic mode decomposition (DMD) \and Magnetohydrodynamics (MHD) \and Bubble flow \and Computational fluid dynamics (CFD) \and Liquid metal}

\section{Introduction}

\textit{Dynamic mode decomposition} (DMD) is a dimensionality reduction algorithm for spectral analysis of periodic or quasi-periodic data. Given a time series of data where each series element is the state of some dynamic system, DMD computes a set of modes each with a fixed oscillation frequency and a decay/growth rate. DMD is effectively a combination of the \textit{principal component analysis} (PCA) and the \textit{Fourier transform} (FT) as it decomposes a series of system states into a number of dominant spatial structures associated with unique frequencies \cite{schmidDynamicModeDecomposition2008}. At the same time, it is also connected to the perturbation theory for partial differential equations \cite{rowleySpectralAnalysisNonlinear2009}.

One of the most common methods for time series analysis is the \textit{discrete Fourier transform} (DFT). If the data exhibits periodic structure, DFT can be used to represent the system in the frequency space which is a more natural representation of the system. However, if the dataset associated with the system is very high-dimensional in space it can be very difficult and inconvenient to analyse its structure because FT does not simplify the system in any way. Another way to decompose time series data is through \textit{proper orthogonal decomposition} (POD) which is equivalent to PCA. POD expresses the time series data in a new orthogonal basis. The basis components are chosen such that they optimally cover the data in terms of energy content captured by successive modes. The time evolution is then analysed in the new reduced coordinate system. While the new coordinate basis is simpler than the original one, it is not guaranteed to be physically meaningful \cite{dmd-theory-and-applications}.

DMD combines aspects of both of these methods. Instead of generating orthogonal basis vectors, a DMD algorithm constructs modes with unique frequencies and growth/decay rates that are not necessarily orthogonal. Each mode's time evolution is independent from other modes. Unlike DFT, DMD generates a \textit{sparse} set of frequencies paired with corresponding spatial modes that account for dominant system dynamics patterns. This enables a simplified yet more physically meaningful representation of the system \cite{schmidDynamicModeDecomposition2008, rowleySpectralAnalysisNonlinear2009, dmd-theory-and-applications}. DMD was originally developed to analyze fluid dynamics systems  \cite{schmidDynamicModeDecomposition2008}. Since its inception it has been used to analyse flow instabilities and vortex shedding in fluids \cite{seenaDynamicModeDecomposition2011, jovanovicSparsitypromotingDynamicMode2014, rojselKoopmanModeAnalysis, rowleySpectralAnalysisNonlinear2009}. DMD has also been used to study different nonlinear systems in meteorology \cite{manningForecastingShorttermDynamics2019}, for video processing \cite{grosekDynamicModeDecomposition2014, ulhaqDynamicModeDecomposition2020}, electrocorticography analysis \cite{bruntonExtractingSpatialtemporalCoherent2014, shiraishiNeuralDecodingElectrocorticographic2020}, sunspot data analysis \cite{albidahProperOrthogonalDynamic2020}, etc. DMD is a contemporary and actively growing field of research with applications spanning multiple disciplines.

Bubble flow in liquid metal is of interest in many applications such as liquid metal stirring, purification, continuous casting, chemical reactors, etc., and these processes can be (and some already are) controlled using applied magnetic field (MF) \cite{birjukovsPhaseBoundaryDynamics2020, baakeNeutronRadiographyVisualization2017, casting-euler-musig, embr-experiment}. While single bubble magnetohydrodynamic (MHD) flow is fairly well studied, many aspects of bubble collective dynamics, especially in presence of MF, are not properly understood \cite{prl-path-instability, natcomms-shape-dynamics, dns-longitudinal-field, imb-transverse-field, zhang-mf-vertical, zhang-mf-simulations, zhang-thesis, uttt-path-instability, udv-review-article, uttt-x-ray-single-bubble, udv-review-article, udv-longitudinal-field, udv-transverse-field, spiral-to-zigzag-explained, shape-and-wake-simulations}. Aside from preventing optimization and efficiency improvements for industrial processes where the underlying physics are unclear, it is also impossible to significantly improve effective models for bubble flow (Euler-Euler and Lagrangian) without insights into how bubbles interact in MHD flow (or without applied MF) \cite{casting-euler-les, casting-euler-musig, casting-lagrange-bubbles, casting-new-collective-dynamics-models}. Through recent developments, however, fundamental investigation of bubble chain systems mimicking industrially relevant conditions is underway \cite{birjukovsArgonBubbleFlow2020, birjukovsPhaseBoundaryDynamics2020, baakeNeutronRadiographyVisualization2017, x-ray-bubble-chain-simulate, x-ray-prime-code, x-ray-bubble-breakup, x-ray-bubble-coalescence, x-ray-validation}. In this context DMD is a prospective method that should enable more meaningful and in-depth data interpretation.

While there are several cases where DMD has been applied to MHD flows \cite{loukopoulosStudyThermomagnetohydrodynamicFlow2020, taylorDynamicModeDecomposition2018} and DMD has been used to study flows containing bubbles \cite{alessandriDynamicModeDecomposition2019, liuDynamicModeDecomposition2019}, there are no cases in literature (to the authors' knowledge at the time of this publication) where DMD was applied to two-phase flow with explicitly resolved bubbles, in particular a chain of bubbles rising due to the buoyancy force, without or with applied MF. In this article we use DMD to extract the main spatial velocity field modes from a dynamic system wherein a chain of deforming bubbles rises through liquid metal in a rectangular vessel. We observe how applied static horizontal MF and gas flow rate influence the mode spectrum and mode spatial configurations both in the static (vessel) reference frame and reference frames of the rising bubbles. To do this we have implemented a custom DMD algorithm that is both noise-resistant and memory efficient due to a special system state correlation strategy (\textit{higher-order DMD}) and the use of a streaming singular value decomposition (SVD) algorithm (\textit{MOSES SVD} \cite{eftekhari2019moses}).

\section{The MHD system}
\label{sec:system-and-cases}

\begin{figure}[htbp]
    \centering
    \includegraphics[width=1\textwidth]{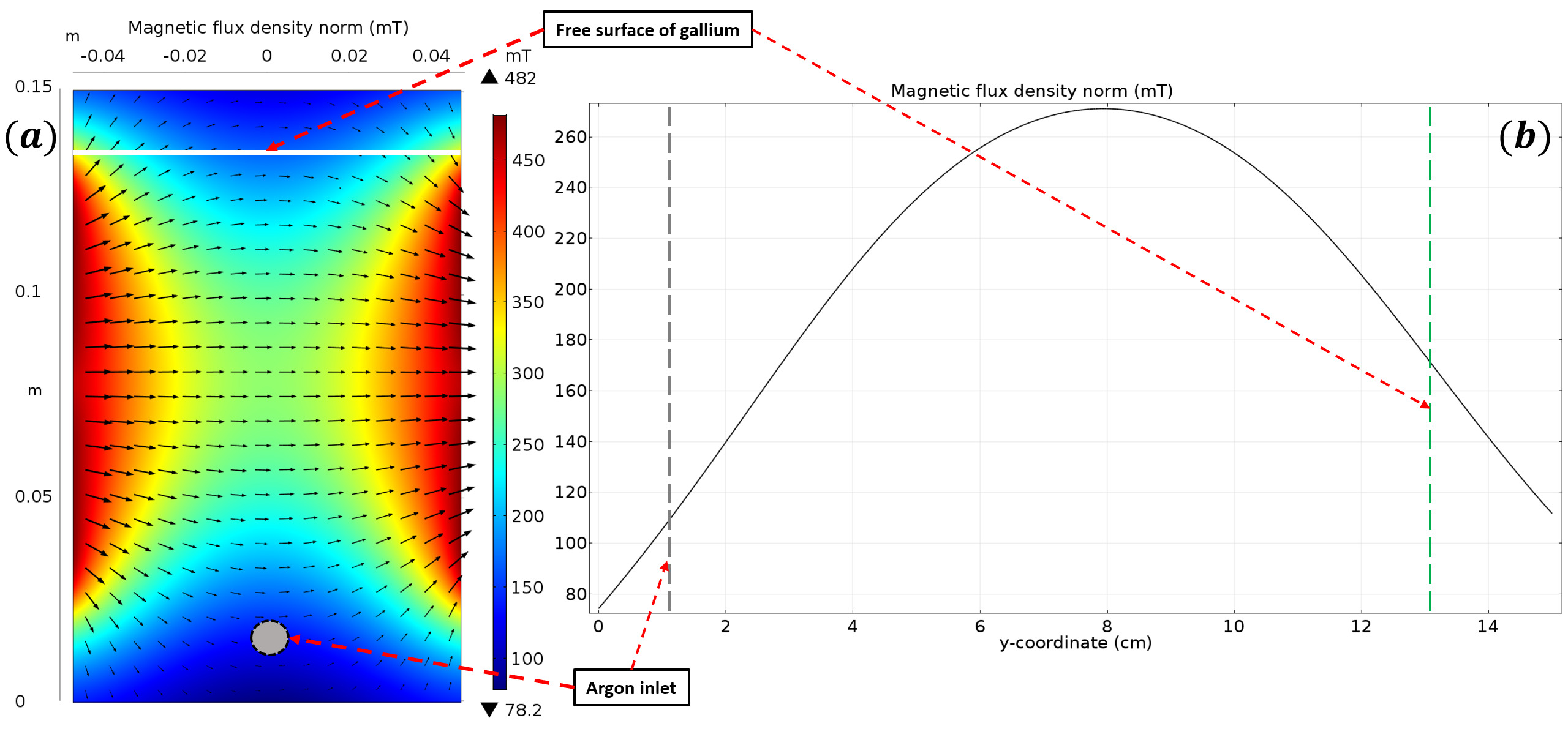}
    \caption{(a) MF in the gallium vessel midplane and (b) MF magnitude over the vessel axis.}
    \label{fig:mf-container}
\end{figure}

The physical system of interest is vertical argon (Ar) bubble chain flow in liquid gallium (Ga) in a rectangular glass vessel, 150 x 90 x 30 $mm$ (Figure \ref{fig:mf-container}). Bubbles are injected vertically/horizontally at the bottom and ascend with acceleration due to buoyancy, exhibiting zigzag trajectories with out-of-plane perturbations. The free surface of Ga is at $130~mm$. Static horizontal MF is applied using a system of permanent magnets and an iron yoke, wherein the liquid metal vessel is placed as shown in \cite{birjukovsArgonBubbleFlow2020,birjukovsPhaseBoundaryDynamics2020} and the resulting MF configuration is illustrated in Figure \ref{fig:mf-container}a.

Four cases are considered to illustrate how DMD can be used to assess the effects of applied MF and varying gas flow rate at the inlet: 

\begin{itemize}
    \item 30 sccm flow rate, no MF
    \item 100 sccm flow rate, no MF
    \item 30 sccm flow rate, $\sim 265~ mT$ horizontal MF in the bubble flow region
    \item 100 sccm flow rate, $\sim 265~ mT$ horizontal MF in the bubble flow region
\end{itemize}
where \textit{sccm} stands for standard cubic centimeters per minute. Note that 30 sccm flow rate will result in an effectively single-bubble flow regime which will serve to demonstrate the differences that collective dynamics introduce into the system at $100~sccm$.

Data for the DMD analysis is generated by simulating the above system under the listed conditions using a numerical model outlined in \cite{birjukovsPhaseBoundaryDynamics2020, birjukovsArgonBubbleFlow2020} (using \textit{EOF-Library} \cite{eof-soft-x}) in the $Rm \ll 1$ (magnetic Reynolds number) approximation. This is done to exclude the induced contribution to the overall MF -- systematic studies of this system including induced MF for higher flow rates is outside the scope of this article and is reserved for future papers. The value intervals of other relevant dimensionless groups are provided in \cite{birjukovsPhaseBoundaryDynamics2020}. One modification to the previously used numerical model is the extension of the \textit{InterFoam} solver with the \textit{isoAdvector} scheme for phase interface compression \cite{iso-advector-1, iso-advector-2} which in this case offers better numerical stability at the cost of a moderate increase in computation time per time step. To improve performance and minimize artefacts in the volume fraction field, \textit{isoFaceTol}, \textit{surfCellTol} and \textit{p\_rghFinal} (tolerance) solver parameters were optimized and the volume fraction field $\alpha \in [0;1]$ was redefined such that $\alpha = 1$ for bubbles and $\alpha = 0$ for gallium. \textit{isoAdvector} also enables to use a rather coarse homogeneous $1~mm$ cubic mesh for the proof-of-concept problems in this article while avoiding artefacts in bubble shape dynamics.

\section{Basic principles of DMD}

\subsection{Koopman mode expansion}

Consider a dynamic system evolving in time on a manifold $M$ such that, for $x_k\in M$,

\begin{equation}
    x_{k+1} = f\qty(x_{k})
    \label{eq:system-map}
\end{equation}

where $f$ is a map from $M$ to itself that governs the time evolution of the system $k \in \mathbb{Z}$. The \textit{Koopman operator} is a linear infinite-dimensional operator $K$ that acts on scalar valued functions on $M$ such that for any scalar-valued function $g : M\rightarrow\rm \mathbb{R}$, $K$ maps $\vec{g}$ to a new function $Kg$:
\begin{equation}
    Kg\qty(x) = g\qty(f\qty(x))
    \label{eq:koopman-map}
\end{equation}

Let $\varphi_i : M \rightarrow \mathbb{R}$ denote eigenfunctions and $\lambda_i \in \mathbb{C}$ denote eigenvalues of the Koopman operator,

\begin{equation}
    K\varphi_i\qty(x) = \lambda_i\varphi_i\qty(x).
\end{equation}

For the MHD system described above, DMD treatment of vector fields (velocity, vorticity, etc.) is of interest. Consider a vector-valued observable $\vec{g}:M\rightarrow \mathbb{R}^p$. If each of the $\vec{g}$ components lies within the span of the eigenfunctions $\varphi_i$, then, as in \cite{mezic-solo-spectral-props}, one may expand $\vec{g}$ in terms of $\varphi_i$ as

\begin{equation}
    \vec{g}\qty(\vec{x}) = \sum_{i=0}^{\infty}\varphi_{i}\qty(\vec{x})\vec{v}_i
    \label{eq:expansion}
\end{equation}

If the components of $\vec{g}$ \textit{do not} lie within the span of $\varphi_i$, one may split $K$ into regular and singular components, and project components of $\vec{g}$ onto the span of the eigenfunctions \cite{mezic-solo-spectral-props}. The expression (\ref{eq:expansion}) is typically viewed as an expansion of $\vec{g}\qty(\vec{x})$ as a linear combination of vectors $\vec{v}_i$, but it is equivalently an expansion of $\vec{g}\qty(\vec{x})$ as a linear combination of $\varphi_i$, where now $\vec{v}_i$ are the vector-valued coefficients of the expansion. In this paper we will refer to the eigenfunctions $\varphi_i$ as Koopman eigenfunctions, and to the corresponding $\vec{v}_i$ in (\ref{eq:expansion}) as the Koopman modes of the map $f$ for the observable $\vec{g}$.

From (\ref{eq:system-map}) and (\ref{eq:koopman-map}), iterates of $\vec{x}_0$ are given by

\begin{equation}
    \vec{g}\qty(\vec{x}_k) =
    \sum_{i=0}^{\infty} K^k\varphi_i\qty(\vec{x}_0)\vec{v}_i =
    \sum_{i=0}^{\infty} \lambda^k_i\varphi_i\qty(\vec{x}_0)\vec{v}_i
\end{equation}

\subsection{The default DMD algorithm}

In its basic implementation DMD computes an approximate linear operator $A$ that represents the analyzed system (potentially nonlinear) such that $A$ advances the system state one step forward in time. In this framework the system states are represented by matrices

\begin{equation}
    X_{m\times n} = \MB{x}, \quad Y_{m\times n} = \MC{x}
    \label{eq:state-matrices}
\end{equation}

where $x_i$ is a column vector that represents the system state for the $i$-th time step. $Y$ columns are system states advanced by one time step from respective state vectors in $X$. This framework also assumes the states are sequential and equally spaced in time. With this the propagation (evolution) equation is given by

\begin{equation}
    AX = Y
    \label{eq:propagation-equation}
\end{equation}

To determine the Koopman mode frequencies one must compute the complex eigenvalues $\lambda_i$ and eigenvectors $\omega_i$ of $A$:

\begin{equation}
    A\omega_i = \lambda_i\omega_i
\end{equation}
where $\omega_i$ form the span of $A$. With the Koopman mode decomposition and the system state correlation via $A$

\begin{equation}
    Ax_i = x_{i+1}
    \label{eq:x_pairs}
\end{equation}
one can express time stepping in terms of $\lambda_i$ and $\omega_i$:

\begin{equation}
    A^{n}x_j = x_{j+n} = \sum_k C_k\lambda_k^n\varphi_k = \sum_k C_k e^{a_k t} e^{i\omega_k t}\varphi_k
\end{equation}

If the system exhibits periodic and/or quasi-periodic processes, $\omega_i$ and $\lambda_i$ represent characteristic time scales within the system. DMD separates dynamics by time scales and the associated system structures (modes) can then be analysed separately.

Typically the matrices $X$ and $Y$ are very large, since they are determined by the (typically fine) resolution of 2D/3D images or measurement point sets from experiments, or simulation grids/meshes. Therefore, solving for $A$ exactly is not feasible. Usually most of the system state information stored in $X$ and $Y$ can be represented by orders of magnitudes smaller amount of data, i.e. it is sufficient to encapsulate the main patterns within the system. This stems from the intuition that systems of interest usually have coherent structures that are much larger than a single point of measurement and therefore much of the fine detail may be discarded without loss of meaningful features.

To compute system eigenvalues, \textit{singular value decomposition} (SVD) is used, which decomposes the input state matrix into three matrices:

\begin{equation}
    X_{m \times n} = U_{m\times k}S_{k \times k}V^*_{k \times n}
\end{equation}
where $UU^* = I$, $VV^* = I$, $S=\textrm{diag}\qty(\sigma_i)$, $k=\textrm{min}\qty(m,n)$. $S$ is a diagonal matrix that contains the singular values $\sigma_i$ of $X$ in descending order. If the data in $X$ has large coherent structures, then $\sigma_i$ values will quickly decay with $i$. This property is used to construct an approximation of $X$ by only taking the first $r$ singular values of the SVD -- this is referred to as the truncated SVD, which will be used throughout the paper:

\begin{equation}
    X_{m \times n} \approx U_{m\times r}S_{r \times r}V^*_{r \times n}
\end{equation}
where $r\ll k$. With the above, (\ref{eq:propagation-equation}) can be rewritten as

\begin{equation}
    AUSV^* = Y
\end{equation}

from which $A$ can be obtained:
    
\begin{equation}
    A = YVS^{-1}U^*
    \label{eq:get-A-matrix}
\end{equation}

Now it is necessary to project $A$ onto the lower-dimensional subspace, which is achieved via the matrix similarity transformation:

\begin{equation}
    \tilde{A}_{r \times r} = U^*_{r \times m} A_{m \times m} U_{m \times r}
    \label{eq:similarity-transform}
\end{equation}
where $\tilde{A}$ is the projection of $A$. This is done because eigenvalue calculation for $\tilde{A}$ is much simpler than for $A$ and by definition their eigenvalues are identical. Rearranging (\ref{eq:similarity-transform}) and substituting it into (\ref{eq:get-A-matrix}) one has:

\begin{equation}
    \tilde{A} = U^* YVS^{-1}
\end{equation}

Now eigenvalues $\lambda_i$ and eigenvectors $\varphi_i$ can be computed

\begin{equation}
    \tilde{A}\tilde{\varphi}_i = \lambda_i\tilde{\varphi}_i
\end{equation}
and then $\varphi_i$ are transformed back to the basis of $A$ via

\begin{equation}
    \varphi_i = U \tilde{\varphi}_i
\end{equation}

To summarize, the default DMD algorithm is as follows:

\begin{algorithm}
    Arrange the data $\MA{x}$ into matrices
    \begin{equation}
        X = \MB{x}, \quad Y = \MC{x}
    \end{equation}
    
    Compute the truncated SVD of $X$
    \begin{equation}
        X_{m\times n} = U_{m\times r}S_{r\times r}{V}^*_{r\times n}
    \end{equation}
    
    Define the matrix
    \begin{equation}
        \tilde A = U^*YVS^{-1}
    \end{equation}
    
    Compute the eigenvalues and eigenvectors of $\tilde{A}$
    \begin{equation}
        \tilde{A}\tilde{\varphi}_i = \lambda_i\tilde{\varphi}_i
    \end{equation}
    
    Compute the full DMD modes of the system
    \begin{equation}
        \varphi_i = U \tilde{\varphi}_i
    \end{equation}
    
    (Optional) Compute the initial amplitudes of the DMD modes \cite{jovanovicSparsitypromotingDynamicMode2014}
    
\caption{Default DMD}
\end{algorithm}

\subsection{Issues with the default algorithm}

The default DMD algorithm is susceptible to noise in the data in that the noise introduces errors in the generated eigenvalues. We notice that for our system and several benchmarks (other systems), if the magnitudes of the expected eigenvalues are close to unity, i.e. they lie or almost lie on the unit circle in the complex plane, then for even relatively high signal-to-noise ratio (SNR) the noise can somewhat reduce the eigenvalue magnitudes and the associated decay rates become very high, quickly damping the modes to near-zero amplitudes.

The SVD algorithm processes all system state data simultaneously. If the input dataset is very large then SVD will require large amounts of computer memory to be feasible. This can make it impossible to process datasets from high-resolution numerical simulations and experiments on computers without significant memory resources.

To mitigate this, we propose an approach that combines the correlation of multiple adjacent (in time) system states with a streaming (online) SVD algorithm.

\section{Proposed DMD implementation}

\subsection{Eigenfrequency \& eigenmode computation}

Equation (\ref{eq:x_pairs}) can be generalised to include more than one preceding system state:

\begin{equation}
    x_{k} = A_{1}x_{k-1} + A_{2}x_{k-2} + \cdots + A_{d}x_{k-d}
    \label{eq:x_pairs_multiple}
\end{equation}

This approach correlates successive system states with multiple preceding states, which can be expressed as

\begin{equation}
    x'_{k+1} = K x'_{k}
\end{equation}

where

\begin{equation}
    x'_{k} =
    \begin{pmatrix}
        x_{k}\\
        x_{k-1}\\
        \cdots \\
        x_{k-d+1}
    \end{pmatrix},\quad 
    K=
    \begin{pmatrix}
        \bm{A}_{1} & \bm{A}_{2} & \bm{A}_{3} & \cdots  & \bm{A}_{d-1} & \bm{A}_{d}\\
        \bm{I} & \bm{0} & \bm{0} & \cdots  & \bm{0} & \bm{0}\\
        \bm{0} & \bm{I} & \bm{0} & \cdots  & \bm{0} & \bm{0}\\
        \cdots  & \cdots  & \cdots  & \cdots  & \cdots  & \cdots \\
        \bm{0} & \bm{0} & \bm{0} & \cdots  & \bm{I} & \bm{0}
    \end{pmatrix}
\end{equation}

and $\bm{I}$ and $\bm{0}$ are $n \times n$ unit and zero matrices, where $n$ is the dimension of $x_i$. Essentially, (\ref{eq:x_pairs_multiple}) is a linear combination of overlapping system state propagations as in (\ref{eq:propagation-equation}). This approach was first proposed in \cite{leclaincheHigherOrderDynamic2017} to examine systems with a low number of spatial dimensions, but is also applicable to systems with a higher amount of spatial dimensions. The overlap implies that $x_k$ is given by a "moving average" of the preceding $d$ states, which serves to filter the noise contained within system state snapshots. This strategy was, for instance, used in \cite{bruntonExtractingSpatialtemporalCoherent2014, shiraishiNeuralDecodingElectrocorticographic2020}.

To introduce this effect, the system state data must be represented ("stacked") appropriately. Let $X_i$ be the following

\begin{equation}
    X_i = \Mi{x}
    \label{eq:Xi}
\end{equation}

Similarly to (\ref{eq:state-matrices}), (\ref{eq:propagation-equation}) the system is expressed as

\begin{equation}
    KM_0 = M_1
\end{equation}

where

\begin{equation}
    M_0 =
    \begin{bmatrix}
    X_{d-1}\\
    X_{d-2}\\
    \vdots \\
    X_{0}
    \end{bmatrix}, \quad
    M_1 = 
    \begin{bmatrix}
    X_{d}\\
    X_{d-1}\\
    \vdots \\
    X_{1}
    \end{bmatrix}
    \label{eq:M01}
\end{equation}

Here matrices $M_0$, $M_1$ are constructed by vertically stacking time-shifted sequences $X_i$ of system states. This means that a single column of $M$ contains information from $d$ different columns of $X$. Thus, instead of correlating only neighbouring snapshots as in (\ref{eq:propagation-equation}), $d$ system states are covered by a moving correlation window. This approach adds extra noise robustness to DMD because larger "correlation radius" for data effectively averages out the noise, if any. We observe that this method can significantly increase the accuracy of the DMD eigenvalues for data with low SNR. Equation (\ref{eq:x_pairs_multiple}) can be interpreted as a discretized linear differential equation of an order up to $d$.

Much like SVD was performed for $X_0$ in the default DMD algorithm, here SVD is performed for $M_0$. Then both $M_0$ and $M_1$ are projected onto the subspace of $U$:

\begin{equation}
    \tilde{M}_0 = U^*M_0, \quad
    \tilde{M}_1 = U^*M_1
\end{equation}

Next the time-forward and time-backward  components of $\tilde{K}$ are computed:

\begin{equation}
    \begin{split}
    \tilde{K}_{+} = \tilde{M}_1\tilde{M}_0^+ \\
    \tilde{K}_{-} = \tilde{M}_0\tilde{M}_1^+
    \end{split}
    \label{eq:Kfb}
\end{equation}

where $\tilde{M}_0^+$, $\tilde{M}_1^+$ are the pseudo-inverses of $\tilde{M}_0$, $\tilde{M}_1$. This allows to compute $\tilde{K}$ (similarity-transformed $K$):

\begin{equation}
    \tilde{K} = \sqrt{\tilde{K}_{+}\tilde{K}_{-}^{-1}}
    \label{eq:combo-propagator}
\end{equation}

While one can also simply set $\tilde{K}=\tilde{K}_{+}$, (\ref{eq:combo-propagator}) yields a minor increase in accuracy for next to no computational cost and is therefore worth implementing \cite{dawsonCharacterizingCorrectingEffect2016}. 

Finally, the eigenvalues of $\tilde{K}$ are computed:

\begin{equation}
    \tilde{K}\tilde{\varphi}_i = \lambda_i\tilde{\varphi}_i
\end{equation}

and the eigenvectors in the original basis are recovered via

\begin{equation}
    \varphi_i = U \tilde{\varphi}_i
\end{equation}

To obtain the modes of $X$ one simply truncates the modes of $M_0$ to the first $m$ elements (vertically), where $m$ is the length of $X$ columns. While one might argue that much of the system information is lost this way, since $M_0$, $M_1$ contain redundant information regarding $X$ due to repeated time-shifted stacking construction, most of the total information can be inferred from the first non-repeating elements. Note also that the partial copies of $X$ evolve in time identically differing only in phase.

To summarize, the proposed approach is as follows:

\begin{algorithm}[H]
    Arrange the data $\MA{x}$ into matrices
    \begin{equation}
        X_i = \Mi{x}
    \end{equation}
    
    Construct matrices $M_0$, $M_1$
    \begin{equation}
    M_0 =
    \begin{bmatrix}
        X_{d-1}\\
        X_{d-2}\\
        \vdots \\
        X_{0}
        \end{bmatrix}, \quad
        M_1 = 
        \begin{bmatrix}
        X_{d}\\
        X_{d-1}\\
        \vdots \\
        X_{1}
        \end{bmatrix}
        \label{eq:M01}
    \end{equation}
    
    Compute the truncated SVD of $M_0$
    \begin{equation}
        M_{m\times n} = U_{m\times r}S_{r\times r}{V}^*_{r\times n}
    \end{equation}
    
         Compute matrices $\tilde{M}_0^+$, $\tilde{M}_1^+$
    \begin{equation}
        \tilde{M}_0 = U^*M_0, \quad
        \tilde{M}_1 = U^*M_1
    \end{equation}
    \caption{Custom DMD}

\end{algorithm}

\clearpage

\begin{algorithm}[H]
\setcounter{AlgoLine}{4}
  Compute matrices $\tilde{K}_{+}$, $\tilde{K}_{-}$
    \begin{equation}
        \begin{split}
        \tilde{K}_{+} = \tilde{M}_1\tilde{M}_0^+ \\
        \tilde{K}_{-} = \tilde{M}_0\tilde{M}_1^+
        \end{split}
    \end{equation}

    Compute $\tilde{K}$
    \begin{equation}
        \tilde{K} = \sqrt{\tilde{K}_{+}\tilde{K}_{-}^{-1}}
    \end{equation}

    Compute the eigenvalues and eigenvectors of $\tilde{K}$
    \begin{equation}
        \tilde{K}\tilde{\varphi}_i = \lambda_i\tilde{\varphi}_i
    \end{equation}
    
    Compute the full DMD modes of the system
    \begin{equation}
        \varphi_i = U \tilde{\varphi}_i
    \end{equation}
    
    Truncate the modes $\varphi_i$ to the first $m$ elements.
    
    (Optional) Compute the initial amplitudes of the DMD modes \cite{jovanovicSparsitypromotingDynamicMode2014}
\end{algorithm}

It is important to note that (\ref{eq:combo-propagator}) does not have a unique solution. To address this, it is recommended to choose a solution that is closest to $\tilde{K}_+$ and $\tilde{K}_-^{-1}$ as the matrix $\tilde{K}$ is expected to be close to these matrices. An alternative approach is to compute $\tilde{K}$ as follows:

\begin{equation}
    \tilde{K} = \frac{1}{2}\qty(\tilde{K}_{+} + \tilde{K}_{-}^{-1})
\end{equation}

It is important to note that there exists a closely related algorithm known as HAVOK \cite{bruntonChaosIntermittentlyForced2017} which uses the same representation as (\ref{eq:M01}) and also uses SVD to extract structures out of the data. HAVOK is closely related to DMD and is capable of extracting the dynamics of highly nonlinear systems.

\subsection{Data pre-processing}
\label{sec:complexify}

DMD decomposes the system into complex oscillating modes:

\begin{equation}
    f'(t) = \sum_{k} C_k \varphi_{k} e^{iz_{k}t} = \sum_{k} C_k \varphi_{k} e^{a_{k}t}e^{ib_{k}t}
\end{equation}

where $C_k \in \mathbb{C}$. In practice, most systems of interest are strictly real-valued:

\begin{equation}
    f(t) = \sum_{k} C'_k \varphi_{k} e^{a_{k}t}\cos\qty(b_{k}t + \phi_k)
\end{equation}


where $C'_k \in \mathbb{R}$. This means that each real mode will be described with two complex modes with conjugate eigenvalues, which effectively makes half of the generated modes redundant.
To address this one can transform the real-valued input into a complex valued input by adding an imaginary $\pi/2$ phase-shifted version of the original input. This ensures that all of the calculated modes are unique. Although this method introduces numerical artefacts at the edges (with respect to time) of system state stacks, it is compensated for by defining cutoff buffers for analyzed datasets. One must be careful, however, since in this case the imaginary mode components do not always "mirror" the real parts.

The above is easily achieved using the Hilbert transform (next page):

\clearpage

\begin{algorithm}[H]
    Compute the DFT ($\mathcal{F}$) of $f\qty(t)$ 
    \begin{equation}
        \tilde{f}\qty(\omega) = \mathcal{F}\qty{f\qty(t)}
    \end{equation}
    
    Introduce a sign function $\theta\qty(\omega)$
    \begin{equation}
        \theta\qty(\omega)=
        \begin{cases}
            1, & \omega \geq 0\\
            -1, & \omega < 0
        \end{cases}
    \end{equation}
    
    Set the Fourier representation of the phase-shifted $f\qty(t)$
    \begin{equation}
        \tilde{g}\qty(\omega) = \tilde{f}\qty(\omega) \theta\qty(\omega)
    \end{equation}
    
    Compute the complex-valued version of $f\qty(t)$ via the inverse DFT of the sum of $\tilde{f}\qty(\omega)$ and $\tilde{g}\qty(\omega)$:
    \begin{equation}
        f'\qty(t) = \mathcal{F}^{-1}\qty{\tilde{f}\qty(\omega) + \tilde{g}\qty(\omega)}
    \end{equation}
    \caption{Real-to-complex domain mapping.}
\end{algorithm}

With this, an imaginary $\pi/2$ phase-shifted copy is added to each real mode.

\clearpage

\section{Results}

Simulations were run for $20~ s$ of flow time for all 4 cases (Section \ref{sec:system-and-cases}), which is sufficient for the system to reach a quasi-stationary state starting from initially zero/stationary gas and fluid velocity fields, i.e. initial conditions are identical to \cite{birjukovsArgonBubbleFlow2020, birjukovsPhaseBoundaryDynamics2020}.

\subsection{Bubble chain flow patterns}

\begin{figure}[H]
    \centering
    \includegraphics[width=0.90\textwidth]{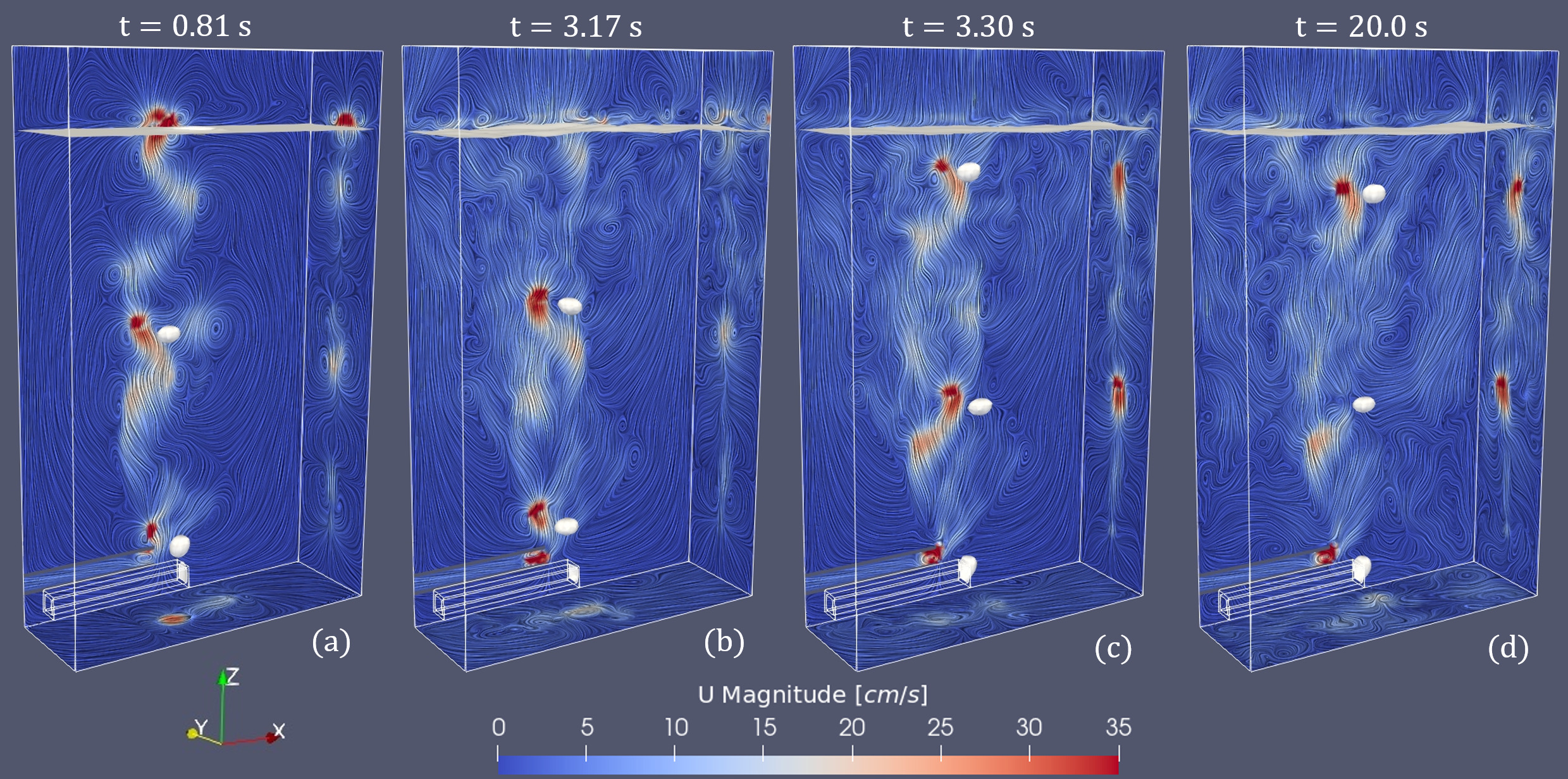}
    \caption{Velocity LIC plots for bubble flow at $30~ sccm$ flow rate without applied MF.}
    \label{fig:vessel-lic-30-sccm-off}
\end{figure}

\begin{figure}[H]
    \centering
    \includegraphics[width=0.90\textwidth]{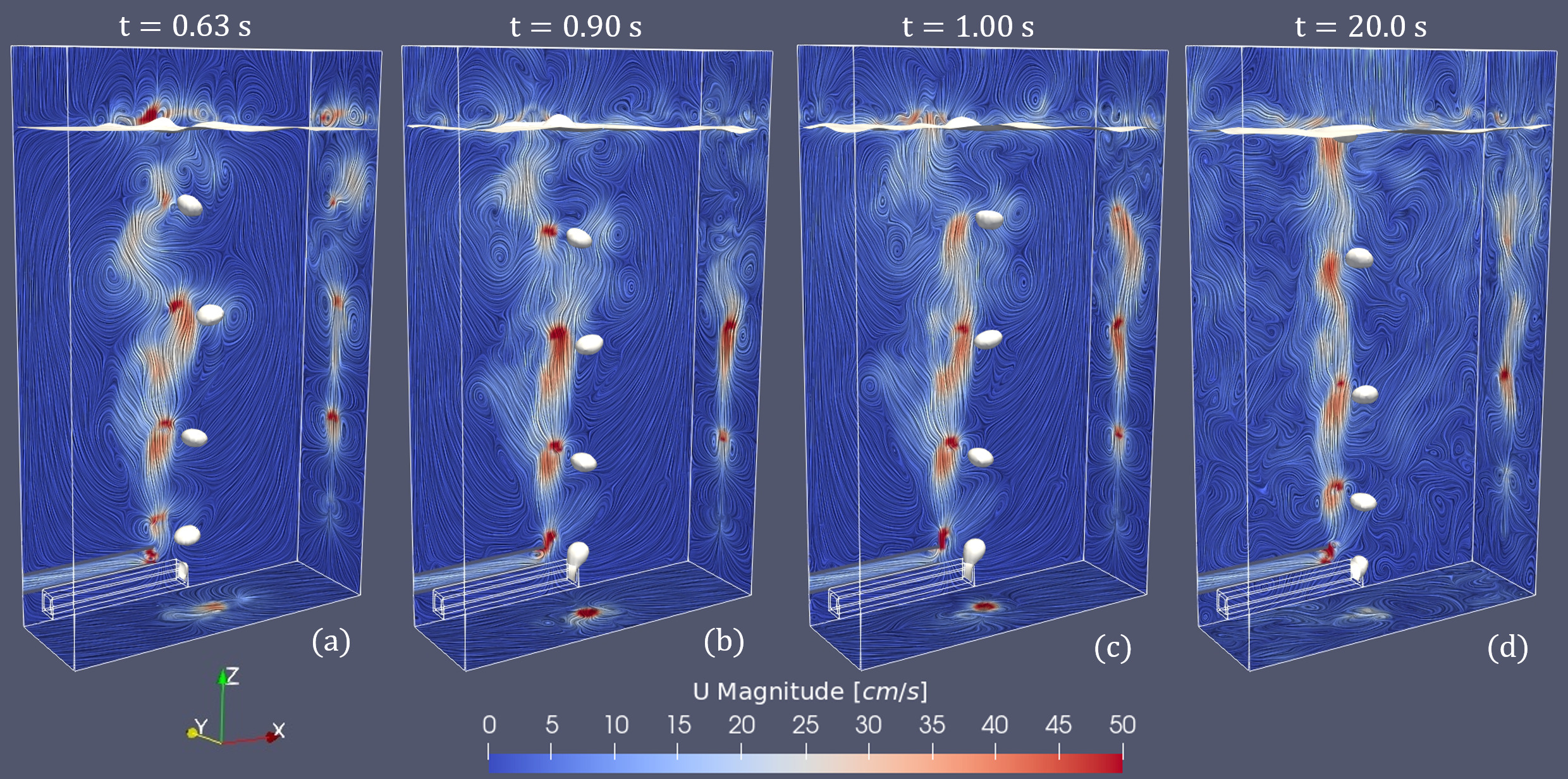}
    \caption{Velocity LIC plots for bubble flow at $100~ sccm$ flow rate without applied MF.}
    \label{fig:vessel-lic-100-sccm-off}
\end{figure}

\begin{figure}[H]
    \centering
    \includegraphics[width=0.46\textwidth]{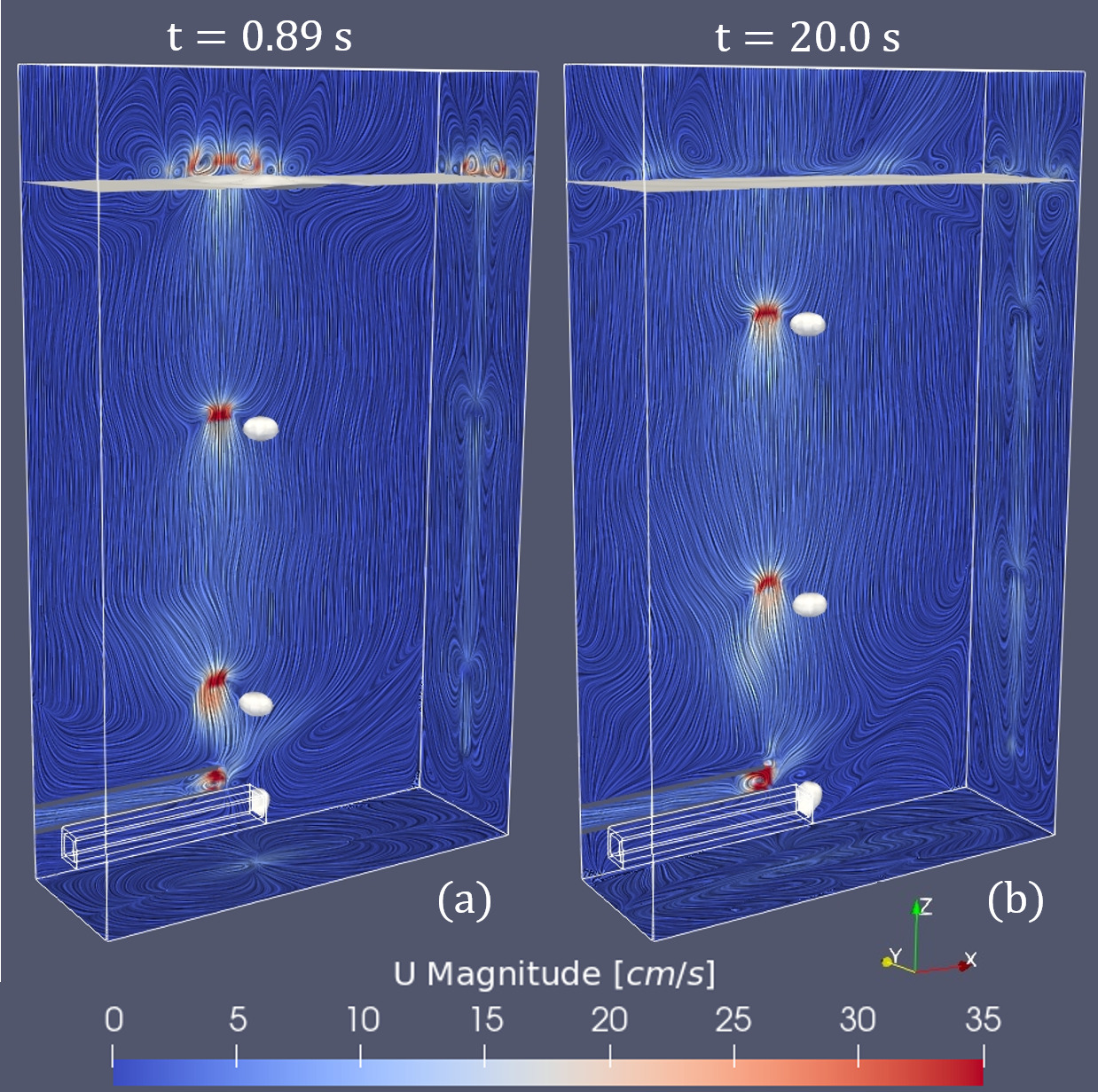}
    \caption{Velocity LIC plots for bubble flow at $30~ sccm$ flow rate with applied MF.}
    \label{fig:vessel-lic-30-sccm-on}
\end{figure}

\begin{figure}[H]
    \centering
    \includegraphics[width=0.46\textwidth]{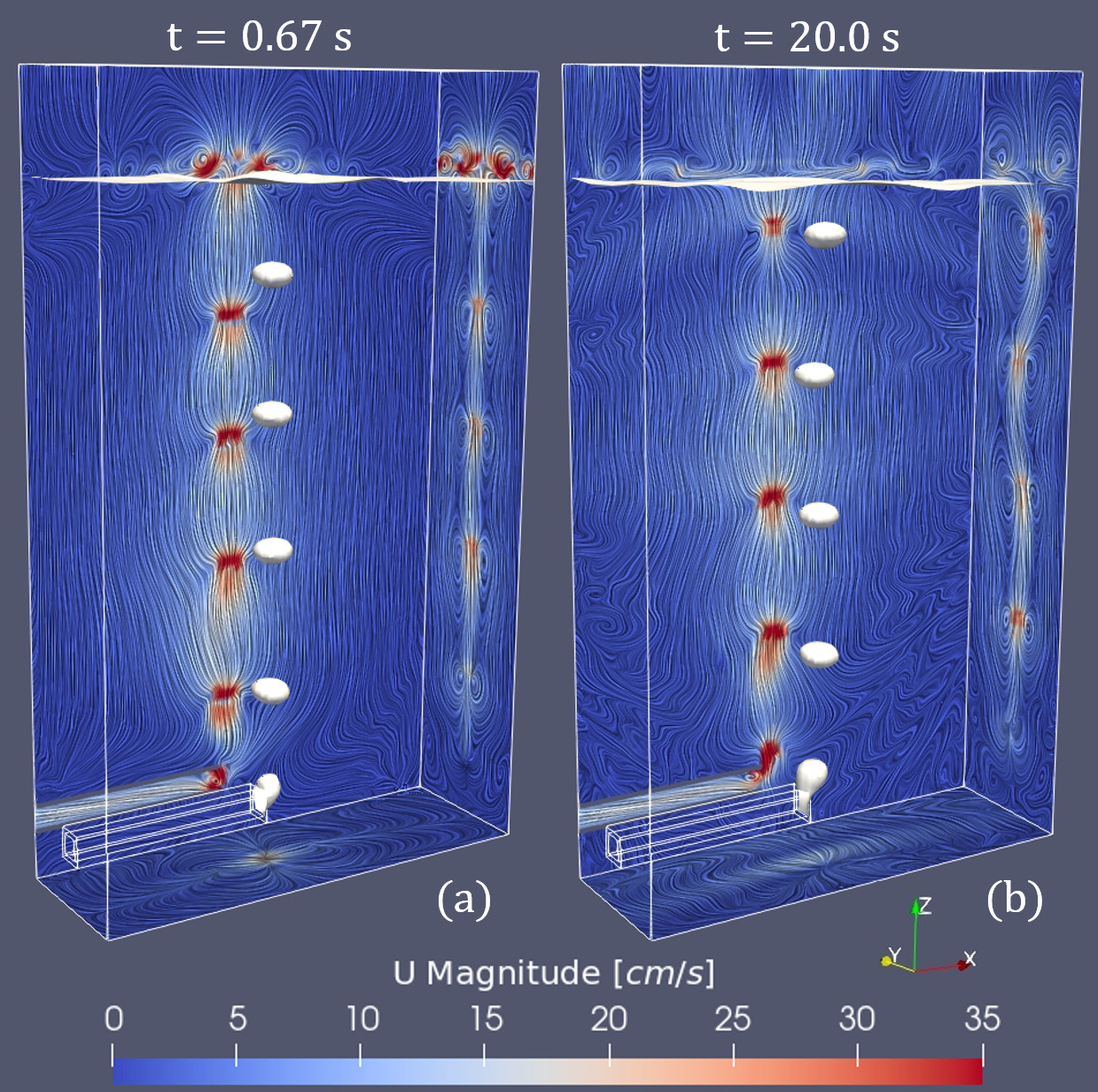}
    \caption{Velocity LIC plots for bubble flow at $100~ sccm$ flow rate with applied MF.}
    \label{fig:vessel-lic-100-sccm-on}
\end{figure}

To provide context for the DMD analysis, we first observe the velocity field and vortex patterns. Characteristic snapshots of velocity fields at different time stamps for the 4 cases considered here are show in Figures \ref{fig:vessel-lic-30-sccm-off}-\ref{fig:vessel-lic-100-sccm-on}. Color maps of the midplane velocity magnitude are computed on the vessel walls and are then processed with the \textit{ParaView} plugin \textit{Surface LIC}, which uses the screen space surface line integral convolution (LIC) to transform the color maps into streamline maps with color-coded velocity magnitude \cite{paraview-surface-lic}. Shaded white-gray overlays indicate bubble interfaces and the free metal surface at the top of the container. The rectangular tube at the bottom of the vessel is the argon inlet.

\clearpage

If no MF is applied, bubble flow exhibits a classic vortex shedding pattern \cite{prl-path-instability, shape-and-wake-simulations, hzdr-bubbles-mf, hzdr-ibm-bubbles-thesis} especially evident at earlier time stamps in Figures \ref{fig:vessel-lic-30-sccm-off}a-\ref{fig:vessel-lic-30-sccm-off}c (\textit{XZ} plane) which later becomes much more disordered, as seen in Figure \ref{fig:vessel-lic-30-sccm-off}d. Note that in the case with $100~ sccm$ without applied MF, larger vortices with greater velocity are shed (Figures \ref{fig:vessel-lic-100-sccm-off}a-c) and, aside from the obvious global increase in velocity magnitude, one can see that the backflow from the free surface and walls is more pronounced near the bottom of the container, although it is difficult to tell from these plots alone how much further down this mixing layer extends at $30$ versus $100~ sccm$. The objective of the DMD analysis will be to determine if there are any special patterns in both of these cases aside from the trivial mean upward flow within the bubble chain and how these patterns differ for the two flow rates.

Applying static horizontal MF results in nearly complete flow laminarization for $30~sccm$ (Figure \ref{fig:vessel-lic-30-sccm-on}) and significant damping for $100~sccm$ (Figure \ref{fig:vessel-lic-100-sccm-on}), since vortex shedding is suppressed as expected \cite{hzdr-bubbles-mf, zhang-mf-simulations, imb-transverse-field, hzdr-ibm-bubbles-thesis}. Note the \textit{YZ} planes in Figure \ref{fig:vessel-lic-30-sccm-on} where the LIC plot hints that bubbles ascend via rectilinear trajectories.

\clearpage

There is an important distinction in the case of $100~sccm$ -- notice that initially rectilinear ascension, as seen in Figure \ref{fig:vessel-lic-100-sccm-on}a, later transitions to a different pattern seen in Figure \ref{fig:vessel-lic-100-sccm-on}b. In the \textit{XZ} plane, standing velocity magnitude waves form after the first $\sim 10~ s$ of flow time while in the \textit{XY} plane one can see that the trajectory becomes slightly oscillatory in the upper half of the vessel, remaining roughly rectilinear in the lower half. DMD will be used to analyze these patterns in detail.

\begin{figure}[H]
    \centering
    \includegraphics[width=0.96\textwidth]{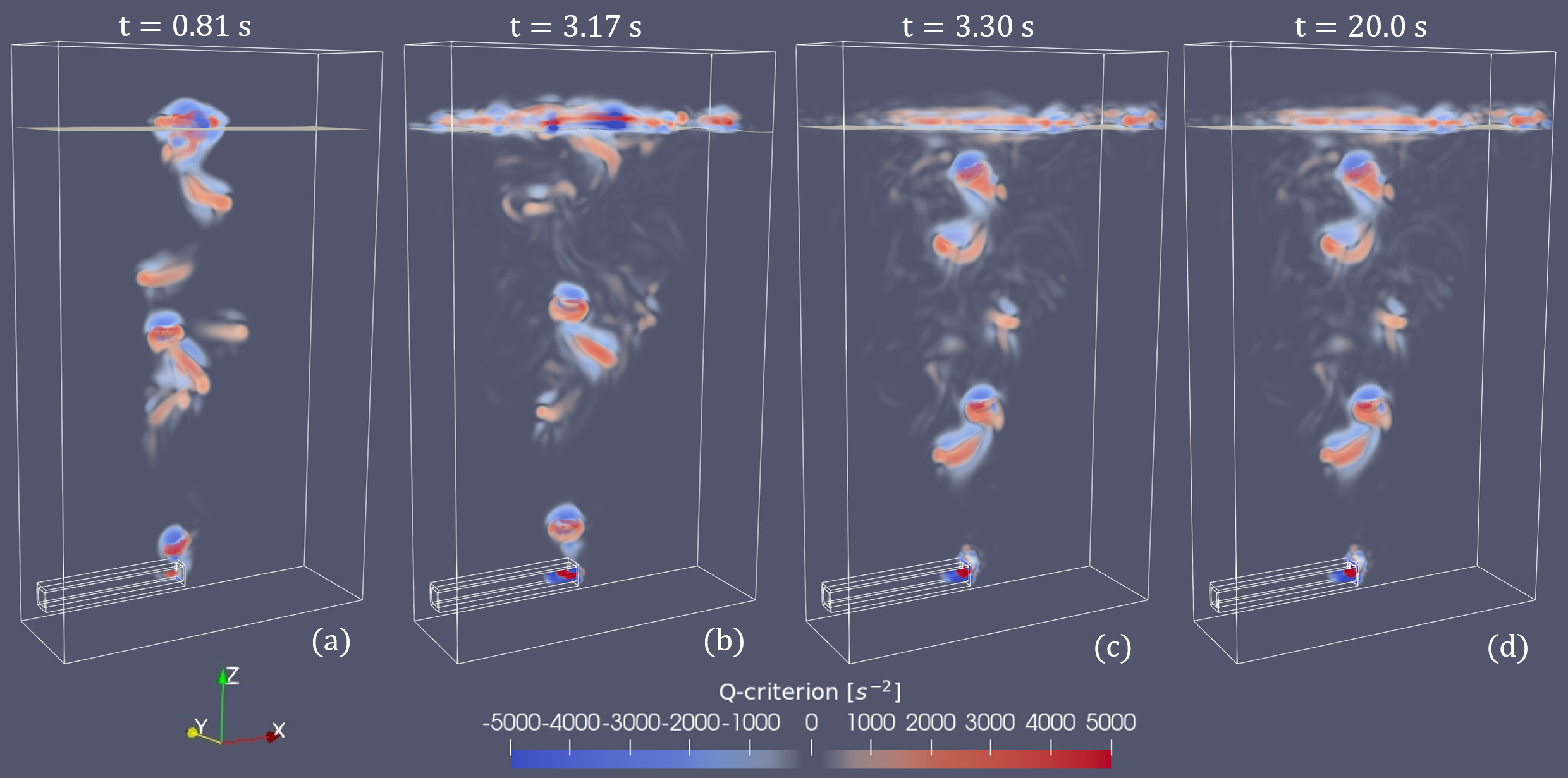}
    \caption{$Q$ plots for bubble flow at $30~ sccm$ flow rate without applied MF.}
    \label{fig:vessel-q-30-sccm-off}
\end{figure}

\begin{figure}[H]
    \centering
    \includegraphics[width=0.96\textwidth]{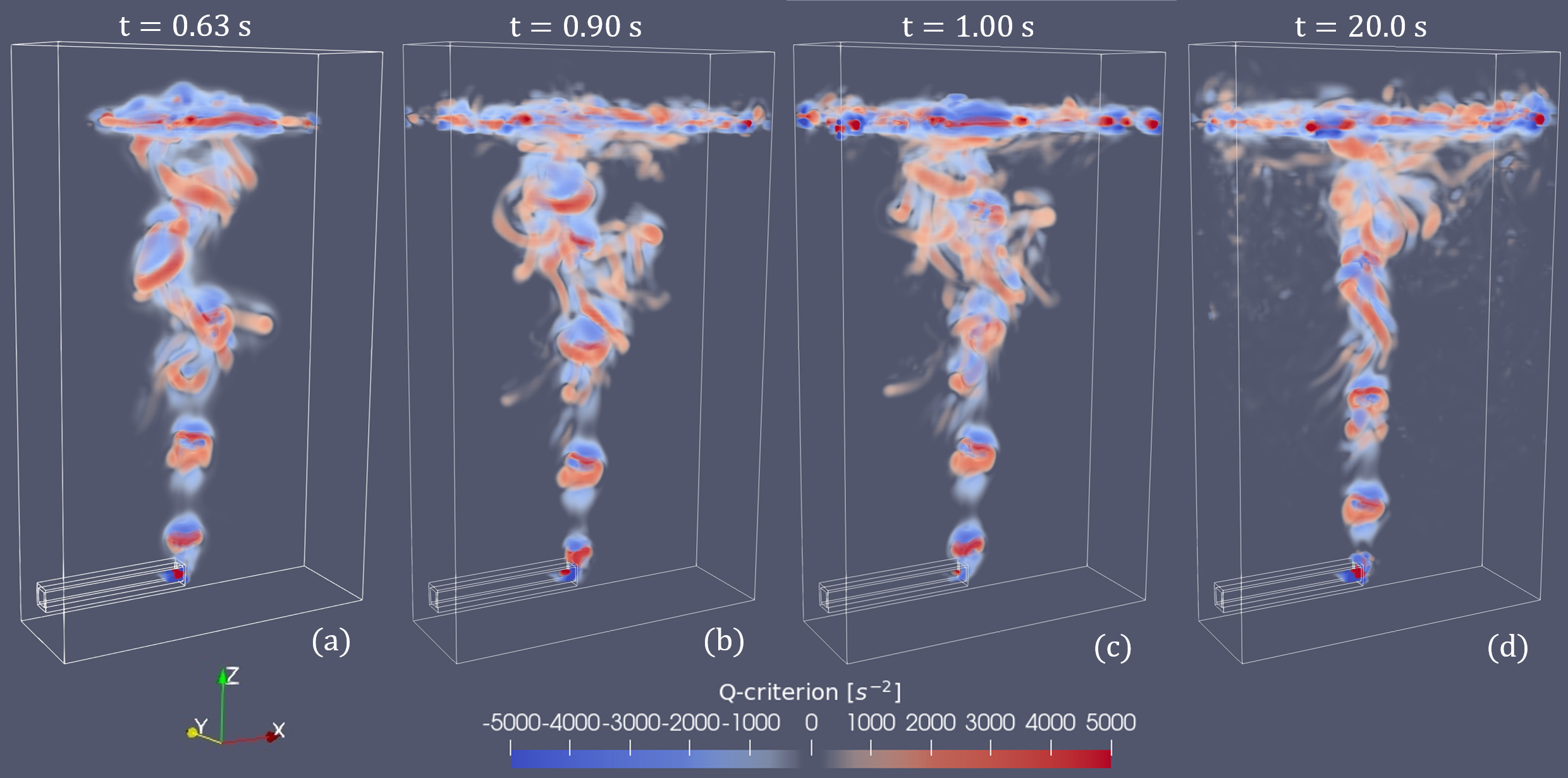}
    \caption{$Q$ plots for bubble flow at $100~ sccm$ flow rate without applied MF.}
    \label{fig:vessel-q-100-sccm-off}
\end{figure}

\begin{figure}[H]
    \centering
    \includegraphics[width=0.47\textwidth]{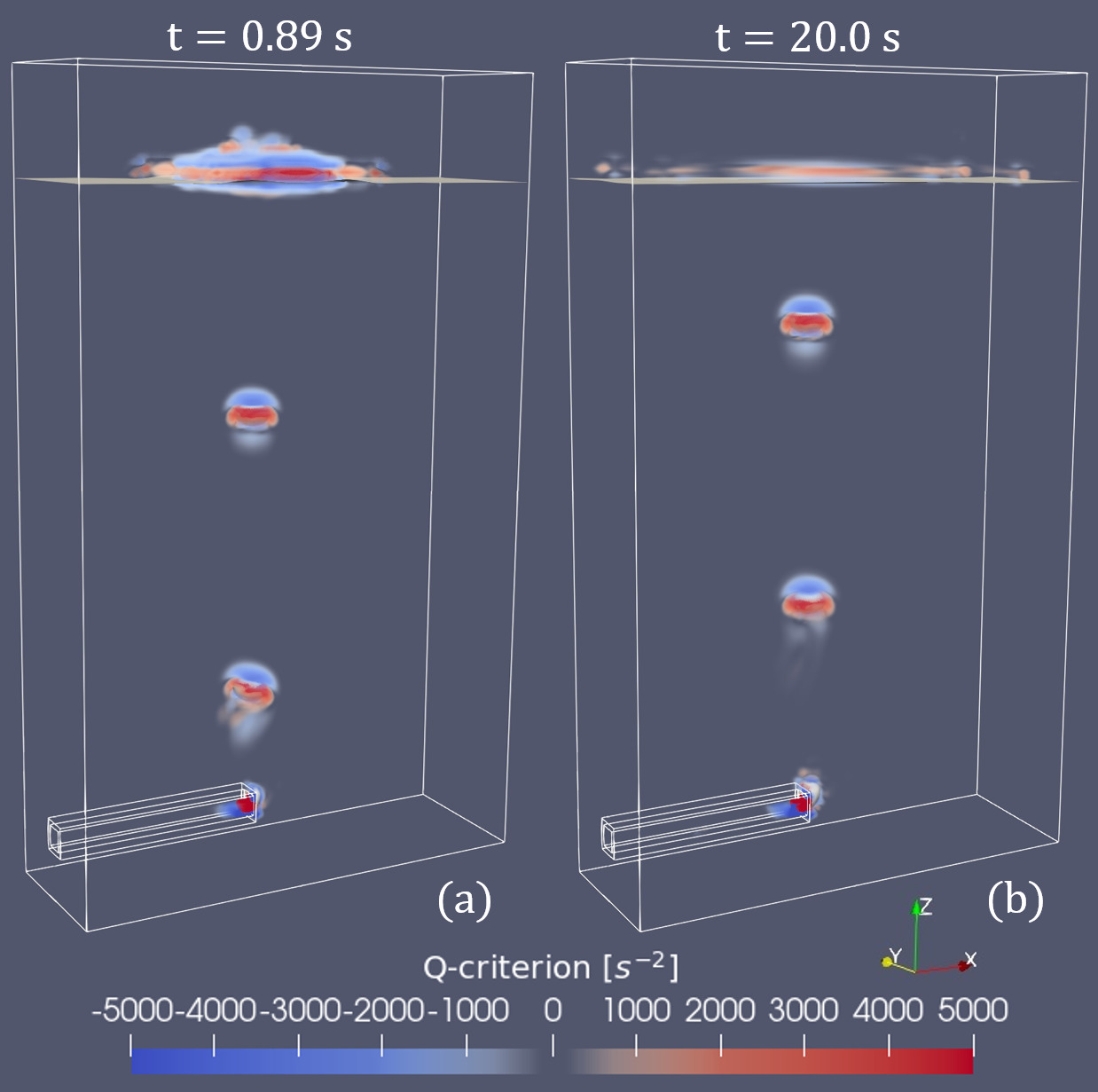}
    \caption{$Q$ plots for bubble flow at $30~ sccm$ flow rate with applied MF.}
    \label{fig:vessel-q-30-sccm-on}
\end{figure}

\begin{figure}[H]
    \centering
    \includegraphics[width=0.47\textwidth]{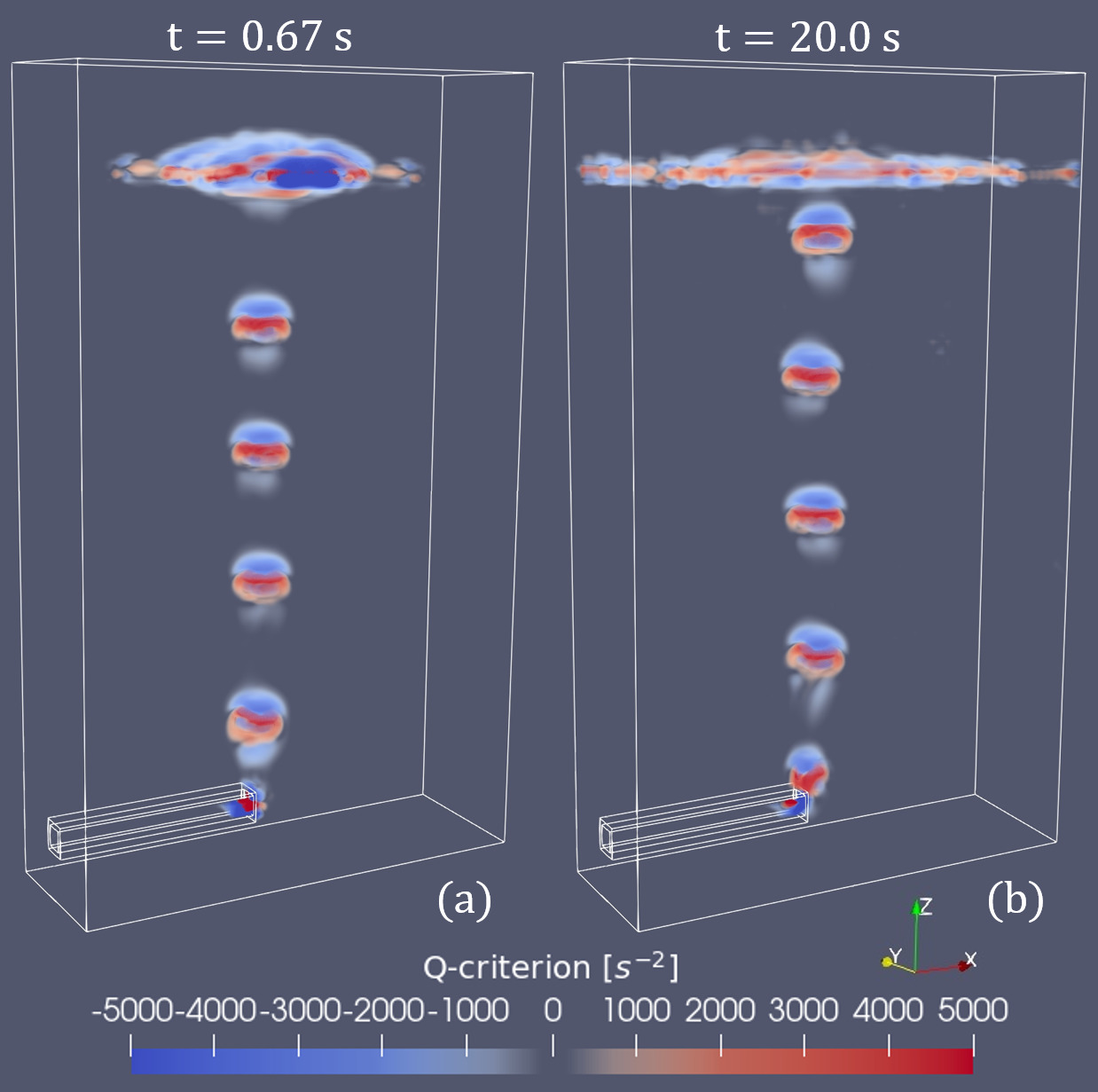}
    \caption{$Q$ plots for bubble flow at $100~ sccm$ flow rate with applied MF.}
    \label{fig:vessel-q-100-sccm-on}
\end{figure}

To assess the flow closer to the bubbles and the bubble chain, it is convenient to use the Q-criterion for vortex detection since bubble wake flow is what determines the trajectories and collective dynamics. The Q-criterion ($Q$ for brevity) represents both vortex cores ($Q>0$) and saddle pattern flow zones ($Q<0$) and is therefore well-suited for assessing the effects of varying the flow rate and MF magnitude \cite{q-factor-og}. Figures \ref{fig:vessel-q-30-sccm-off}-\ref{fig:vessel-q-100-sccm-on} show the volume rendering of $Q$ for the above treated cases. $Q$ is computed in \textit{ParaView} using the \textit{VTK gradient of unstructured data set} filter.

For $30~ sccm$, Figure \ref{fig:vessel-q-30-sccm-off} indicates that bubble wakes exhibit a classic cofiguration with pairs of elongated "hairpin" vortices \cite{prl-path-instability, shape-and-wake-simulations, hzdr-ibm-bubbles-thesis}. One can see that indeed, as suggested in Section \ref{sec:system-and-cases}, $30~ sccm$ corresponds to a quasi single-bubble regime, as trailing bubbles do not run into pronounced vortices behind leading bubbles owing to sufficient spacing that results in vortex decay/relaxation before their collisions with bubbles can take place. This is especially the case when flow stabilizes (Figures \ref{fig:vessel-q-30-sccm-off}b-c -- note the time stamps: flow stabilization occurs much faster than for $100~ sccm$) and is only violated a few times at the beginning of the simulation, a good example of which can be seen in Figure \ref{fig:vessel-q-30-sccm-off}a. For $100~sccm$ the situation differs radically -- note how vortex cores in Figures \ref{fig:vessel-q-100-sccm-off} intertwine and collide with trailing bubbles that enter wakes, resulting in a much more disordered flow.

When MF is applied, however, its virtually eliminates vortices outside of the near-bubble zones, which is evident from Figures \ref{fig:vessel-q-30-sccm-on} and \ref{fig:vessel-q-100-sccm-on}. Notice how, while short vortex core trails are visible in Figures \ref{fig:vessel-q-30-sccm-on} and \ref{fig:vessel-q-100-sccm-on} near the inlet, they completely disappear shortly after bubble detachment from the inlet. Note also that wake vortices are visibly fainter for $30~sccm$.

\clearpage

\subsection{Flow modes in the metal vessel}

\begin{figure}[H]
    \centering
    \includegraphics[width=0.67\textwidth]{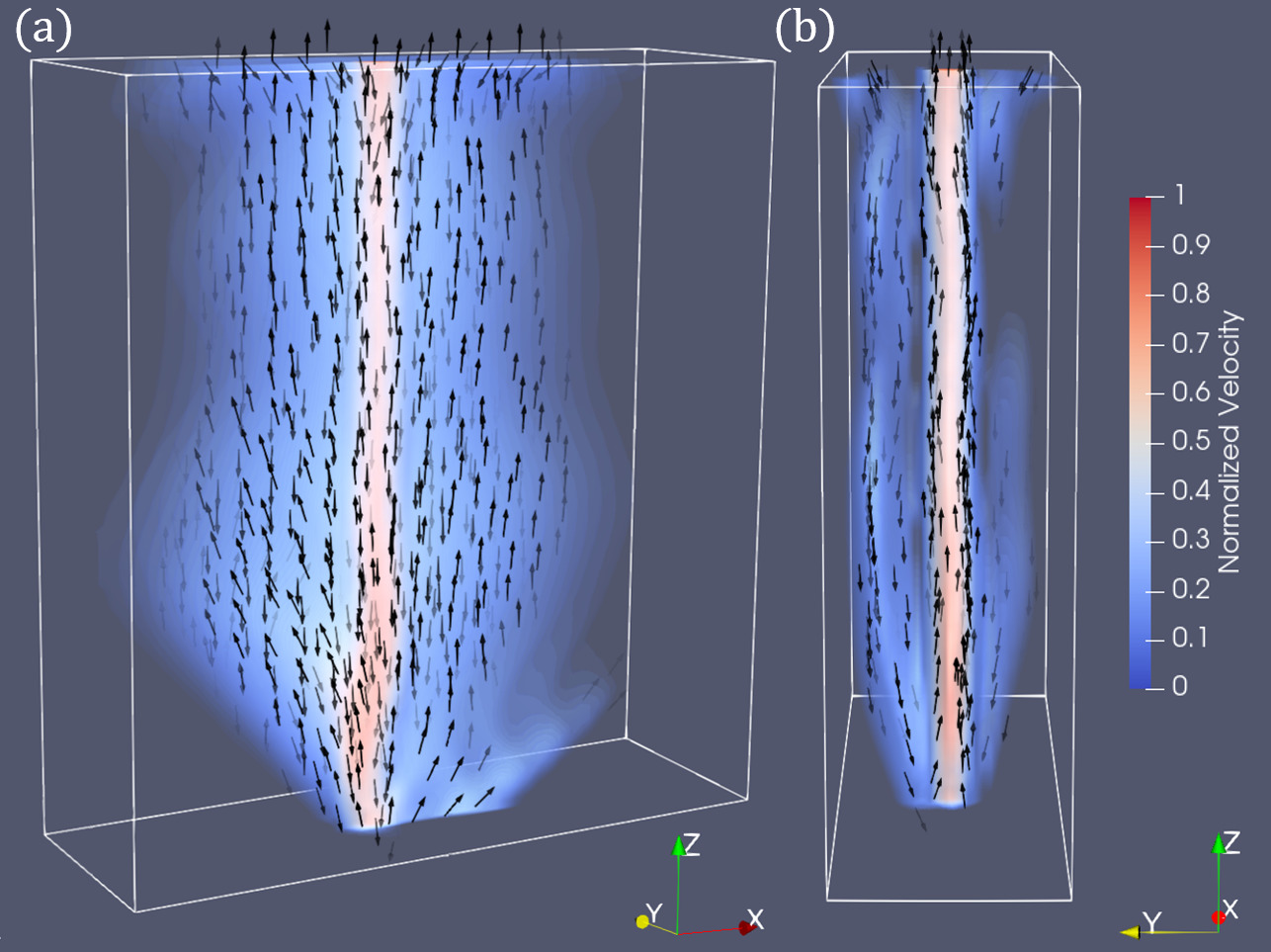}
    \caption{The zeroth DMD velocity field mode for $30~ sccm$ with applied MF.}
    \label{fig:vessel-mode-30-sccm-on-m0}
\end{figure}

\begin{figure}[H]
    \centering
    \includegraphics[width=0.67\textwidth]{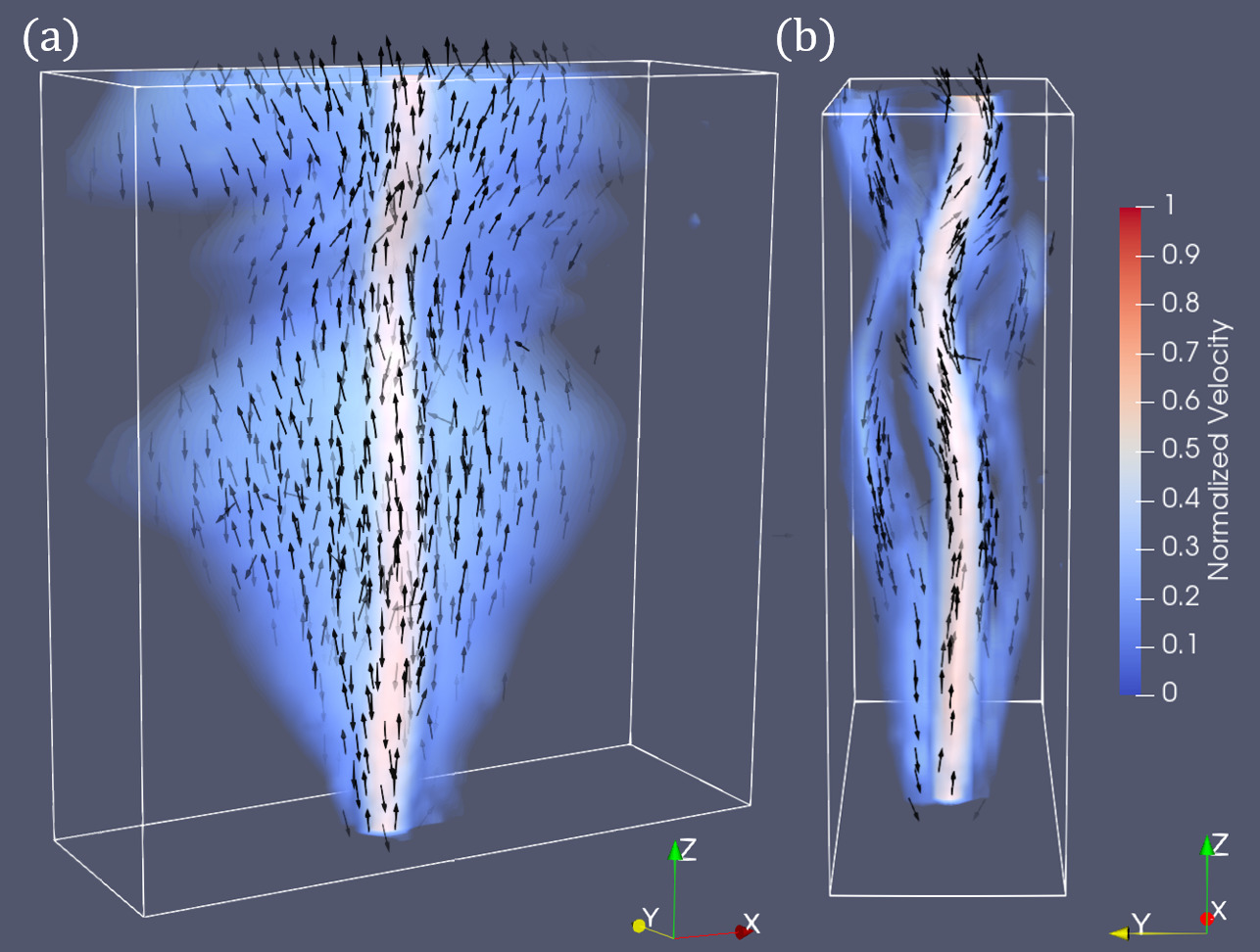}
    \caption{The zeroth DMD velocity field mode for $100~ sccm$ with applied MF.}
    \label{fig:vessel-mode-100-sccm-on-m0}
\end{figure}

\begin{figure}[H]
    \centering
    \includegraphics[width=0.60\textwidth]{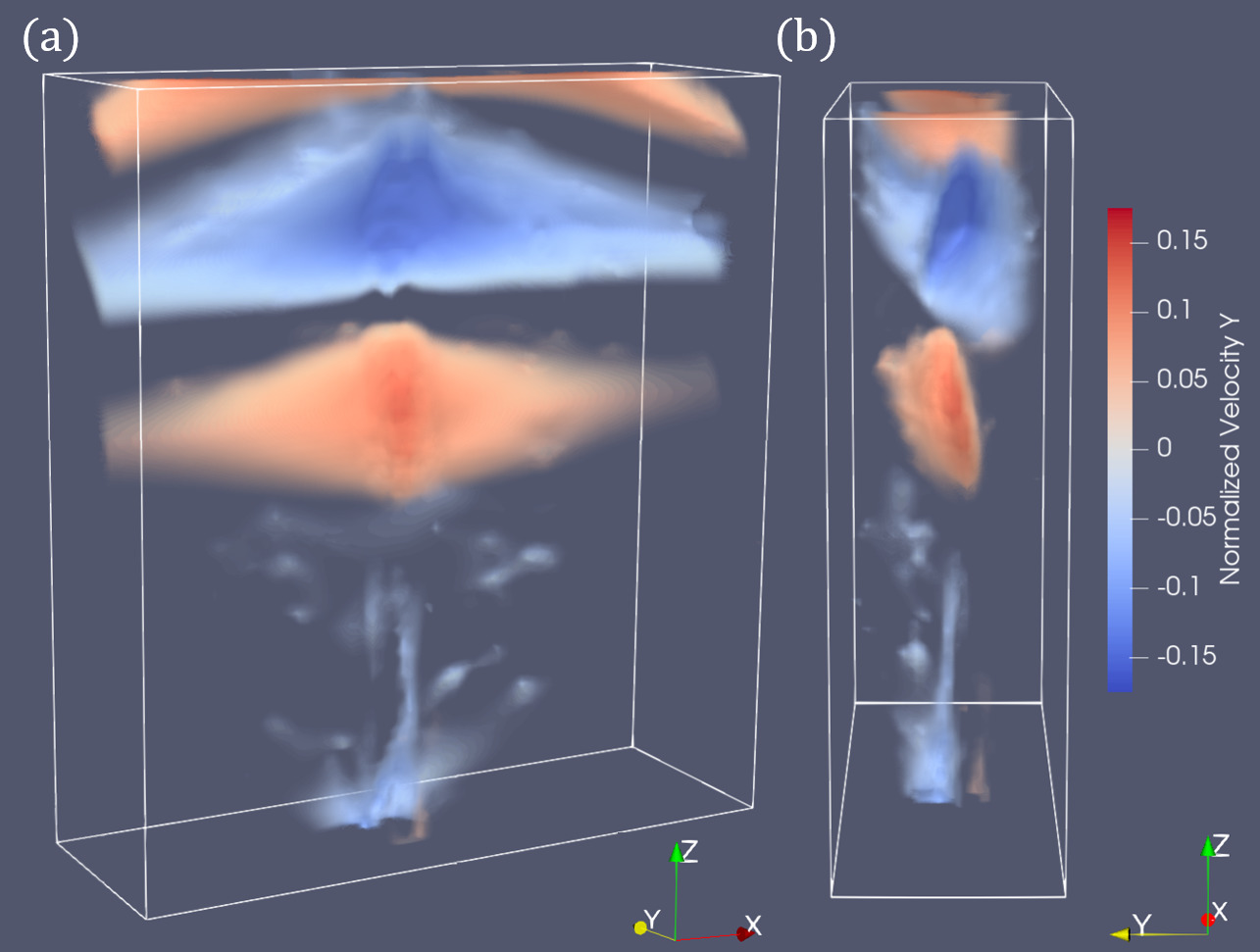}
    \caption{The \textit{Y} component of the zeroth DMD velocity field mode for $100~ sccm$ with applied MF, normalized with respect to the mode magnitude.}
    \label{fig:vessel-mode-100-sccm-on-m0-y}
\end{figure}

\begin{figure}[H]
    \centering
    \includegraphics[width=0.90\textwidth]{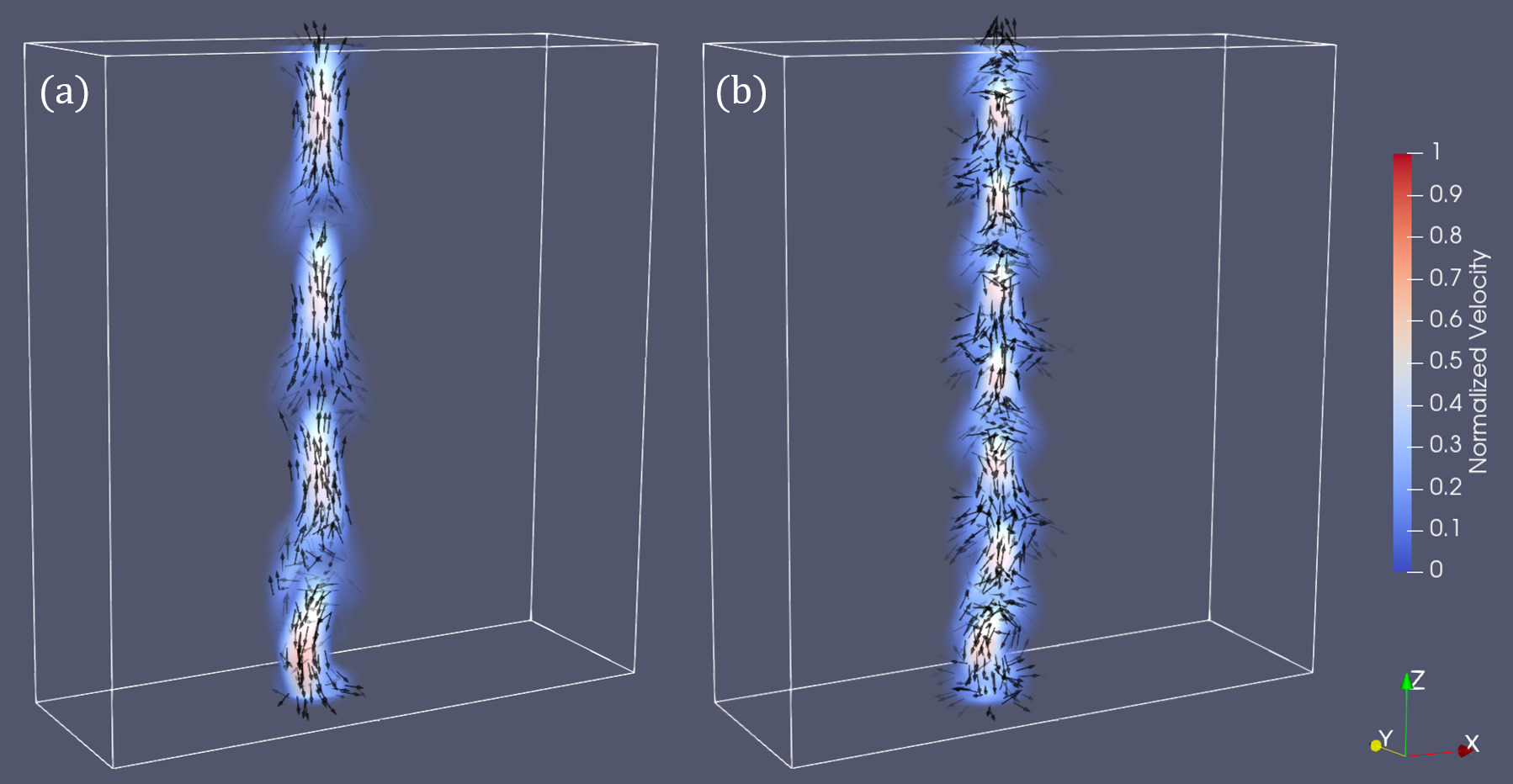}
    \caption{(a) The 1-st and (b) 2-nd DMD velocity field modes for $30~ sccm$ with applied MF.}
    \label{fig:vessel-mode-30-sccm-on-m1-m2}
\end{figure}

To compute the DMD modes for the velocity field in the liquid metal vessel, the last $600$ frames ($6$ seconds) of the simulation output with the argon and the free surfaces cropped out are fed to the real-to-complex mapper after which a $50$-frame buffer is removed from each end of the $600$-frame sampling interval to avoid artefacts (see Section \ref{sec:complexify}). DMD is therefore performed for $500$ system states. The procedure is identical for all of the above cases. Given the flow patterns seen in Figures \ref{fig:vessel-lic-30-sccm-off}-\ref{fig:vessel-q-100-sccm-on} it is expected that DMD modes for cases with applied MF are going to be simpler/more ordered. Therefore it makes sense to start with these as they are easier to interpret and then compare to the cases without applied MF. The dominant modes for $30$ and $100~sccm$ when MF is applied are shown in Figures \ref{fig:vessel-mode-30-sccm-on-m0}-\ref{fig:vessel-mode-30-sccm-on-m1-m2}.

Starting with $30~sccm$ and ranking the modes in terms of relative amplitudes, the zeroth (strongest) velocity field mode is shown in Figure \ref{fig:vessel-mode-30-sccm-on-m0}. This mode has a very low frequency ($\omega_0 \sim 0.15~ mHz$) and a negligible growth rate ($a_0 \sim -1.9 \cdot 10^{-3}~ s^{-1} $), and can be considered stationary. Note the pronounced bubble chain flow region in Figure \ref{fig:vessel-mode-30-sccm-on-m0}a where the maximum normalized velocity is concentrated. Interestingly, as seen in Figure \ref{fig:vessel-mode-30-sccm-on-m0}b, metal flow in the \textit{Y} direction is organized in three sheets -- one with upwards metal flow about the bubble chain region, extending over the \textit{XZ} plane (Figure \ref{fig:vessel-mode-30-sccm-on-m0}a), and two counter-flow sheets to its left and right in the \textit{YZ} plane. This mode exhibits a great deal of symmetry with respect to the \textit{XZ} mid-plane in addition to the \textit{YZ} plane.

For $100~sccm$, on the other hand, while a similar pattern appears for the zeroth mode ($\omega_0 \sim 0.3~ mHz$, $a_0 \sim 1.7 \cdot 10^{-4}~ s^{-1}$), its symmetry in the \textit{YZ} plane is broken in the upper part of the liquid metal vessel as seen in Figure \ref{fig:vessel-mode-100-sccm-on-m0}b, and the counter-flow sheets are disordered. Standing waves in the \textit{Y} component of the mode form in the upper part of the vessel which is illustrated in Figure \ref{fig:vessel-mode-100-sccm-on-m0-y}.

\begin{figure}[H]
    \centering
    \includegraphics[width=0.90\textwidth]{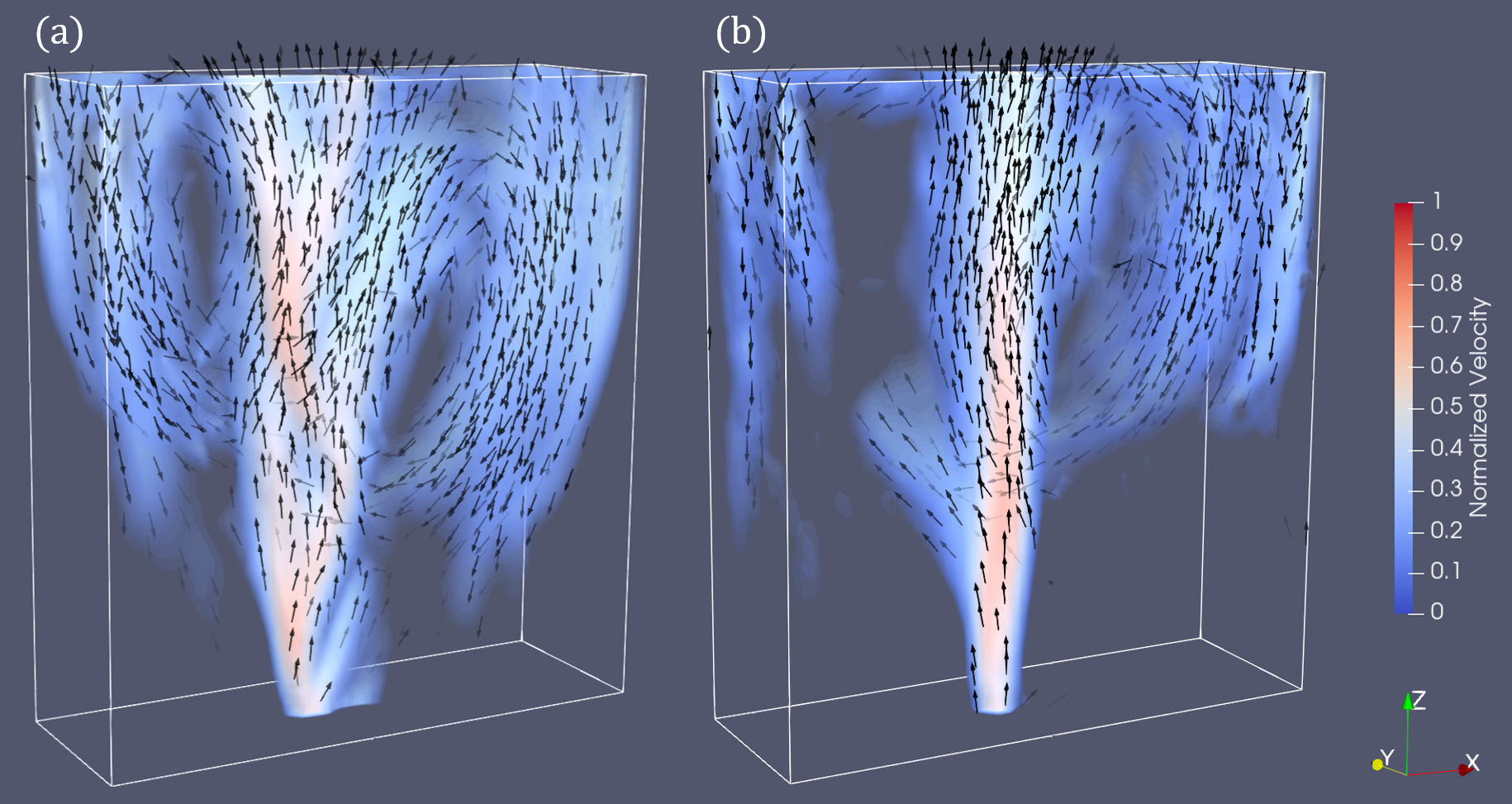}
    \caption{The zeroth (stationary) velocity field modes for (a) $30$ and (b) $100~ sccm$ without applied MF. Normalization is separate for each case.}
    \label{fig:vessel-mode-30-and-100-sccm-off-m0}
\end{figure}

\begin{figure}[H]
    \centering
    \includegraphics[width=0.90\textwidth]{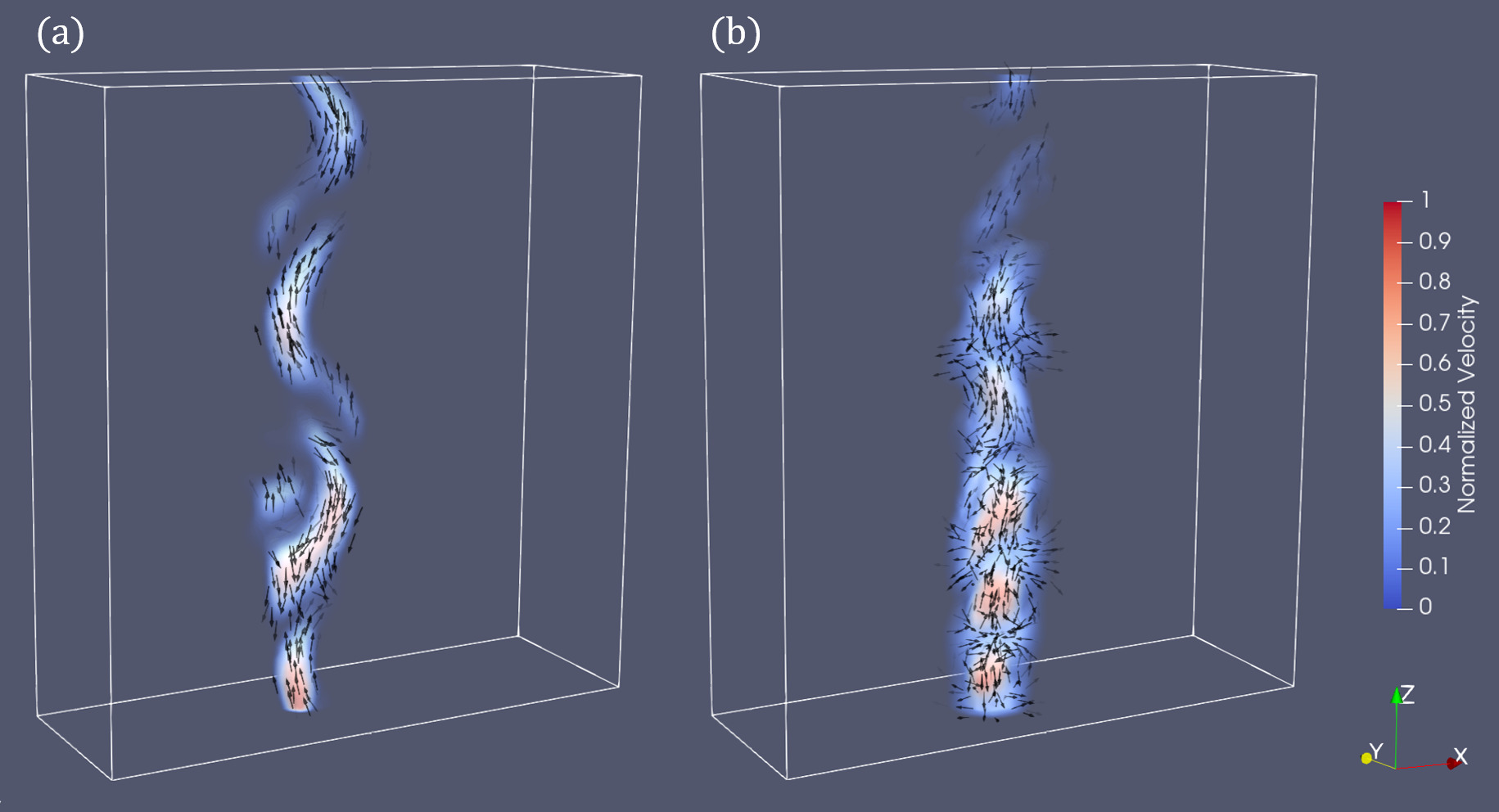}
    \caption{The (a) 2-nd velocity field mode for $30~sccm$ and (b) the 6-th mode for $100~ sccm$ without applied MF. Normalization is separate for each case.}
    \label{fig:vessel-mode-30-m2-100-m6-off}
\end{figure}

The 1-st and 2-nd modes for the velocity field for $30~sccm$ with applied MF are shown in Figures \ref{fig:vessel-mode-30-sccm-on-m1-m2}a and \ref{fig:vessel-mode-30-sccm-on-m1-m2}b. These modes are, unlike the zeroth modes for $30$ and $100~sccm$, non-stationary and their flow patterns oscillate at their respective eigenfreqencies ($\omega_1 \sim 4.2~ Hz$, $\omega_2 \sim 8.5~ Hz$ with $a_1 \sim  -6.7 \cdot 10^{-3} ~ s^{-1}$ and $a_2 \sim -6.4 \cdot 10^{-3} ~ s^{-1}$). These modes, as well as their weaker higher-order spatial harmonics (not shown here), can be interpreted as a measure of coherence of motion within the bubble chain. In this case strictly periodic vertical patterns can be seen, indicating that applied MF enforces highly ordered bubble chains with stable rectilinear trajectories. For $100~sccm$ these harmonics associated with periodic bubble motion within the chain are very similar but exhibit greater \textit{XY} components. It is then interesting to compare the zeroth, stationary modes for $30$ and $100~sccm$ without applied MF against the above cases -- this is shown in Figures \ref{fig:vessel-mode-30-and-100-sccm-off-m0} and \ref{fig:vessel-mode-30-m2-100-m6-off}.

Figures \ref{fig:vessel-mode-30-and-100-sccm-off-m0}a and \ref{fig:vessel-mode-30-and-100-sccm-off-m0}b show the zeroth modes for $30$ ($\omega_0 \sim 3.1~ mHz$ with $a_0 \sim -2 \cdot 10^{-3} ~ s^{-1}$) and $100~ sccm$ ($\omega_0 \sim 2.1~ mHz$ with $a_0 \sim 8.3 \cdot 10^{-3} ~ s^{-1}$), respectively, without applied MF. The zeroth mode in Figure \ref{fig:vessel-mode-30-and-100-sccm-off-m0}a covers much more space about the bubble chain \textit{core} where its magnitude is relatively very high, whereas in \ref{fig:vessel-mode-30-and-100-sccm-off-m0}b the core zone is much thinner, indicating that some other modes are dominant further away; in other words, metal flow about the ascending bubble chain is much more coherent for $30~sccm$. The other thing to note is that the $30~sccm$ mode is much more symmetric about the \textit{YZ} mid-plane (though much less than with applied MF). This makes sense given the lower flow rate corresponding to Figure \ref{fig:vessel-mode-30-and-100-sccm-off-m0}a, but what is significant here is that, comparing especially Figures \ref{fig:vessel-mode-30-sccm-on-m0} and \ref{fig:vessel-mode-30-and-100-sccm-off-m0}a, there are no longer two symmetry planes. It should therefore be of interest to later study the transition from 1 to 2 symmetry planes with ordered flow sheets for a fixed flow rate as the MF magnitude is swept from zero upwards, as well as how symmetry is disrupted as flow rate is increased (e.g. at a fixed MF magnitude).

The modes presented in Figure \ref{fig:vessel-mode-30-m2-100-m6-off} for the cases without MF, meanwhile, are noteworthy for several reasons. First, note that they no longer come in first in terms of amplitude, but rather second for $30$ (Figure \ref{fig:vessel-mode-30-m2-100-m6-off}a, $\omega_2 \sim 4.3~ Hz$, $a_2 \sim 1.6 \cdot 10^{-2}~ s^{-1}$) and sixth for $100~ sccm$ (Figure \ref{fig:vessel-mode-30-m2-100-m6-off}b, $\omega_6 \sim 10.3~ Hz$, $a_6 \sim 6.0 \cdot 10^{-2}~ s^{-1}$), indicating that, as expected, the overall coherence of bubble motion is much lower than in the cases with applied MF. Moreover, comparing Figure \ref{fig:vessel-mode-30-m2-100-m6-off}a against Figure \ref{fig:vessel-mode-30-m2-100-m6-off}b, one can conclude that, while overall weaker compared to the other flow patterns, Figure \ref{fig:vessel-mode-30-m2-100-m6-off}a shows coherent zig-zag motion extending throughout the gallium vessel, whereas in Figure \ref{fig:vessel-mode-30-m2-100-m6-off}b one can see that the mode magnitude falls off dramatically above a certain elevation threshold, indicating a \textit{coherence length}. A criterion could be defined to measure it which would enable to quantify the effects of varying the flow rate and MF magnitude on flow stability.

\begin{figure}[H]
    \centering
    \includegraphics[width=0.60\textwidth]{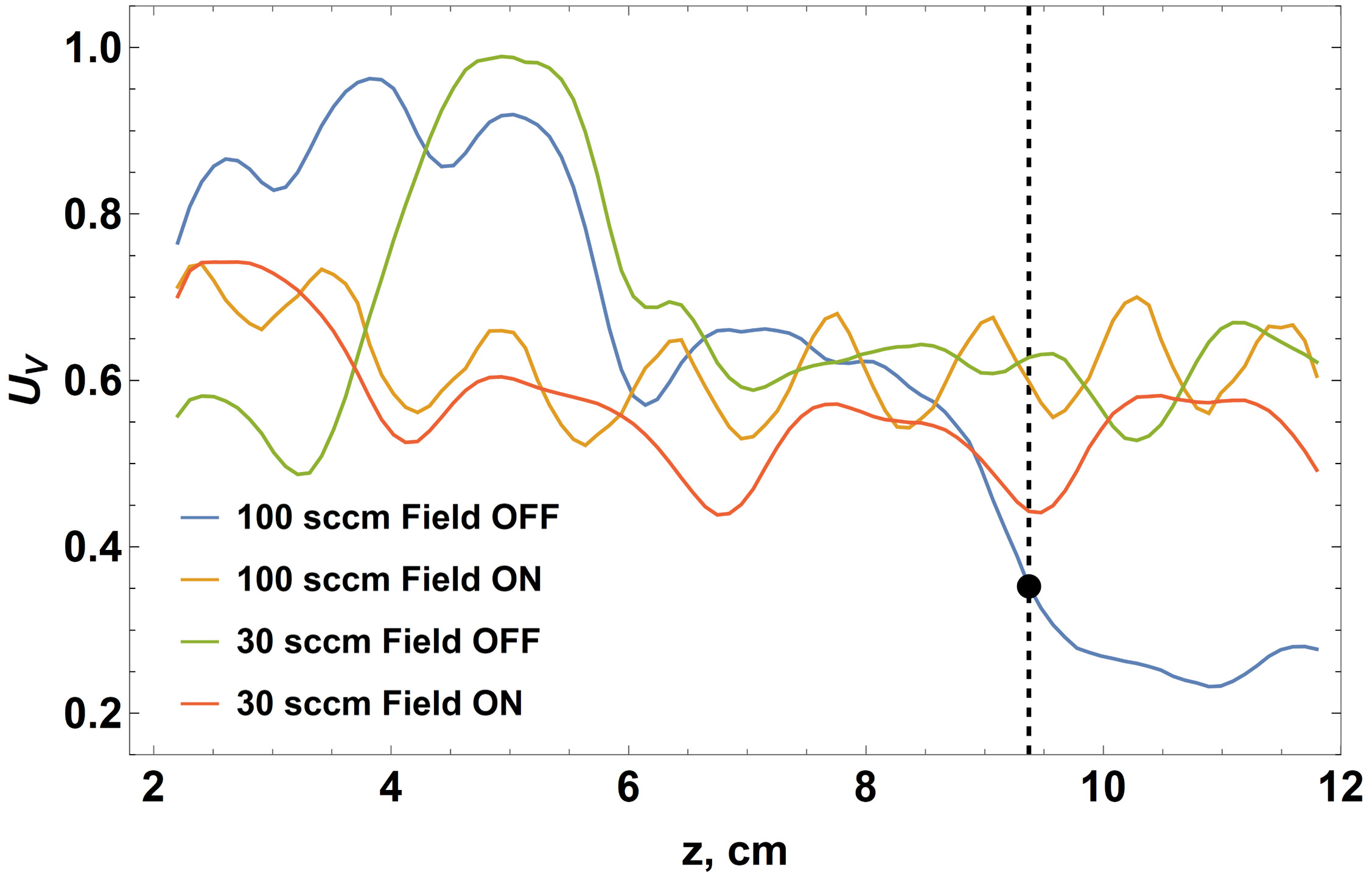}
    \caption{Normalized velocity magnitude integrals $U_V$ for the strongest velocity field modes associated with bubble motion patterns within the bubble chain. Integrals are computed over elliptic cross sections at different height ($z$) containing velocity magnitude values above a threshold that is identical in all cases.}
    \label{fig:vessel-correlation-length}
\end{figure}

One simple way to introduce the coherence length is to fit (with a velocity magnitude threshold) volumes defined by an elliptic cross section extruded over the \textit{Z} dimension of the vessel to velocity fields of bubble chain modes, such as in Figures \ref{fig:vessel-mode-30-sccm-on-m1-m2} and \ref{fig:vessel-mode-30-m2-100-m6-off}, and compute velocity magnitude integrals over elliptic cross sections along \textit{Z}. As an example, consider the strongest modes associated with bubble motion in the chain for each case -- the output of the above procedure is shown in Figure \ref{fig:vessel-correlation-length}. A cutoff threshold of 50\% of the velocity value nearest to the inlet is used to define the coherence length -- one can see that the coherence lengths of bubble chains for $30$ and $100~sccm$ with applied MF and $30~sccm$ without MF extend all the way to the free surface (potentially beyond), while the coherence length for $100~sccm$ without MF is $\sim 7.35~ cm$ (the inlet is below the $2~cm$ mark). Indeed, Figure \ref{fig:vessel-mode-30-m2-100-m6-off}b suggests that bubble motion within the chain becomes incoherent above the 3/4 of the inlet to free surface distance. This is also consistent with what is seen in Figure \ref{fig:vessel-q-100-sccm-off}, especially near the end of the simulation time interval. Note also that Figure \ref{fig:vessel-correlation-length} is representative of bubble spatial frequency within the chains for cases with applied MF.

\begin{figure}[H]
    \centering
    \includegraphics[width=0.90\textwidth]{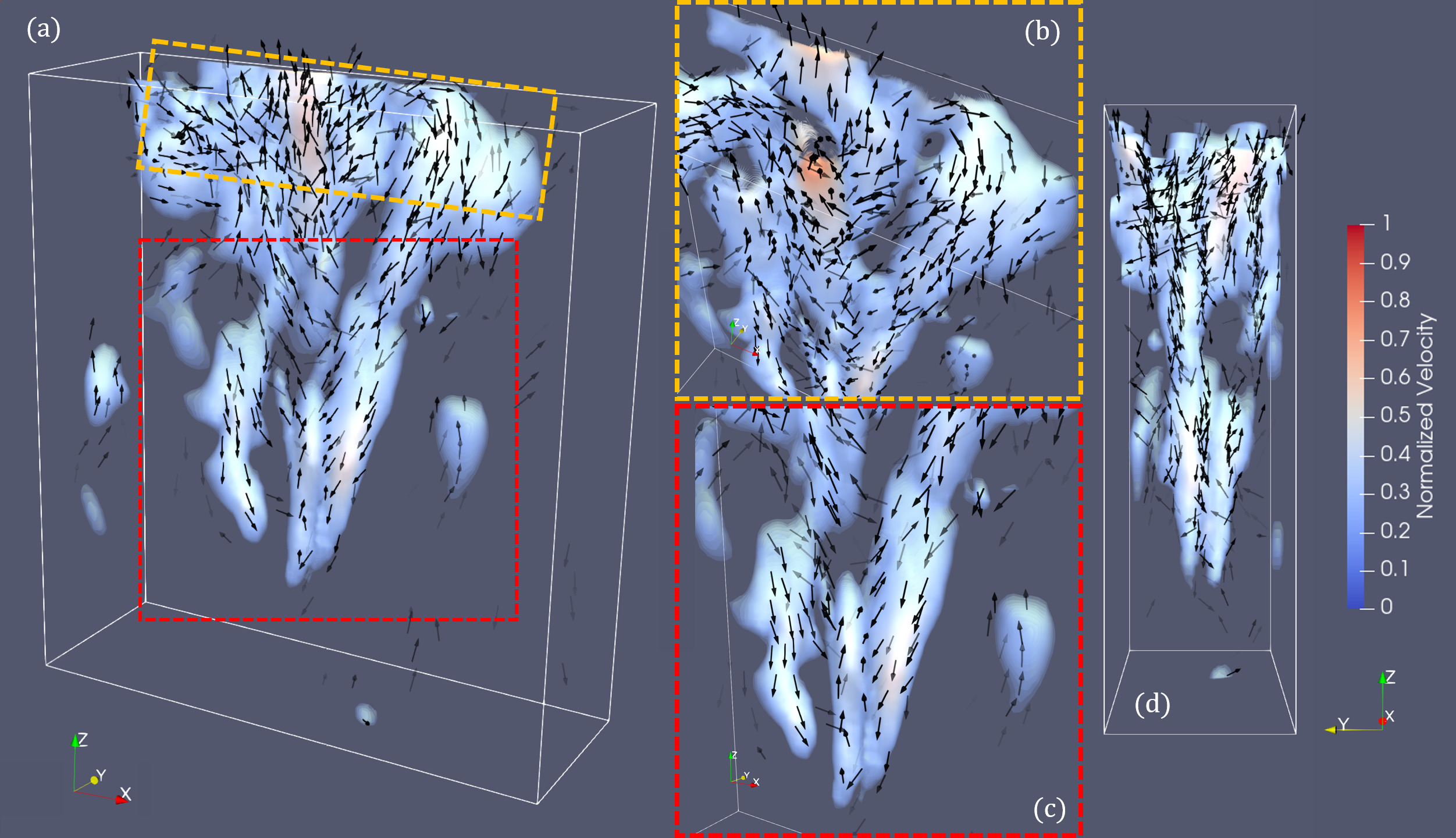}
    \caption{The first velocity field mode for $30~sccm$ without applied MF.}
    \label{fig:vessel-mode-30-m1-off}
\end{figure}

\begin{figure}[H]
    \centering
    \includegraphics[width=0.92\textwidth]{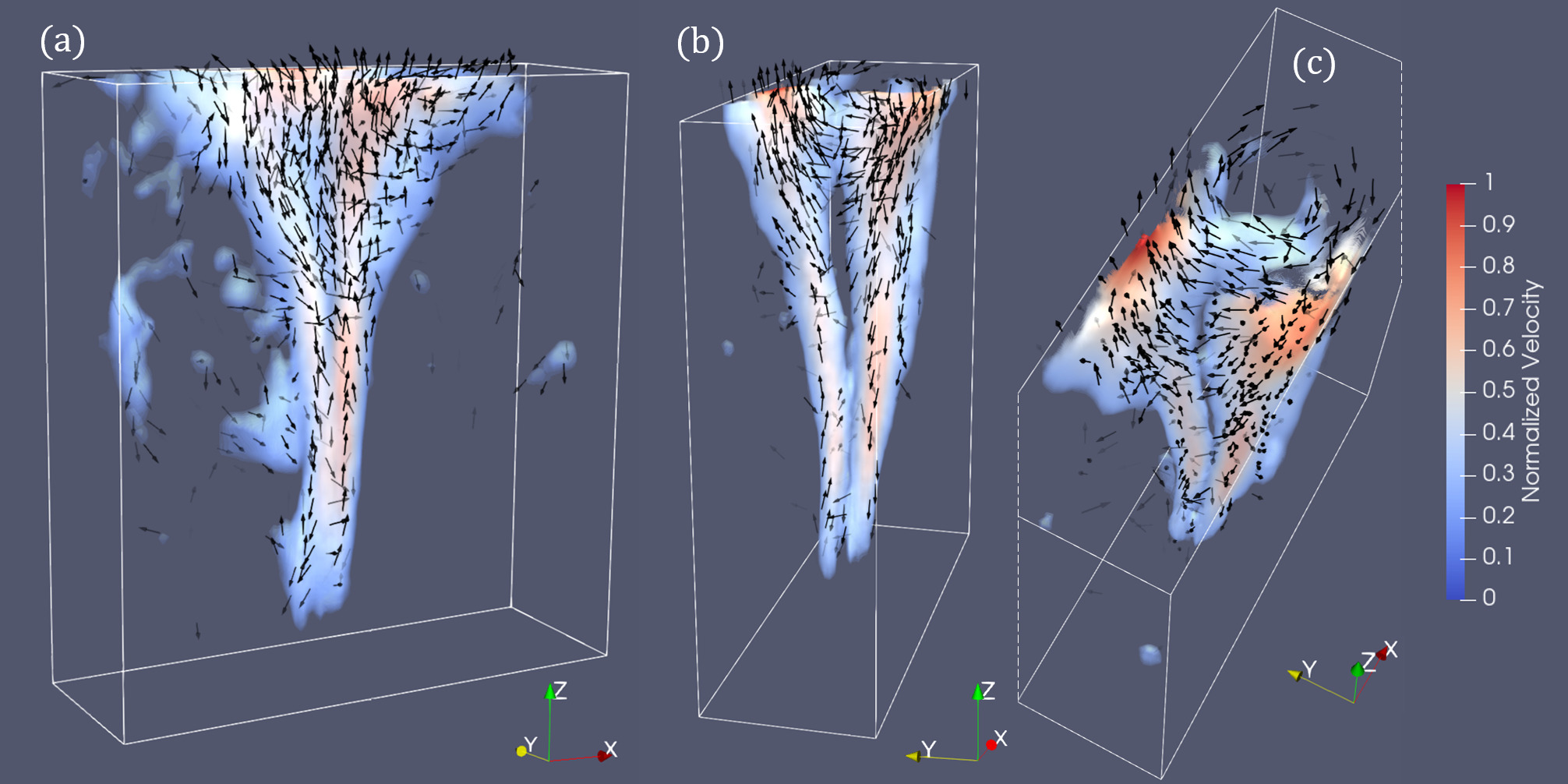}
    \caption{The first velocity field mode for $100~sccm$ without applied MF.}
    \label{fig:vessel-mode-100-m1-off}
\end{figure}

\begin{figure}[H]
    \centering
    \includegraphics[width=0.7\textwidth]{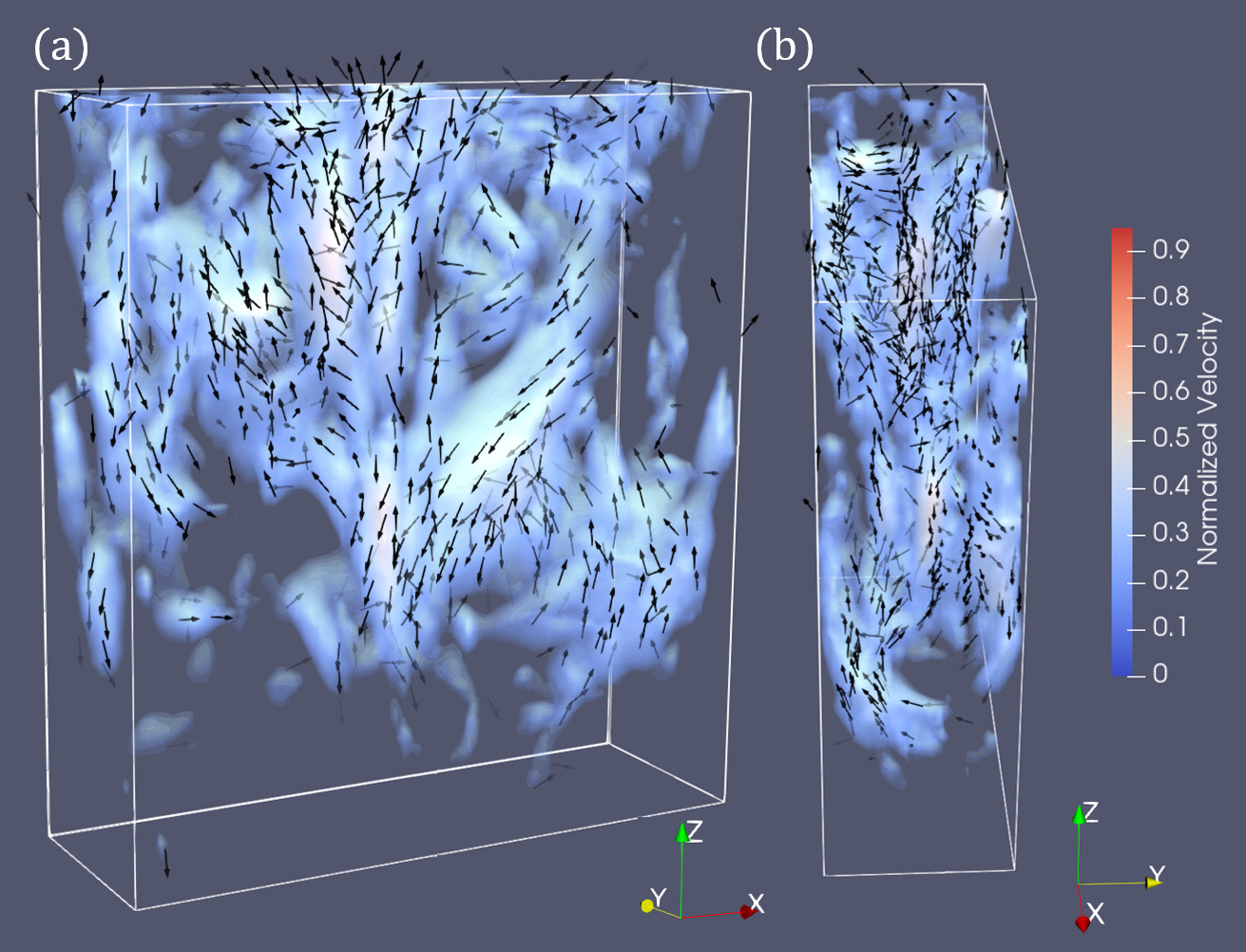}
    \caption{The 2-nd velocity field mode for $100~sccm$ without applied MF.}
    \label{fig:vessel-mode-100-m2-off}
\end{figure}

It is also interesting to examine modes for $30$ and $100~sccm$ without applied MF that capture flow turbulence, such as the modes shown in Figure \ref{fig:vessel-mode-30-m1-off}-\ref{fig:vessel-mode-100-m2-off}. Consider the 1-st velocity field mode for $30~sccm$ without applied MF ($\omega_1 \sim 0.18~ Hz$, $a_1 \sim 4.6 \cdot 10^{-2}~ s^{-1}$) seen in Figure \ref{fig:vessel-mode-30-m1-off}: Figure \ref{fig:vessel-mode-30-m1-off}a indicates, when viewed alongside the zeroth mode in Figure \ref{fig:vessel-mode-30-and-100-sccm-off-m0}a, that mode 1 occupies the space about the central core (with respect to the vessel and the bubble chain) of the zeroth mode. It is also comprised of counter-flowing vertical jets, better seen in Figures \ref{fig:vessel-mode-30-m1-off}c and \ref{fig:vessel-mode-30-m1-off}d, the latter showing that the counter-flowing regions seem to be delimited (roughly) by the \textit{XZ} mid-plane, although there is no discernible symmetry like in the zeroth modes with applied MF (Figures \ref{fig:vessel-mode-30-sccm-on-m0}b and \ref{fig:vessel-mode-100-sccm-on-m0}b). In Figure \ref{fig:vessel-mode-30-m1-off}b one can also see what looks like swirl flow in the upper region of the mode.

In comparison, Figure \ref{fig:vessel-mode-100-m1-off} shows the 1-st velocity field mode for $100~sccm$ without applied MF ($\omega_1 \sim 0.36~ Hz$, $a_1 \sim -1.2 \cdot 10^{-2}~ s^{-1}$). Here the noteworthy feature is that, unlike its $30~sccm$ counterpart, this mode exhibits two clearly separated regions with counter-flow, evident from Figure \ref{fig:vessel-mode-100-m1-off}b. Again, the mode, like in the $30~sccm$ case, occupies the space about the core of the zeroth mode (Figure \ref{fig:vessel-mode-30-and-100-sccm-off-m0})b, this time with a much clearer symmetry about the \textit{XZ} mid-plane. Note also that Figure \ref{fig:vessel-mode-100-m1-off}c indicates, similarly to Figure \ref{fig:vessel-mode-30-m1-off}b, swirl flow in the upper region of the mode, but here it is more ordered. Also, unlike in Figure \ref{fig:vessel-mode-30-m1-off}b, one can see in Figure \ref{fig:vessel-mode-100-m1-off}b that the swirl-like flow seems to extend further below the free surface than in the $30~sccm$ case.

However, Figures \ref{fig:vessel-lic-30-sccm-off} and \ref{fig:vessel-lic-100-sccm-off}, as well as Figures \ref{fig:vessel-q-30-sccm-off} and \ref{fig:vessel-q-100-sccm-off} suggest that the velocity field is more disordered in the $100~sccm$ case. The reason is that there exists a pronounced 2-nd mode ($\omega_2 \sim 0.19~ Hz$, $a_2 \sim -4.9 \cdot 10^{-2}~ s^{-1}$), shown in Figure \ref{fig:vessel-mode-100-m2-off}, that spans most of the vessel volume. This mode does not seem to exhibit any discernible symmetry and likely determines the finer disordered velocity field structure seen in Figures \ref{fig:vessel-lic-100-sccm-off} and \ref{fig:vessel-q-100-sccm-off}. A similar mode for $30~sccm$ has a much smaller amplitude, lower than that of the 2-nd mode in Figure \ref{fig:vessel-mode-30-m2-100-m6-off}, which should explain the observations from the velocity and $Q$ plots.

There are also higher-order modes for the cases without applied MF, but these have lower amplitudes and exhibit patterns similar to the one in Figure \ref{fig:vessel-mode-100-m2-off} with the characteristic scales of their spatial structures decreasing with mode order. They should be taken with a grain of salt for two reasons: first, the $1~mm$ cube mesh should not accurately capture the finer flow structures in the higher-order modes; second, especially for the $100~sccm$ case, longer sampling time is most likely required to capture representative flow field configurations. As such, further modes for the gallium vessel are not shown in this paper.

\begin{figure}[H]
    \centering
    \includegraphics[width=1\textwidth]{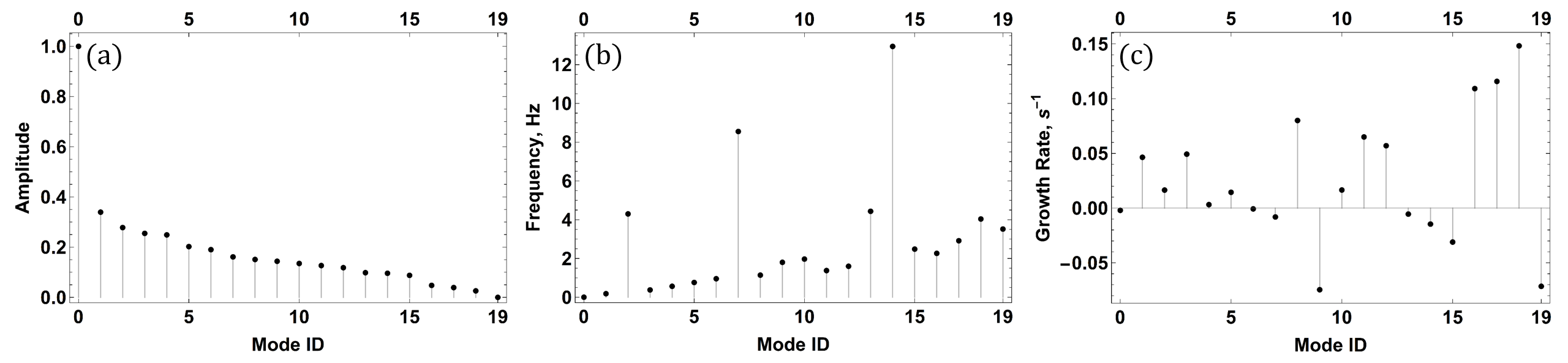}
    \caption{(a) Initial normalized amplitudes, (b) frequencies and (c) growth rates for modes at $30~ sccm$ without applied MF.}
    \label{fig:stats-vessel-modes-30-off}
\end{figure}

\begin{figure}[H]
    \centering
    \includegraphics[width=1\textwidth]{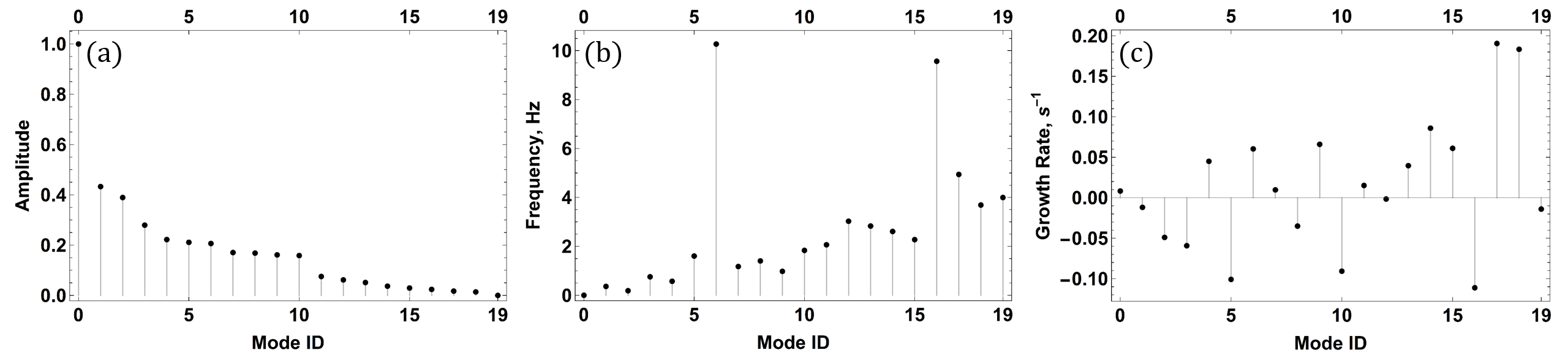}
    \caption{(a) Initial normalized amplitudes, (b) frequencies and (c) growth rates for modes at $100~ sccm$ without applied MF.}
    \label{fig:stats-vessel-modes-100-off}
\end{figure}

\begin{figure}[H]
    \centering
    \includegraphics[width=1\textwidth]{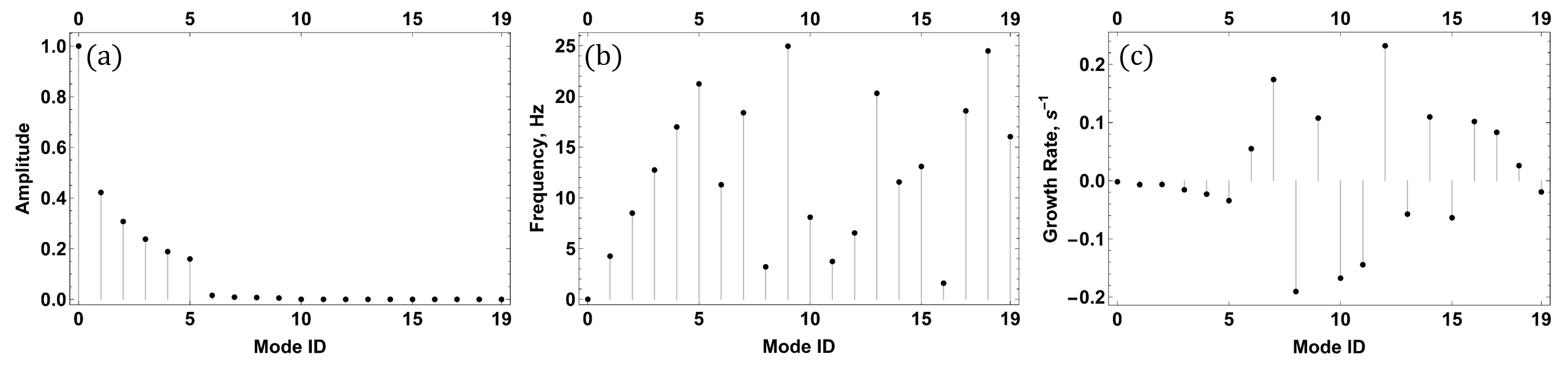}
    \caption{(a) Initial normalized amplitudes, (b) frequencies and (c) growth rates for modes at $30~ sccm$ with applied MF.}
    \label{fig:stats-vessel-modes-30-on}
\end{figure}

\begin{figure}[H]
    \centering
    \includegraphics[width=1\textwidth]{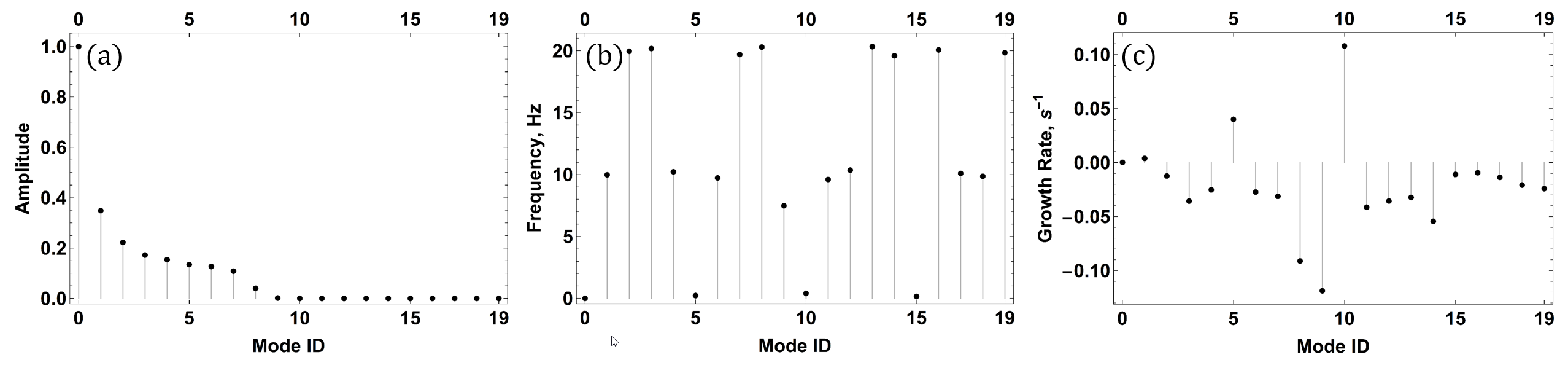}
    \caption{(a) Initial normalized amplitudes, (b) frequencies and (c) growth rates for modes at $100~ sccm$ with applied MF.}
    \label{fig:stats-vessel-modes-100-on}
\end{figure}

While many modes may have initially large or conversely very small amplitudes, it is important to consider their growth rates over the DMD sampling time interval ($5$ seconds). It is also of interest to see what frequencies appear, and what the respective amplitudes and growth rates are in each of the cases considered herein -- this is shown in Figures \ref{fig:stats-vessel-modes-30-off}-\ref{fig:stats-vessel-modes-100-on}.

In all cases the zeroth stationary modes have more than double initial amplitudes compared with all other modes. One can note several patterns for the cases without applied MF (Figures \ref{fig:stats-vessel-modes-30-off} and \ref{fig:stats-vessel-modes-100-off}). First, modes with higher temporal frequencies generally have lower amplitudes, aside from a few outlier modes that are different for $30$ and $100~sccm$ -- note that the lower frequency peak seen in Figure \ref{fig:stats-vessel-modes-30-off}b is not present in Figure \ref{fig:stats-vessel-modes-100-off}b. Another characteristic feature is that the normalized amplitudes decrease with mode order more gradually for $30$ than for $100~sccm$ (Figures \ref{fig:stats-vessel-modes-30-off}a and \ref{fig:stats-vessel-modes-100-off}a) where in the latter case the amplitude slope for modes 1-4 is steeper and there is a slight step-down from mode 10 to 11. It is also of interest that, while the amplitude does not vary too much for higher-order modes despite higher growth coefficients due to their lower initial amplitudes, modes 1-3 and 5 for $100~sccm$  have rather significant negative growth rates (Figure \ref{fig:stats-vessel-modes-100-off}c), meaning that their dominance over higher frequency modes will diminish over time. Note also that mode 6 has a notable positive growth rate and the highest frequency (Figure \ref{fig:stats-vessel-modes-100-off}b). Meanwhile for $30~sccm$ one can see in Figure \ref{fig:stats-vessel-modes-30-off}c that the lower frequency modes have either significant positive, or near-zero growth rates and the two modes with the highest frequencies have negative growth rates. This would suggest that the higher flow rate promotes finer flow structures (characteristic for high order modes when no MF is applied) with smaller time scales, and the DMD provides the means of quantifying this. In principle, for simulations with a higher grid resolution, it should be feasible to measure characteristic length scales for DMD mode flow structures via image processing and/or spectral methods for a more in-depth analysis.

When MF is applied (Figures \ref{fig:stats-vessel-modes-30-on} and \ref{fig:stats-vessel-modes-100-on}), several key differences appear. First, there is now no clear pattern for frequency versus mode order as seen in Figures \ref{fig:stats-vessel-modes-30-on}b and \ref{fig:stats-vessel-modes-100-on}b. Second, Figures \ref{fig:stats-vessel-modes-30-on}a and \ref{fig:stats-vessel-modes-100-on}a indicate that, unlike the cases without MF, there is a very sharp mode amplitude cutoff after mode 5 for $30~sccm$ and mode 7 form $100~sccm$ beyond which the amplitudes are so insignificant that even the high positive growth rates in the $30~sccm$ case (Figure \ref{fig:stats-vessel-modes-30-on}) make no difference to the overall dynamics. Also, this positive growth for more than half of the modes beyond mode 5 in the $30~sccm$ is in stark contrast with what is seen in Figure \ref{fig:stats-vessel-modes-100-on}c for $100~sccm$ where one can see that most of the modes have significant negative growth rate. Figures \ref{fig:stats-vessel-modes-30-on}b and \ref{fig:stats-vessel-modes-100-on}b also indicate an overall increase in mode temporal frequency -- this can be attributed to spatial harmonics associated with the bubble chain like the ones in Figure \ref{fig:vessel-mode-30-sccm-on-m1-m2}. All of these are quantitative indicators -- including the smaller number of modes required to encapsulate the system with applied MF -- that can be used so systematically study how MF stabilizes and regularizes the flow field and how this is disrupted at higher flow rates. Note that the mode growth rates in Figure \ref{fig:stats-vessel-modes-100-on}c would also explain the gradual transition from rectilinear to \textit{YZ} plane zig-zag trajectories for $100~sccm$ with applied MF (Figure \ref{fig:vessel-mode-100-sccm-on-m0}) -- the zeroth mode becomes more pronounced over time versus the other decaying modes.

\begin{figure}[H]
    \centering
    \includegraphics[width=0.85\textwidth]{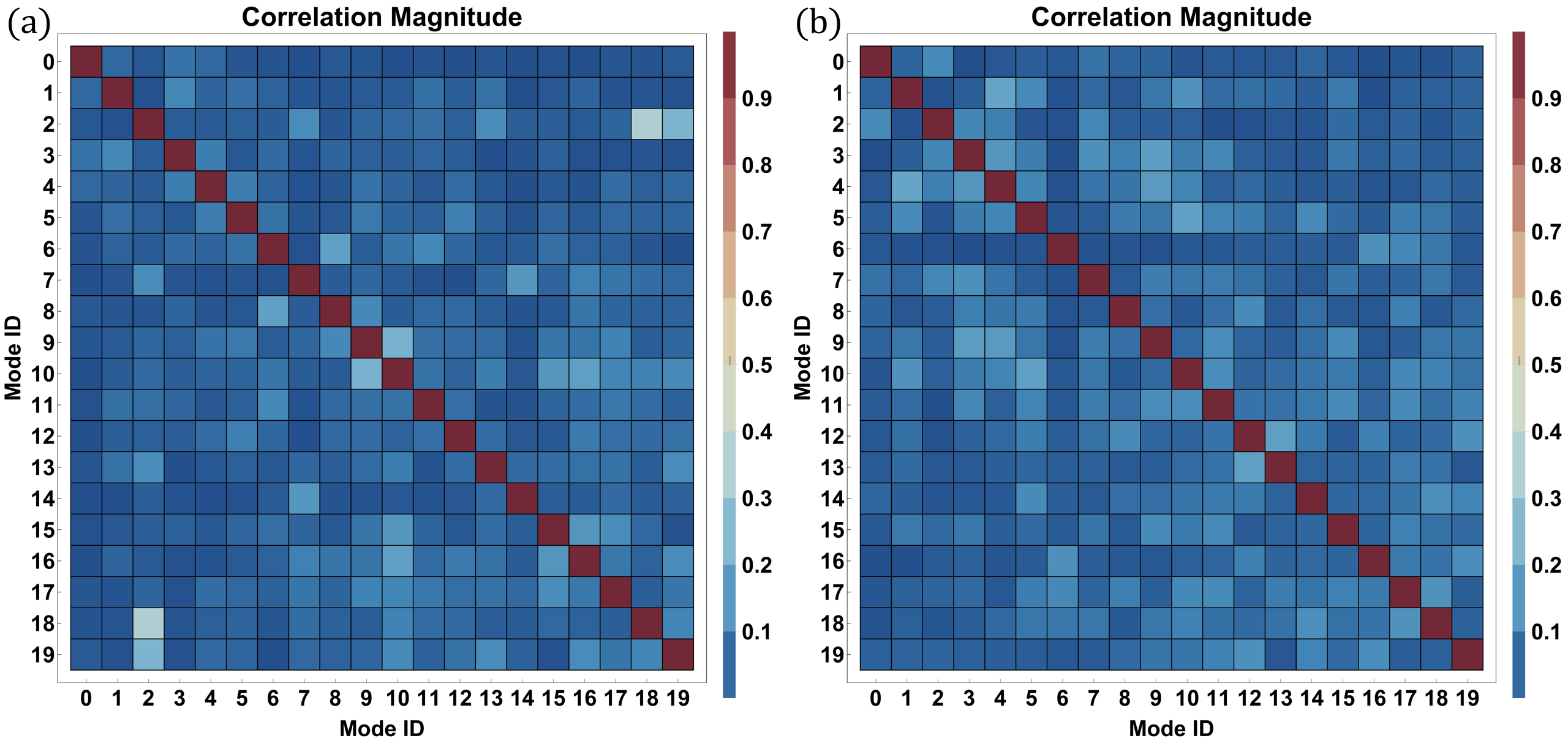}
    \caption{Correlation matrices for the vessel velocity field modes for (a) $30~sccm$ and (b) $100~sccm$ without applied MF.}
    \label{fig:correlations-vessel-U-off}
\end{figure}

\begin{figure}[H]
    \centering
    \includegraphics[width=0.85\textwidth]{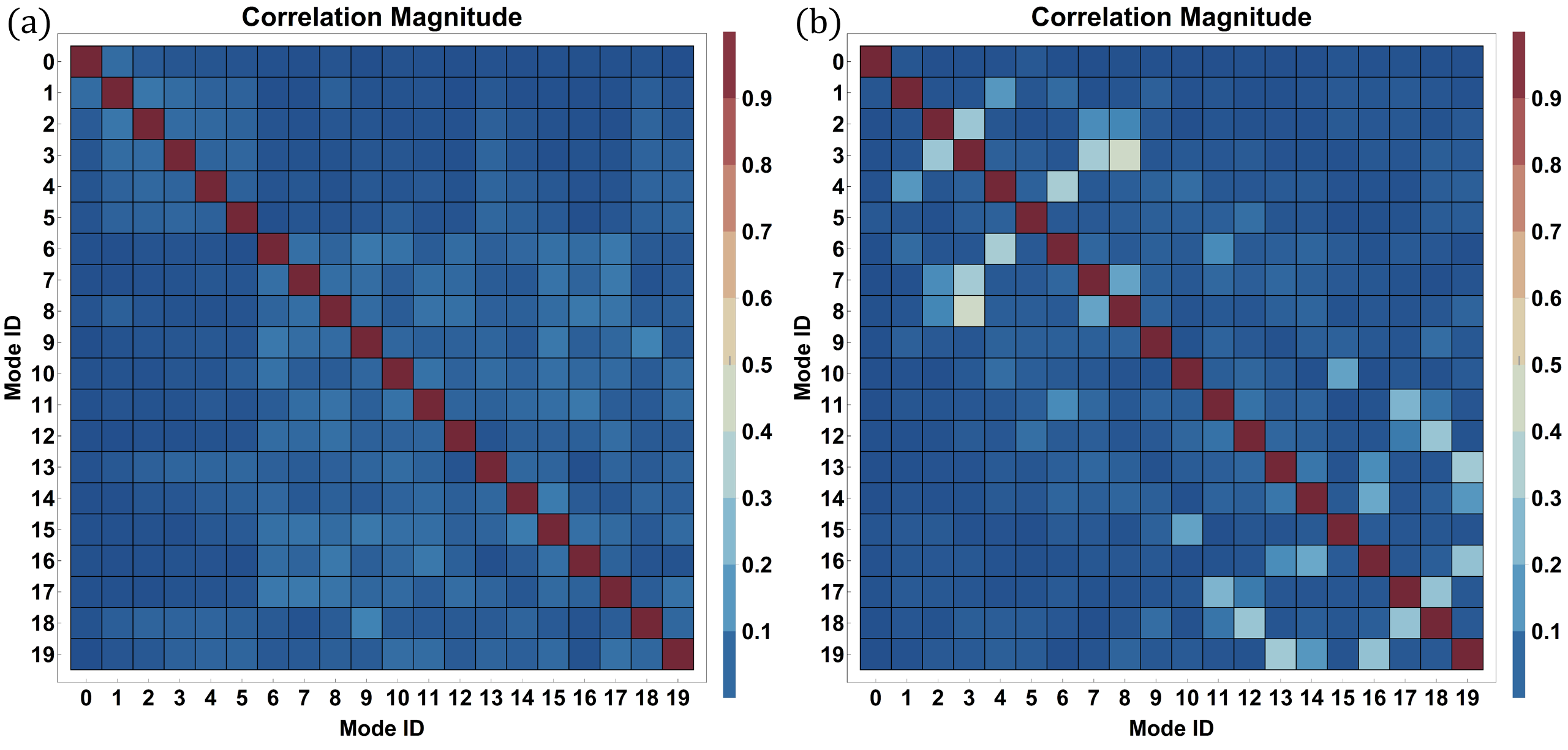}
    \caption{Correlation matrices for the vessel velocity field modes for (a) $30~sccm$ and (b) $100~sccm$ with applied MF.}
    \label{fig:correlations-vessel-U-on}
\end{figure}

Finally, it is important to analyze the degree of spatial correlation between the DMD modes -- mode correlation matrices are presented in Figures \ref{fig:correlations-vessel-U-off} and \ref{fig:correlations-vessel-U-on} where one can see that the modes are rather weakly correlated with the exception of several modes in the $100~sccm$ case with applied MF (Figure \ref{fig:correlations-vessel-U-on}b). The reason why some of the mode pairs are less orthogonal than others is that the oscillations of the argon velocity field within bubbles are captured by the DMD. At the current resolution these oscillations are under-resolved and are essentially correlated noise, hence whatever modes contain these noise patterns are correlated to some degree. Generally higher-order modes are more susceptible to this effect because argon velocity field noise exhibits fine length scales. One may notice that the overall off-diagonal correlation magnitude is less for $30~sccm$ with MF than it is for $100~sccm$ (Figure \ref{fig:correlations-vessel-U-on}).

Meanwhile, for the cases without MF (Figure \ref{fig:correlations-vessel-U-off}), one can see that for $30~sccm$ there are a few more pronounced correlations, but overall the values are lower than in the $100~sccm$ where no significant peaks are present, i.e. the off-diagonal values are more diluted. The latter can be explained by the fact that most of the modes in the $100~sccm$ are much more disordered and exhibit finer spatial structures than in the $30~sccm$ case with spatial/temporal timescales often slightly overlapping with the correlated noise, hence the relative homogeneity of the off-diagonal matrix elements. In the former cases where MF is applied flow is largely laminarized and virtually the only fine temporal/spatial structures on scales similar to the noise are found in some of the spatial harmonics associated with the bubble chain, i.e. higher-order modes similar to the ones in Figure \ref{fig:vessel-mode-30-sccm-on-m1-m2} -- the overlap seems to be stronger for $100~sccm$. 

However, the fact that overall the modes, especially the strongest of the lower-order, are very much orthogonal means that it is quite safe to physically attribute the significant modes to various momentum transfer mechanisms (and spatial harmonics thereof) occurring within the vessel.

\clearpage

\subsection{Bubble reference frame: trajectories \& flow patterns}

To analyze bubble wake flow in greater detail one must first transition to the bubble reference frame by tracking individual bubbles, reconstructing their centroid trajectories and performing velocimetry, wherefrom the relative velocity field can be computed. This is done as follows:

\begin{itemize}
    \item Apply the \textit{VTK marching cubes} algorithm to extract phase boundaries.
    \item Remove the free surface and the bubbles at the inlet that have not yet detached.
    \item Compute bubble centroids and volume using the \textit{STL Python libarary}.
    \item Perform trajectory reconstruction and velocimetry using our \textit{MHT-X} tracing code (open-source) \cite{zvejnieks-mhtx-arxiv}.
\end{itemize}

\begin{figure}[H]
    \centering
    \includegraphics[width=0.81\textwidth]{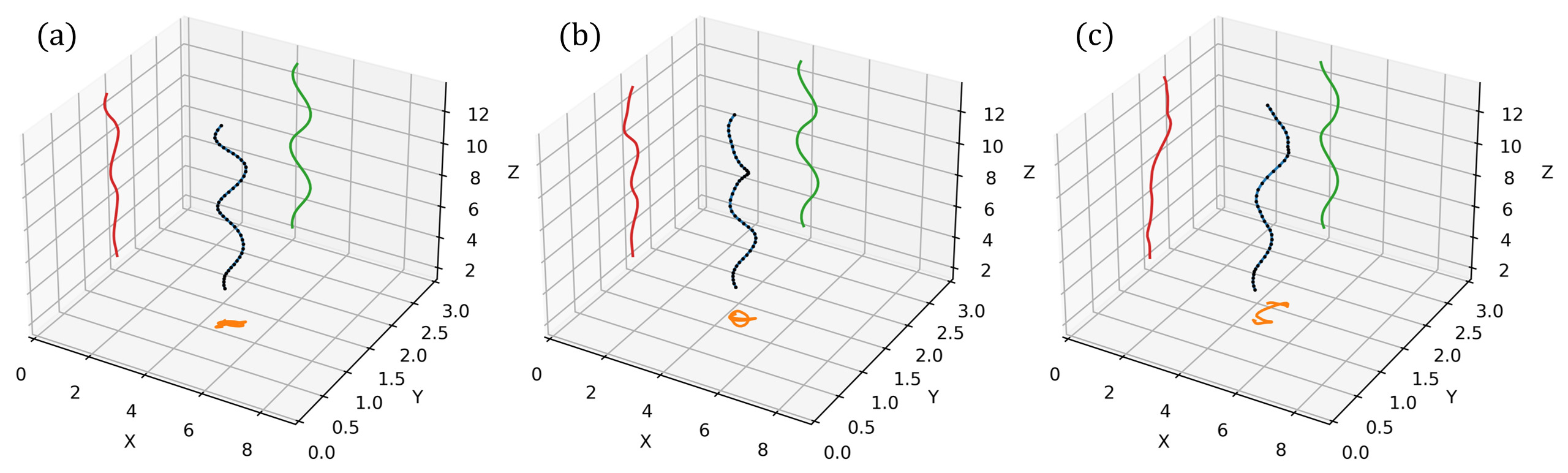}
    \caption{Representative trajectories (blue lines) for $30~sccm$ without applied MF. Trajectories (a-b) are ordered in time. Plot axes correspond to the vessel axes. Orange, green and red lines are trajectory projections onto \textit{XY}, \textit{XZ} and \textit{YZ} planes, respectively. The black dots along the blue lines are bubble centroids used to reconstruct the trajectories.}
    \label{fig:trajectories-30-sccm-field-off}
\end{figure}

\begin{figure}[H]
    \centering
    \includegraphics[width=0.81\textwidth]{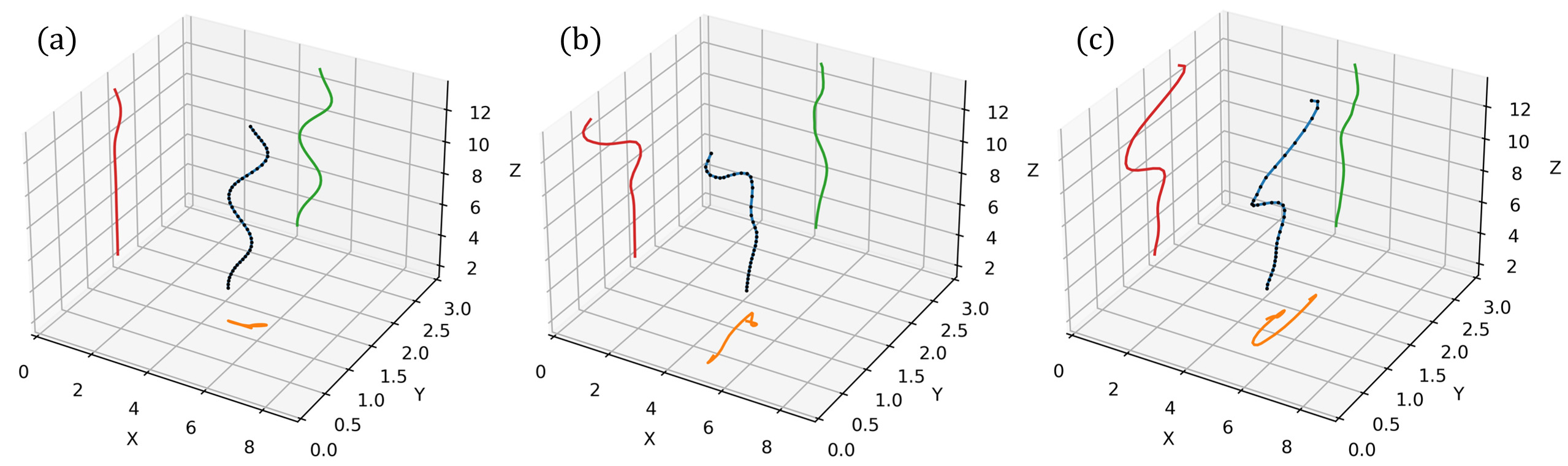}
    \caption{Representative trajectories for $100~sccm$ without applied MF. Trajectories (a-b) are ordered in time.}
    \label{fig:trajectories-100-sccm-field-off}
\end{figure}

\begin{figure}[H]
    \centering
    \includegraphics[width=0.81\textwidth]{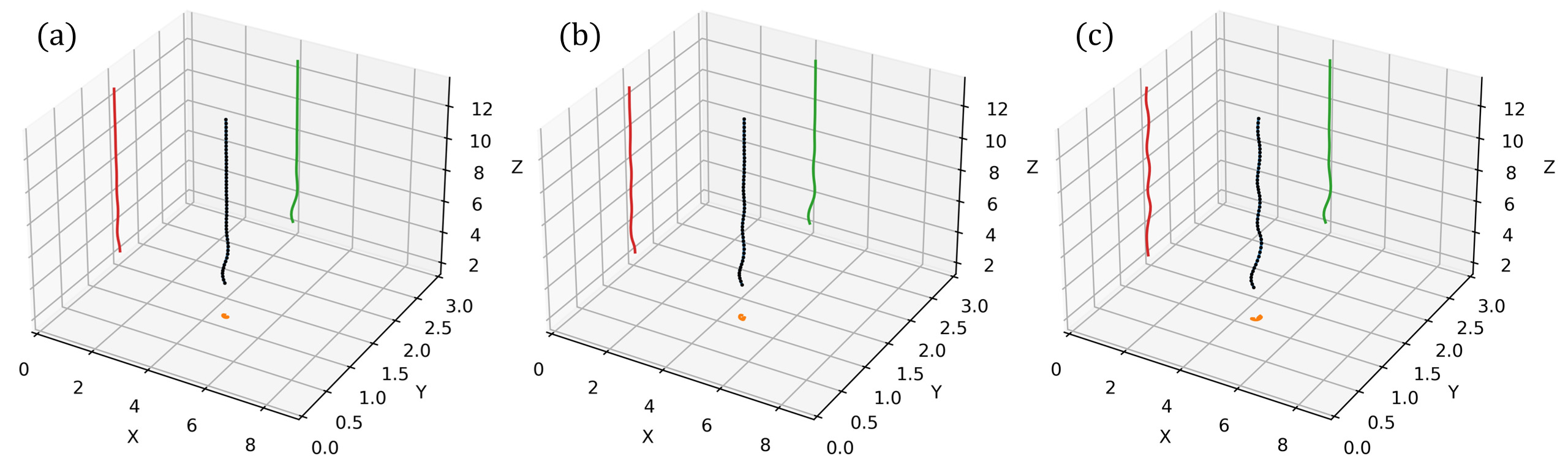}
    \caption{Representative trajectories for $30~sccm$ with applied MF. Trajectories (a-b) are ordered in time.}
    \label{fig:trajectories-30-sccm-field-on}
\end{figure}

\begin{figure}[H]
    \centering
    \includegraphics[width=0.81\textwidth]{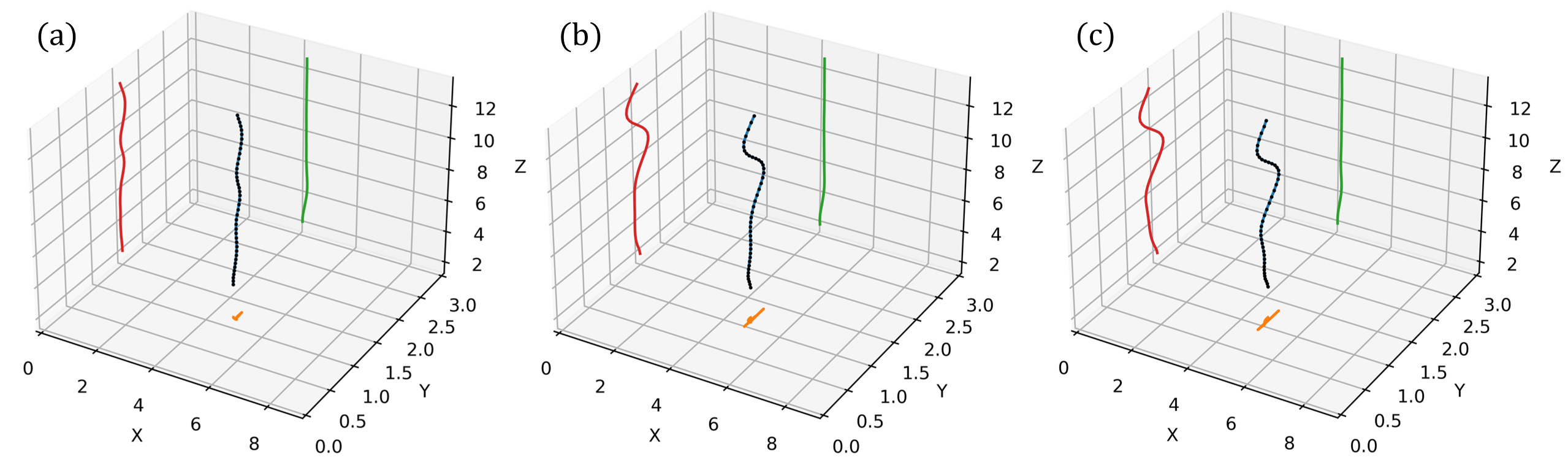}
    \caption{Representative trajectories for $100~sccm$ with applied MF. Trajectories (a-b) are ordered in time.}
    \label{fig:trajectories-100-sccm-field-on}
\end{figure}

\begin{figure}[H]
    \centering
    \includegraphics[width=0.92\textwidth]{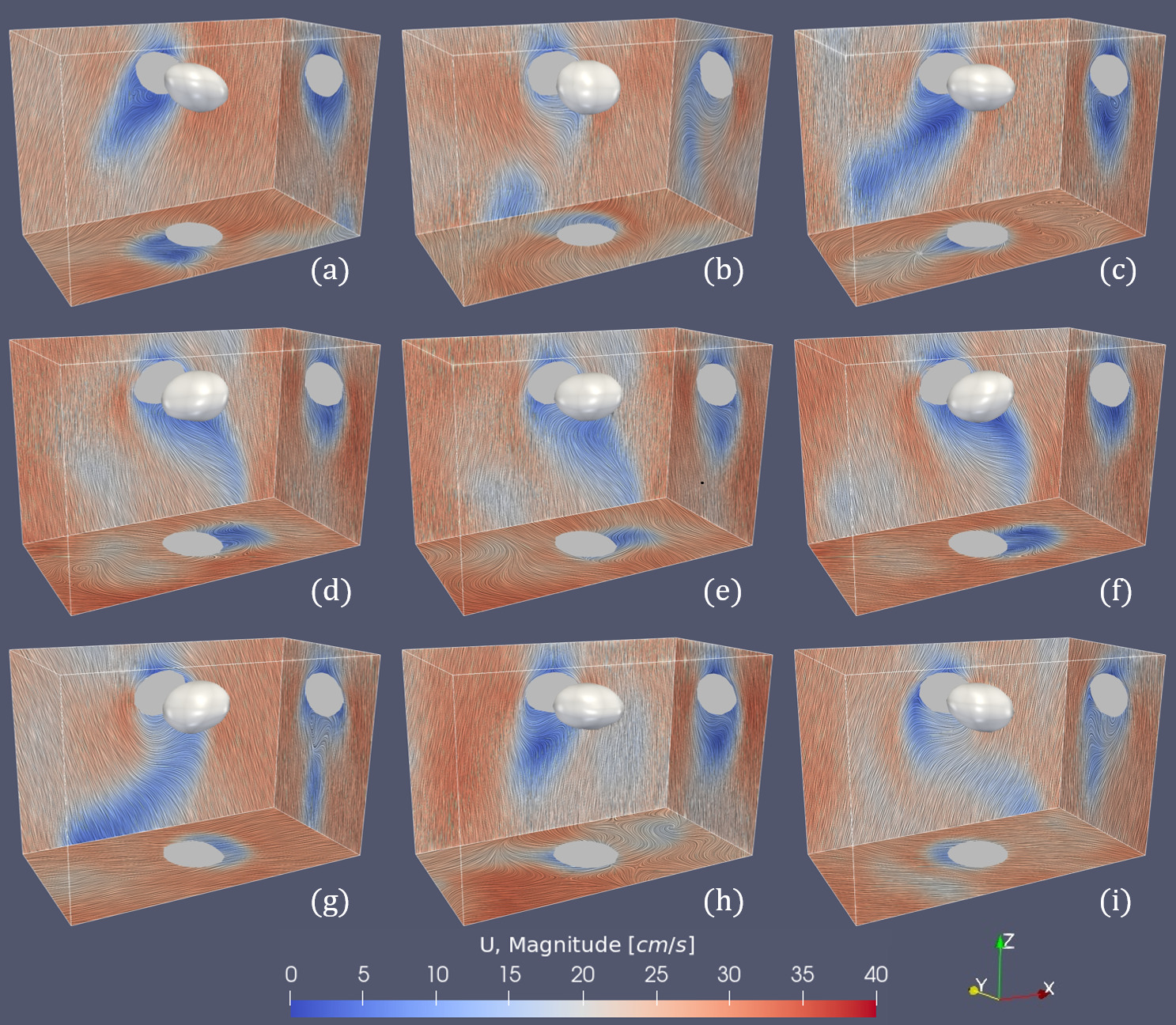}
    \caption{$30~sccm$ without applied MF: LIC plots of the relative velocity field about the bubbles for representative time stamps along bubble trajectories.}
    \label{fig:bubbles-lic-30-sccm-off}
\end{figure}

\begin{figure}[H]
    \centering
    \includegraphics[width=0.92\textwidth]{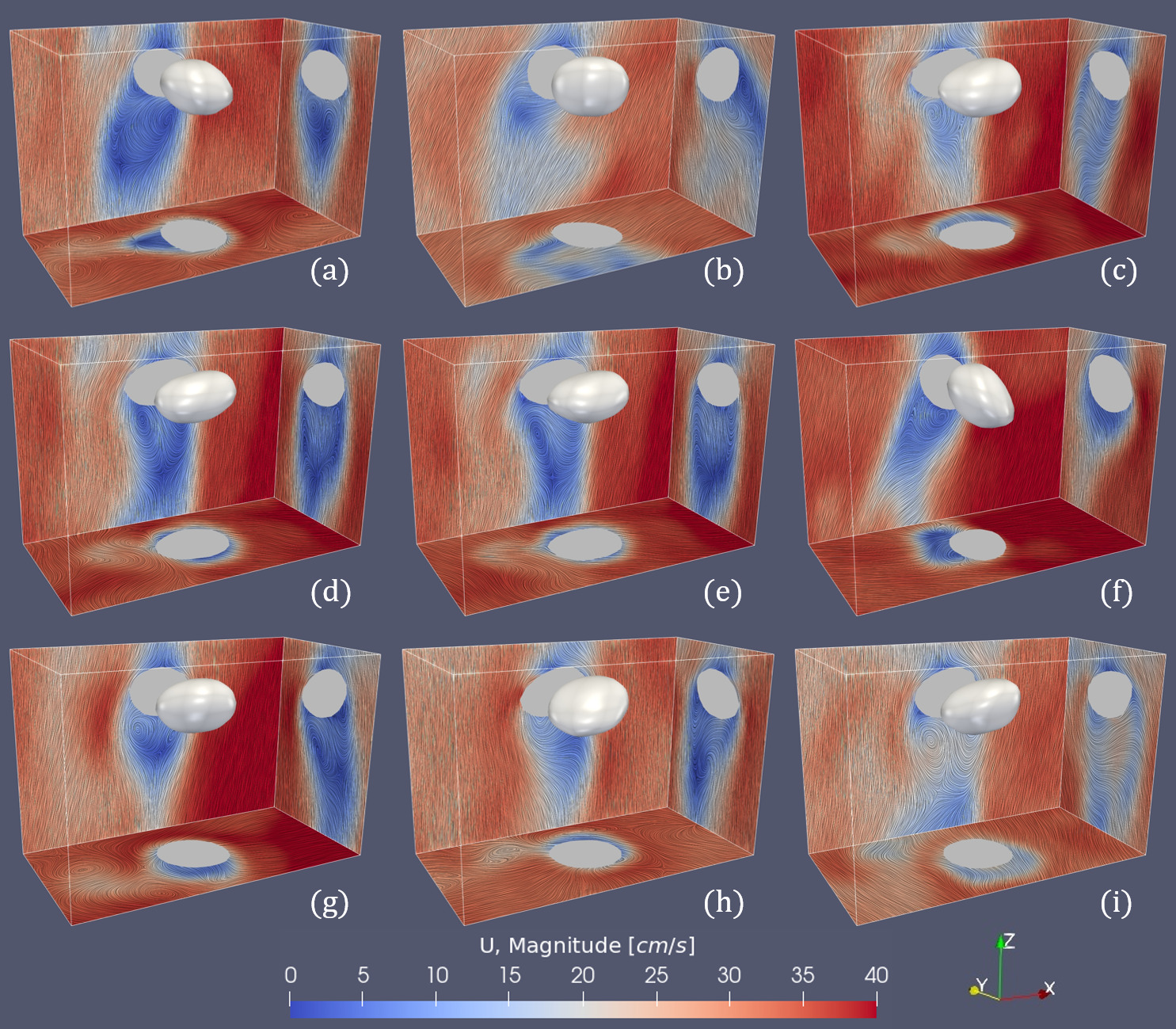}
    \caption{$100~sccm$ without applied MF: LIC plots of the relative velocity field about the bubbles for representative time stamps along bubble trajectories.}
    \label{fig:bubbles-lic-100-sccm-off}
\end{figure}

\begin{figure}[H]
    \centering
    \includegraphics[width=0.90\textwidth]{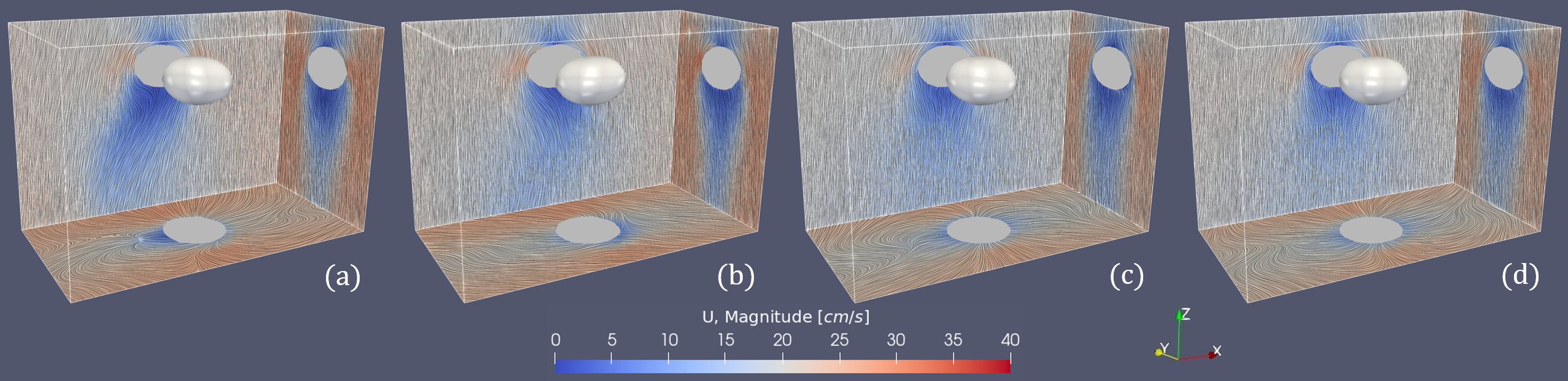}
    \caption{$30~sccm$ with applied MF: LIC plots of the relative velocity field about the bubbles for representative time stamps along bubble trajectories.}
    \label{fig:bubbles-lic-30-sccm-on}
\end{figure}

\begin{figure}[H]
    \centering
    \includegraphics[width=0.90\textwidth]{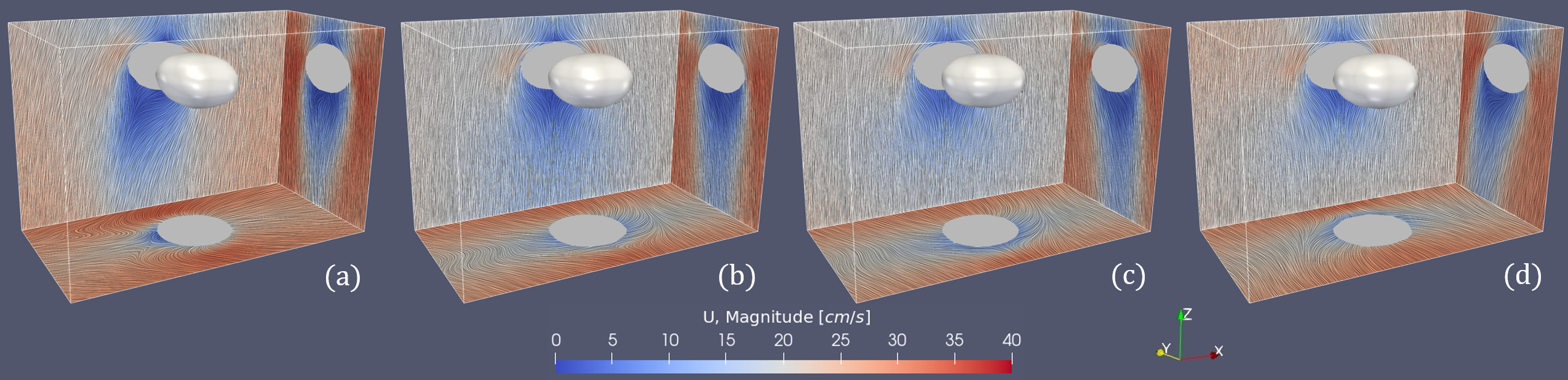}
    \caption{$100~sccm$ with applied MF: LIC plots of the relative velocity field about the bubbles for representative time stamps along bubble trajectories.}
    \label{fig:bubbles-lic-100-sccm-on}
\end{figure}

\begin{figure}[H]
    \centering
    \includegraphics[width=0.75\textwidth]{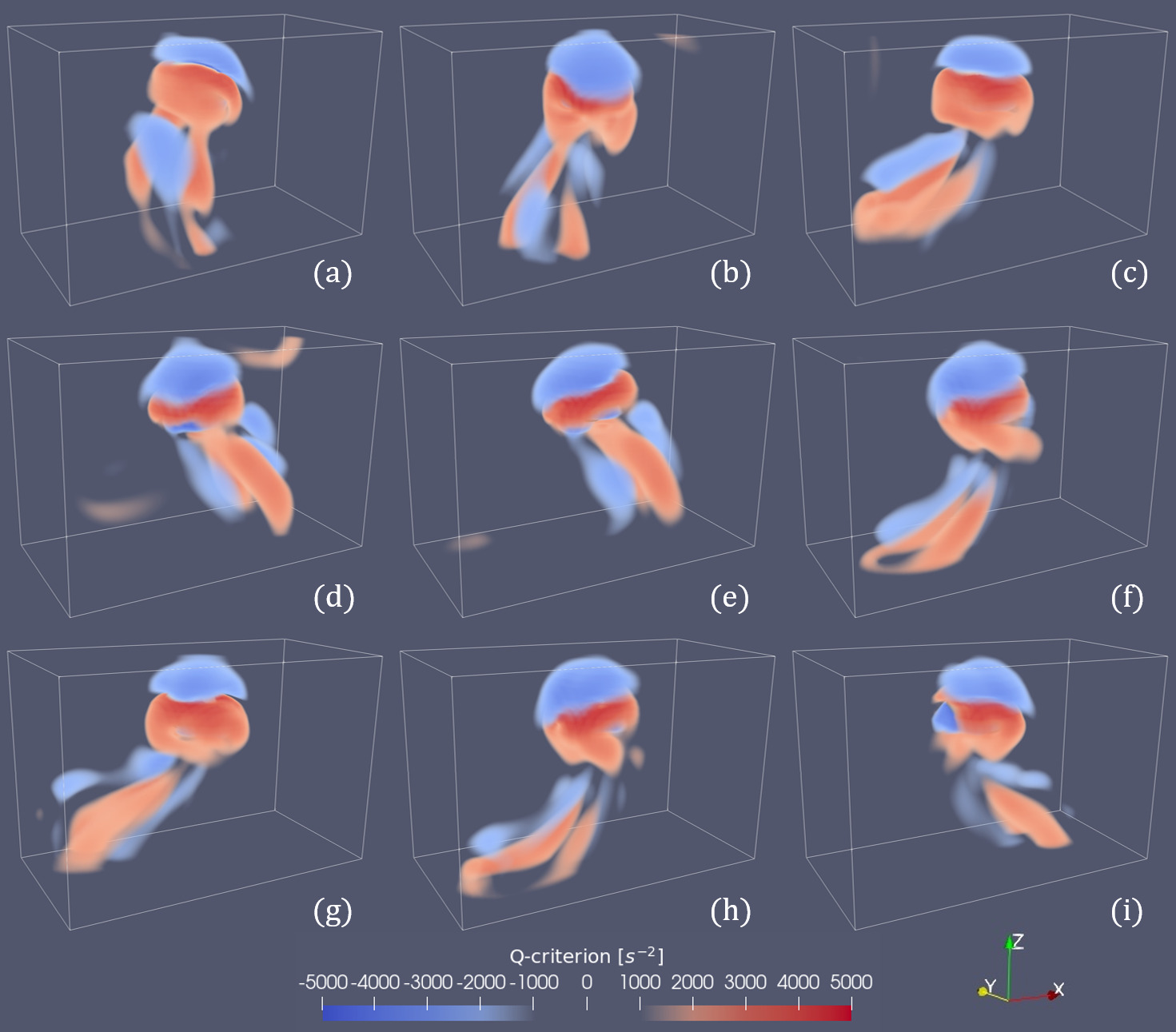}
    \caption{$30~sccm$ without applied MF: $Q$ field about the bubbles for representative time stamps along bubble trajectories.}
    \label{fig:bubbles-q-30-sccm-off}
\end{figure}

\begin{figure}[H]
    \centering
    \includegraphics[width=0.75\textwidth]{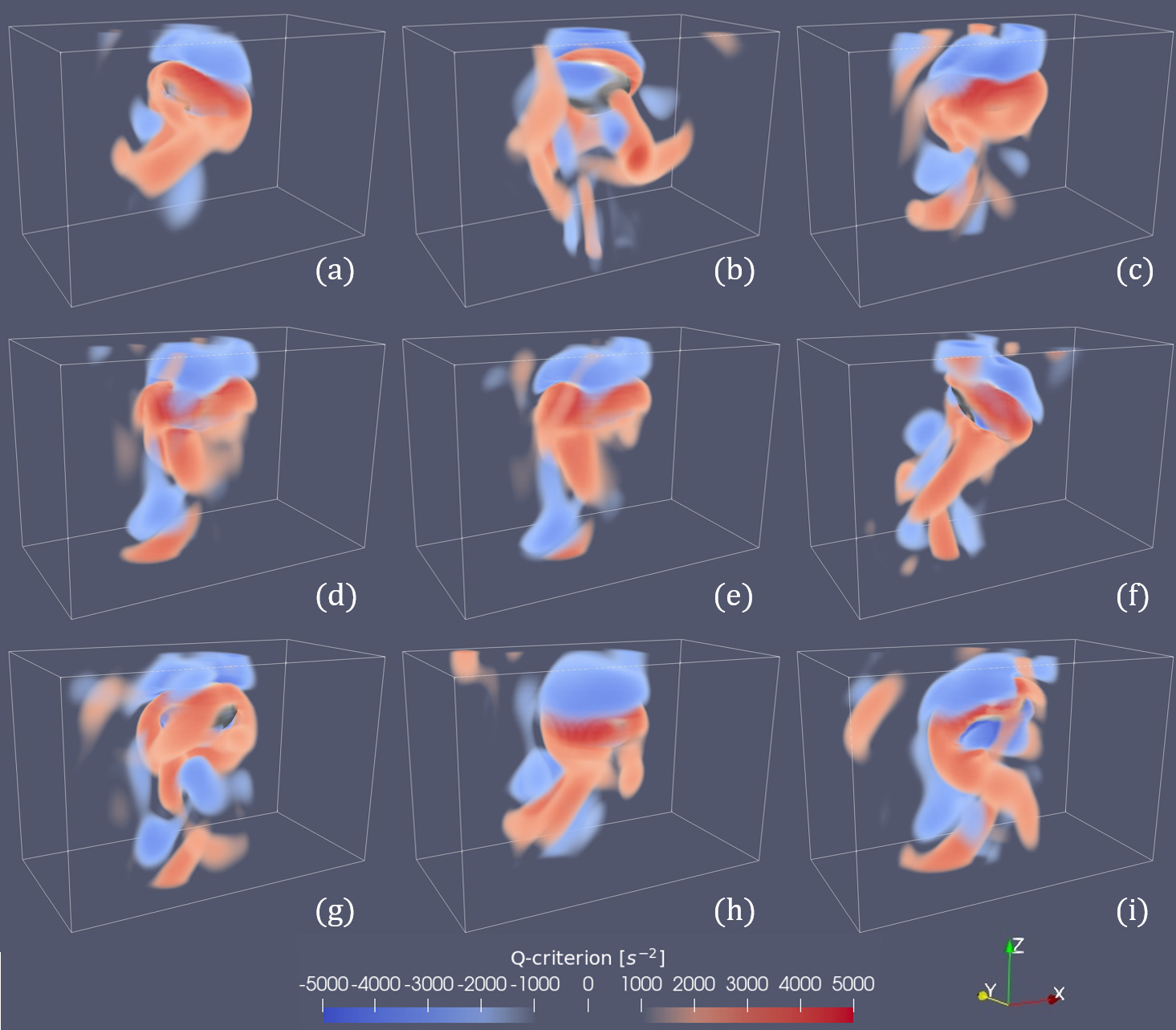}
    \caption{$100~sccm$ without applied MF: $Q$ field about the bubbles for representative time stamps along bubble trajectories}
    \label{fig:bubbles-q-100-sccm-off}
\end{figure}

\begin{figure}[H]
    \centering
    \includegraphics[width=0.9\textwidth]{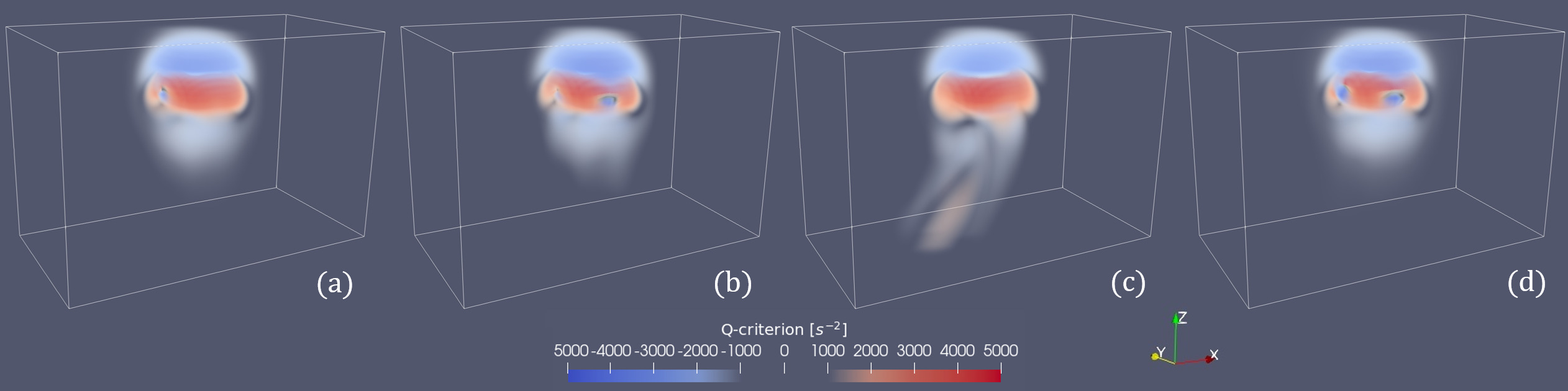}
    \caption{$30~sccm$ with applied MF: $Q$ field about the bubbles for representative time stamps along bubble trajectories.}
    \label{fig:bubbles-q-30-sccm-on}
\end{figure}

\begin{figure}[H]
    \centering
    \includegraphics[width=0.9\textwidth]{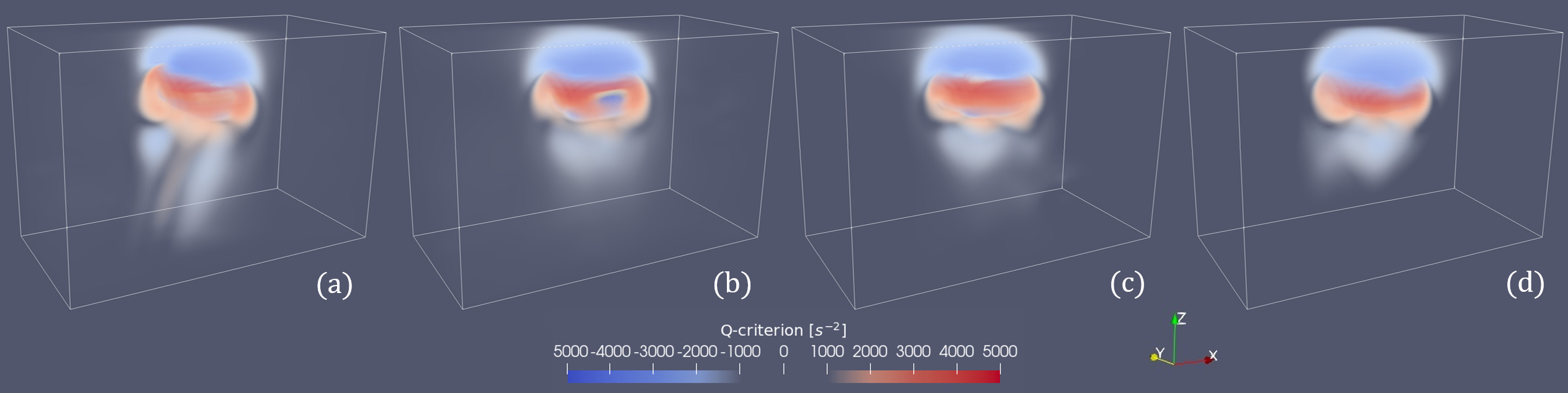}
    \caption{$100~sccm$ with applied MF: $Q$ field about the bubbles for representative time stamps along bubble trajectories.}
    \label{fig:bubbles-q-100-sccm-on}
\end{figure}

Figures \ref{fig:trajectories-30-sccm-field-off}-\ref{fig:trajectories-100-sccm-field-on} show the representative computed trajectories. With this, the relative velocity fields for bubbles were computed for the 4 cases of interest (Section \ref{sec:system-and-cases}). These and the respective local $Q$ fields (invariant to the Galilean transformation) about the bubbles are shown in Figures \ref{fig:bubbles-lic-30-sccm-off}-\ref{fig:bubbles-q-100-sccm-on}.

First, note the difference between the trajectories without MF at $30$ and $100~sccm$ -- while there are perturbations, $30~sccm$ corresponds to rather regular classic zig-zags that persist over time (Figure \ref{fig:trajectories-30-sccm-field-off}), while at $100~sccm$ bubbles exhibit disordered zig-zag paths with greater out-of-plane (\textit{XZ}) deviations (Figure \ref{fig:trajectories-100-sccm-field-off}). The cases with applied MF are interesting in that trajectories are initially almost perfectly rectilinear, but over time very small periodic perturbations are developed for $30~sccm$ (Figure \ref{fig:trajectories-30-sccm-field-on}). For $100~sccm$, as seen in Figure \ref{fig:trajectories-100-sccm-field-on}, a significant out-of-plane perturbation develops in the upper part of the trajectories and after some time becomes stationary (Figure \ref{fig:trajectories-30-sccm-field-off}c) -- this is explained by the zeroth velocity field mode shown in Figure \ref{fig:vessel-mode-100-sccm-on-m0}.

Regarding the velocity field patterns in the bubble reference frame, characteristic snapshots of the velocity field about ascending bubbles at $30~sccm$ without applied MF are shown in Figure \ref{fig:bubbles-lic-30-sccm-off}. Note the wake flow beneath the bubble that undergoes periodic oscillations mostly within the \textit{XZ} plane and only slight oscillations are observable in the \textit{YZ} plane. Note also that, while vortices develop in the bubble wake (e.g. Figures \ref{fig:bubbles-lic-30-sccm-off}a and \ref{fig:bubbles-lic-30-sccm-off}i) and pronounced wake asymmetrization occurs (Figures \ref{fig:bubbles-lic-30-sccm-off}b, \ref{fig:bubbles-lic-30-sccm-off}c, \ref{fig:bubbles-lic-30-sccm-off}g-\ref{fig:bubbles-lic-30-sccm-off}i). However, the vortices that are formed in the wake are rather small and the wake region itself is relatively small, as opposed to the $100~sccm$ case -- velocity field instances are shown in Figure \ref{fig:bubbles-lic-100-sccm-off} -- where one can see that the bubble wake often extends all the way through or beyond the sampling volume. Note that vortices are also much larger as seen in Figures \ref{fig:bubbles-lic-100-sccm-off}a, \ref{fig:bubbles-lic-100-sccm-off}d-g and \ref{fig:bubbles-lic-100-sccm-off}i. The wake area also oscillates with a greater amplitude in the \textit{YZ} plane. Looking at the \textit{XY} planes in Figures \ref{fig:bubbles-lic-30-sccm-off} and \ref{fig:bubbles-lic-100-sccm-off} one may also notice that the wake several bubble diameters below the bubble position is much more disordered.

When MF is applied, bubble wake is laminarized and its size decreases considerably as seen in Figures \ref{fig:bubbles-lic-30-sccm-on} and \ref{fig:bubbles-lic-100-sccm-on} for $30$ and $100~sccm$, respectively. One can see that in both cases vortices do not form or detach beneath the bubble. Note, however, an important difference between the two cases besides the overall velocity magnitude increase in the $100~sccm$ case: there are considerable wake oscillations in the \textit{YZ} plane (e.g. Figures \ref{fig:bubbles-lic-100-sccm-on}c and \ref{fig:bubbles-lic-100-sccm-on}d). These, again, are due to the zeroth vessel velocity field mode (Figure \ref{fig:vessel-mode-100-sccm-on-m0}). Note Figures \ref{fig:bubbles-lic-30-sccm-on}a and \ref{fig:bubbles-lic-100-sccm-on}a -- wake inclination in the \textit{XZ} plane occurs during bubble detachment from the horizontally (positive \textit{X} direction) oriented inlet.

\clearpage

Looking at the $Q$ field within bubble wakes for $30~sccm$ without applied MF (Figure \ref{fig:bubbles-q-30-sccm-off}) one can see the usual pattern of pairs of hairpin-shaped vortices as indicated by their highlighted cores. Note that the cores are rather ordered and the $Q$ field does not exhibit a high density of smaller scale vortices about the main wake cores, as well as the fact that there is no significant interference from leading bubbles above the sampling volumes. 

This is in contrast with the $100~sccm$ case shown in Figure \ref{fig:bubbles-q-100-sccm-off} where one can see that vortex structures left over in the wake of leading bubbles do not dissipate before trailing bubbles interact with them. The combination of this factor and more intense turbulent pulsations within bubble wakes act to disrupt the otherwise more regular vortex cores below the bubbles. The strong influence of leading bubble wakes is especially evident in Figures \ref{fig:bubbles-q-100-sccm-off}c, \ref{fig:bubbles-q-100-sccm-off}g, \ref{fig:bubbles-q-100-sccm-off}i. The bubble in Figure \ref{fig:bubbles-q-100-sccm-off}a is one of the earlier bubbles to be released, hence its more ordered wake and surroundings.

If MF is applied, electromagnetic damping of the wake flow results in wake vortex suppression as seen in Figure \ref{fig:bubbles-q-30-sccm-on} were vortex cores below the bubbles are not present and only the vortex ($Q>0$) and saddle-type flow ($Q<0$) envelopes above the bubble and about its interface are pronounced.

\clearpage

For $100~sccm$ with applied MF (Figure \ref{fig:bubbles-q-100-sccm-on}) the $Q$ field patterns are rather similar to the $30~sccm$ case, with only slightly more pronounced vortex and saddle-type flow envelopes and slightly higher $Q$ magnitudes directly beneath the bubbles. Note the bubble wakes in Figures \ref{fig:bubbles-q-30-sccm-on}c and \ref{fig:bubbles-q-100-sccm-on}a -- these correspond to wake snapshots after bubble detachment from the inlet, which is accompanied by a strong perturbation and vorticity generation within the wake which is then almost immediately laminarized and damped by the Lorentz force \cite{birjukovsPhaseBoundaryDynamics2020}.

\clearpage

\subsection{Bubble reference frame: DMD of the wake velocity field}

\begin{figure}[H]
    \centering
    \includegraphics[width=0.25\textwidth]{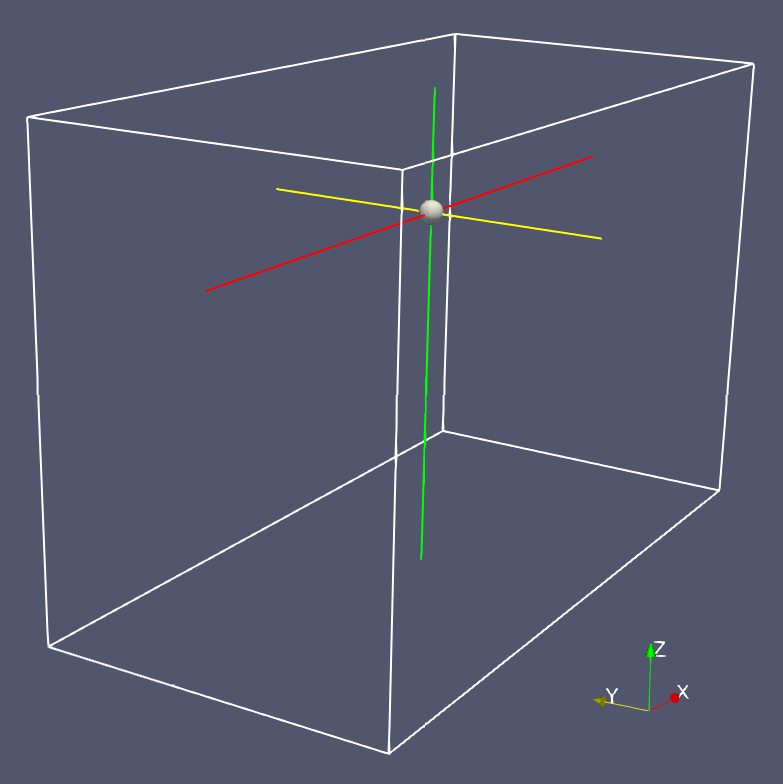}
    \caption{The sampling box for the velocity field in the bubble reference frame. The gray dot at the axes' intersection is the bubble centroid. The box dimensions are: $x^-=x^+=15~mm$; $y^-=y^+=7.5~mm$; $z^+=5~mm$; $z^-=-15~mm$.}
    \label{fig:sampling-box-velocity}
\end{figure}

To compute the DMD modes for the wake velocity field in the bubble reference frame, we use the bubbles with trajectories within the time window of the vessel DMD analysis and remove trajectories shorter than 10 frames. The remaining trajectories are trimmed to include bubbles in $z \in [40~ mm, 110~ mm]$. This is done to avoid sharp field transients near the inlet and near the top. The transition to the bubble reference frame is performed as follows: bubble trajectories are interpolated with cubic splines over time and instantaneous bubble velocities are calculated by taking the position time derivative along the splines; the velocity field is resampled to a rectangular box (Figure \ref{fig:sampling-box-velocity}) that follows bubble centroids and the bubble centroid velocity is subtracted from the resampled velocity field. The sampling box is extended in the $z^-$ direction to capture more of the bubble wake velocity field.

A spherical mask with a $4.5~mm$ radius centered at the bubble centroid is applied to the sampled velocity field. This is done to mask the argon velocity field variations within bubbles. Real-to-complex domain mapping is applied to each field grid point over the time range of a trajectory. Field snapshot stacking (\ref{eq:M01}) is applied to trajectories separately yielding $M_0$ and $M_1$ matrices for every trajectory. Matrices $M_0$ are trimmed in time on either end to prevent trajectory overlap due to time-delay stacking, then concatenated in order of their trajectory appearance in time to form $M'_0$. The same is done for $M_1$ producing $M'_1$, respectively. DMD is then performed with $M'_0$ and $M'_1$ yielding modes for the entire ensemble of trajectories. Initial amplitudes are calculated for each trajectory separately as in \cite{jovanovicSparsitypromotingDynamicMode2014}. As was the case for the vessel, the first 20 modes were computed for the bubble wake DMD.

\begin{figure}[H]
    \centering
    \includegraphics[width=0.4\textwidth]{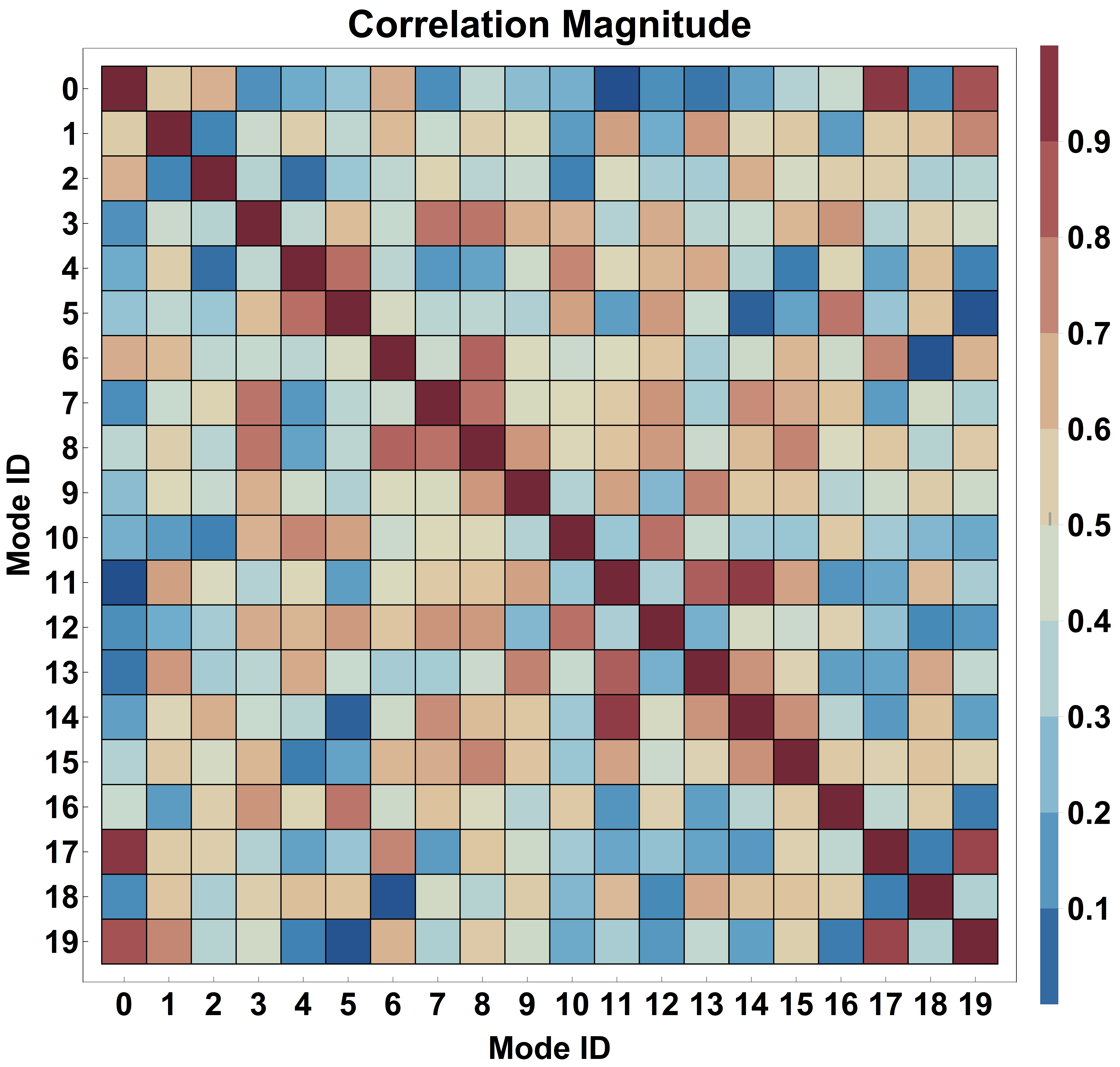}
    \caption{The bubble DMD mode correlation magnitude matrix for $30~sccm$ with applied MF.}
    \label{fig:u-bubble-mode-correlations-30-on}
\end{figure}

Starting with $30~sccm$ with applied MF, Figure \ref{fig:u-bubble-mode-correlations-30-on} shows that in the case of the modes for the bubble reference frame velocity field in the sampling box (Figure \ref{fig:sampling-box-velocity}) -- referred to as simply "bubble modes" for brevity -- the modes are rather strongly correlated. This means that, unlike the vessel velocity field modes, these must be interpreted jointly and a more in-depth analysis is required.

\clearpage

\begin{figure}[H]
    \centering
    \includegraphics[width=1\textwidth]{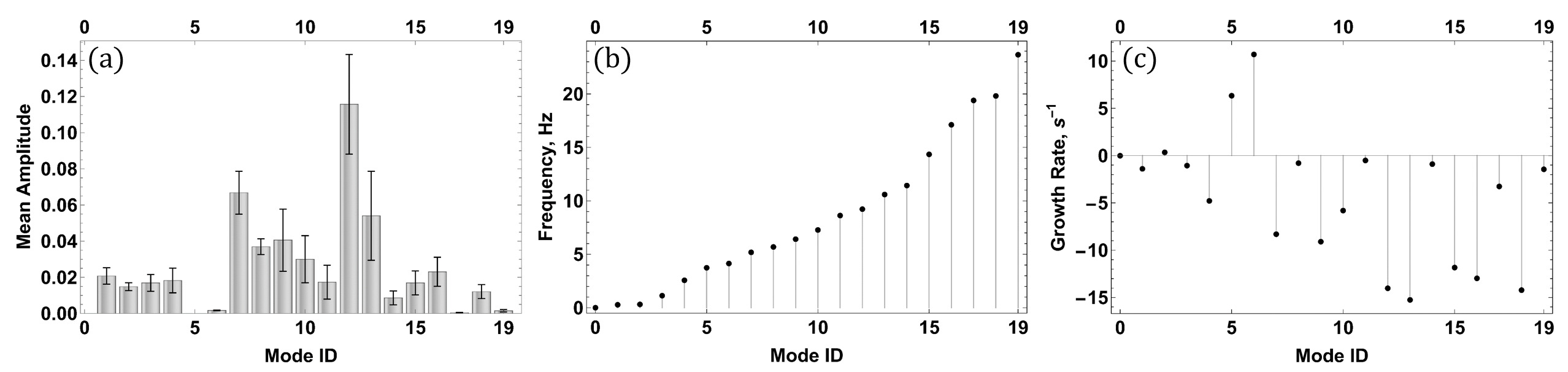}
    \caption{$30~sccm$ with applied MF: bubble velocity field mode (a) normalized initial amplitudes, (b) frequencies and (c) growth rates. Normalization with respect to the zeroth mode.}
    \label{fig:u-bubble-mode-stats-30-on}
\end{figure}

\begin{figure}[H]
    \centering
    \includegraphics[width=1\textwidth]{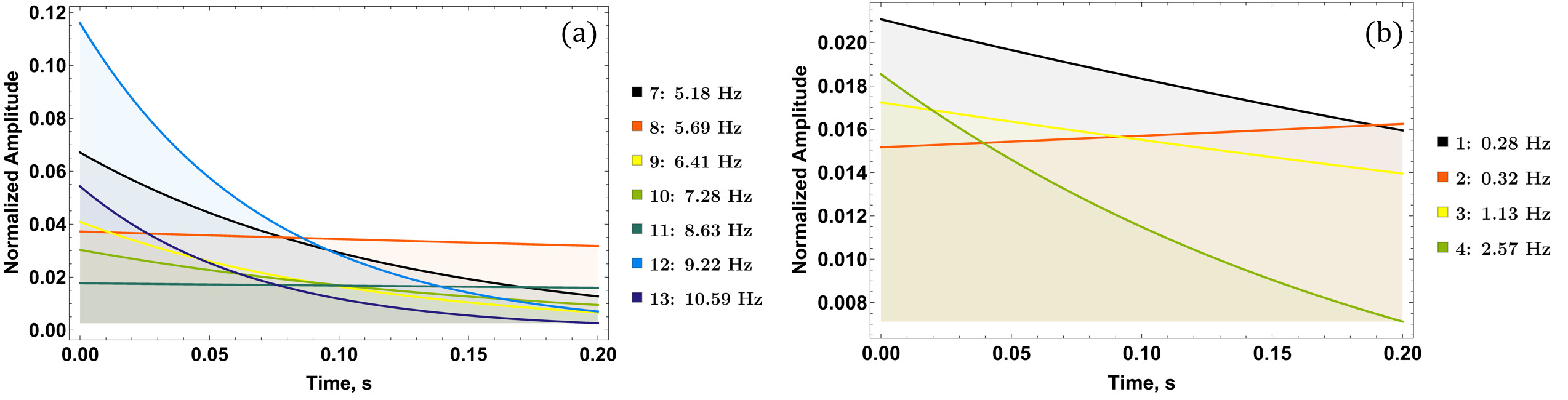}
    \caption{$30~sccm$ with applied MF: dynamics of normalized amplitudes over the mean trajectory time for significant modes. Normalization with respect to the zeroth mode. Legend: mode IDs and frequencies.}
    \label{fig:u-bubble-mode-growth-30-on}
\end{figure}

\begin{figure}[H]
    \centering
    \includegraphics[width=0.8\textwidth]{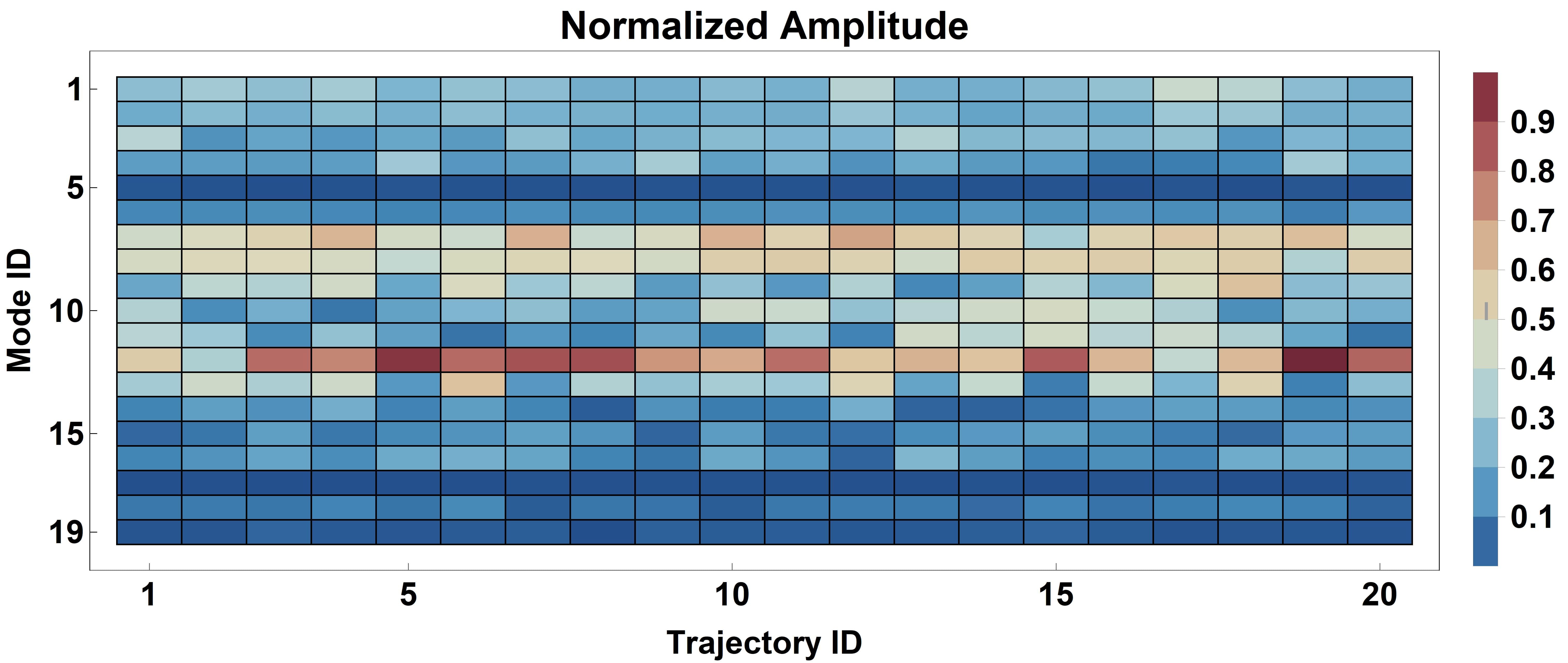}
    \caption{$30~sccm$ with applied MF: normalized root mean square (RMS) amplitudes for modes over all processed trajectories. Normalization performed with modes 1 to 19.}
    \label{fig:u-bubble-mode-rms-amps-30-on}
\end{figure}

To this end, in addition to mean initial amplitudes (as before, normalized to the zeroth DMD mode) with standard deviations, frequencies and growth rates (Figure \ref{fig:u-bubble-mode-stats-30-on}), one must also carefully examine the dynamics of normalized mode amplitudes (Figure \ref{fig:u-bubble-mode-growth-30-on}), as well as normalized root mean square (RMS) mode amplitudes for every analyzed trajectory (Figure \ref{fig:u-bubble-mode-rms-amps-30-on}). Note that in the latter case normalization is performed for modes 1 to 19, for visual purposes -- this is because the zeroth mode's amplitude is roughly an order of magnitude greater than that of the second-highest, in this case mode 12. This pattern for these three types of figures is used for the other three flow cases as well.

\begin{figure}[H]
    \centering
    \includegraphics[width=1\textwidth]{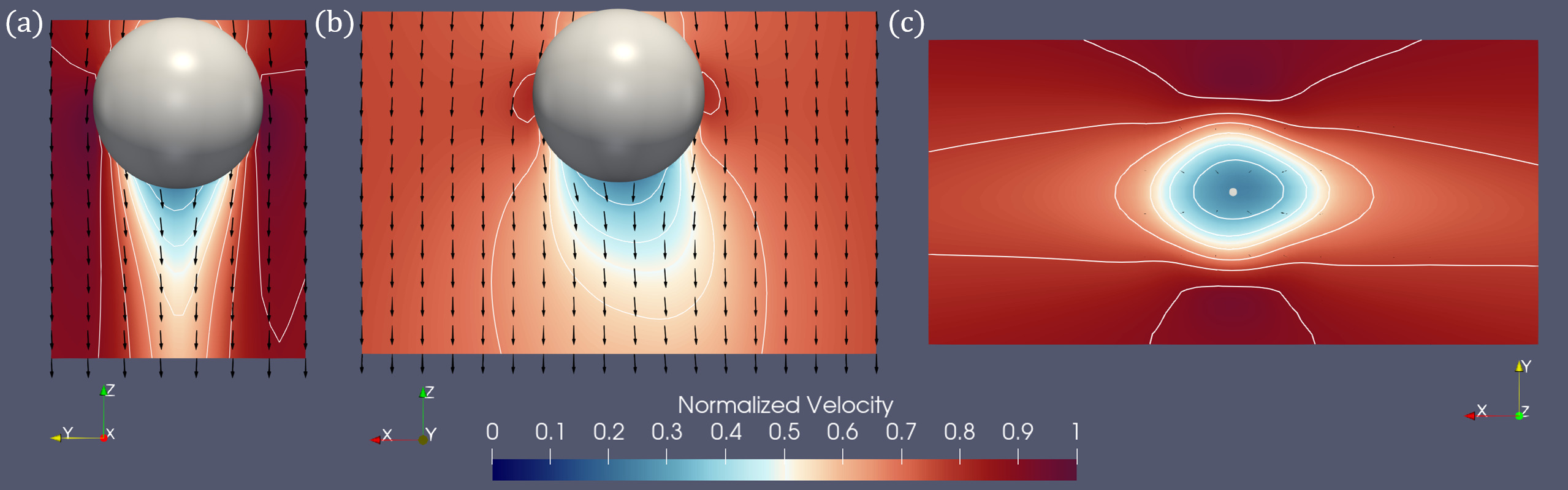}
    \caption{The zeroth bubble velocity field mode for $30~sccm$ with applied MF: (a) \textit{YZ} plane, (b) \textit{XZ} plane and (c) \textit{XY} plane. The gray specular sphere in (a) and (b) is the mask used within the DMD sampling box (Figure \ref{fig:sampling-box-velocity}). Note: the gray dot in (c) in the center of the plane is where the mask intersects the \textit{XY} plane, i.e. the plane is right below the bubble. The \textit{YZ} and \textit{XZ} planes contain the center of the spherical mask. The in-plane velocity field vector lengths in (a-c) are scaled by projecting a grid of equally spaced vectors of equal lengths onto the planes.}
    \label{fig:u-bubble-30-on-m0}
\end{figure}

The strongest bubble velocity field mode is the zeroth mode which is essentially a mean flow mode with $\omega_0 \sim 2.6~ mHz$ and $a_0 \sim -9.4 \cdot 10^{-3} ~ s^{-1}$, shown in Figure \ref{fig:u-bubble-30-on-m0}. Its relative magnitude is consistently about unity for all trajectories, with a $\sim 0.26\%$ standard deviation. This mode aside, as seen in Figures \ref{fig:u-bubble-mode-stats-30-on} and \ref{fig:u-bubble-mode-growth-30-on}a, the initially dominant modes are 12, 7 and 13 with 9, 8, 10 and 11 having smaller, but still significant amplitudes; other modes of note are 1 to 4, the lower frequency modes, their amplitude dynamics shown in Figure \ref{fig:u-bubble-mode-growth-30-on}b. Note that modes 12, 7, 13, 9 and 10 decline very quickly over trajectory time, while the amplitudes of modes 8 and 11 are almost unchanged. To put this in perspective, consider that initially the sum of amplitudes of the modes shown in Figure \ref{fig:u-bubble-mode-growth-30-on}a amounts to $\sim 0.38$ of the zeroth mode amplitude, while by the end of the trajectory time interval their contribution is reduced to $\sim 0.1$.

\begin{figure}[H]
    \centering
    \includegraphics[width=1\textwidth]{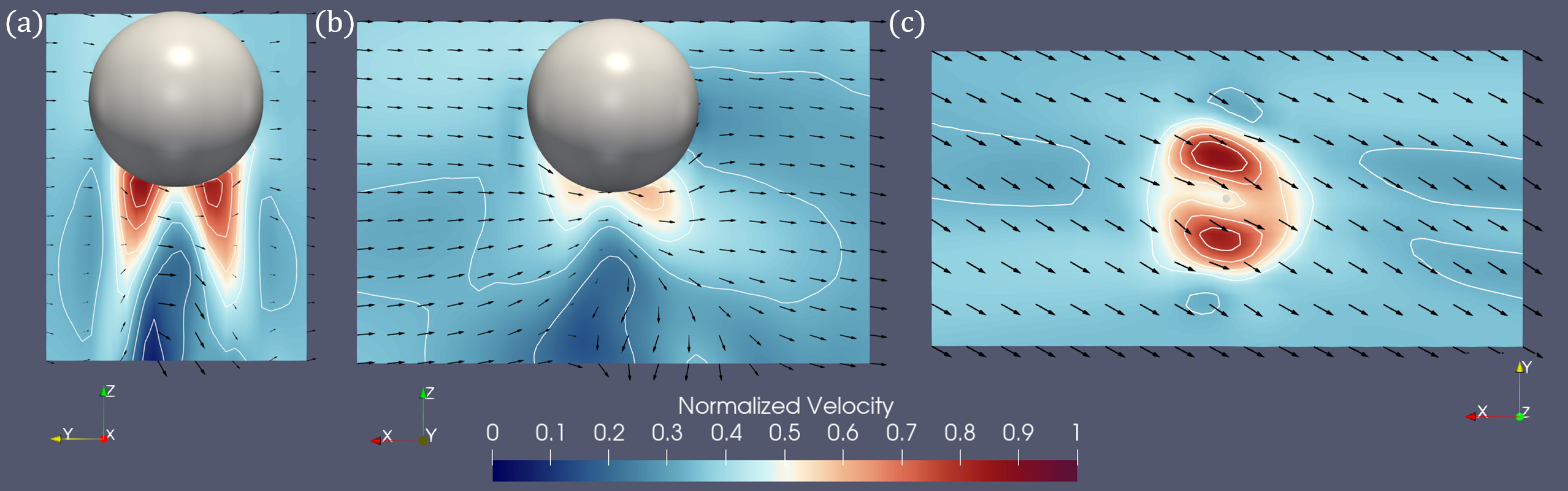}
    \caption{The 12-th bubble velocity field mode for $30~sccm$ with applied MF: (a) \textit{YZ} plane, (b) \textit{XZ} plane and (c) \textit{XY} plane.}
    \label{fig:u-bubble-30-on-m12}
\end{figure}

\begin{figure}[H]
    \centering
    \includegraphics[width=1\textwidth]{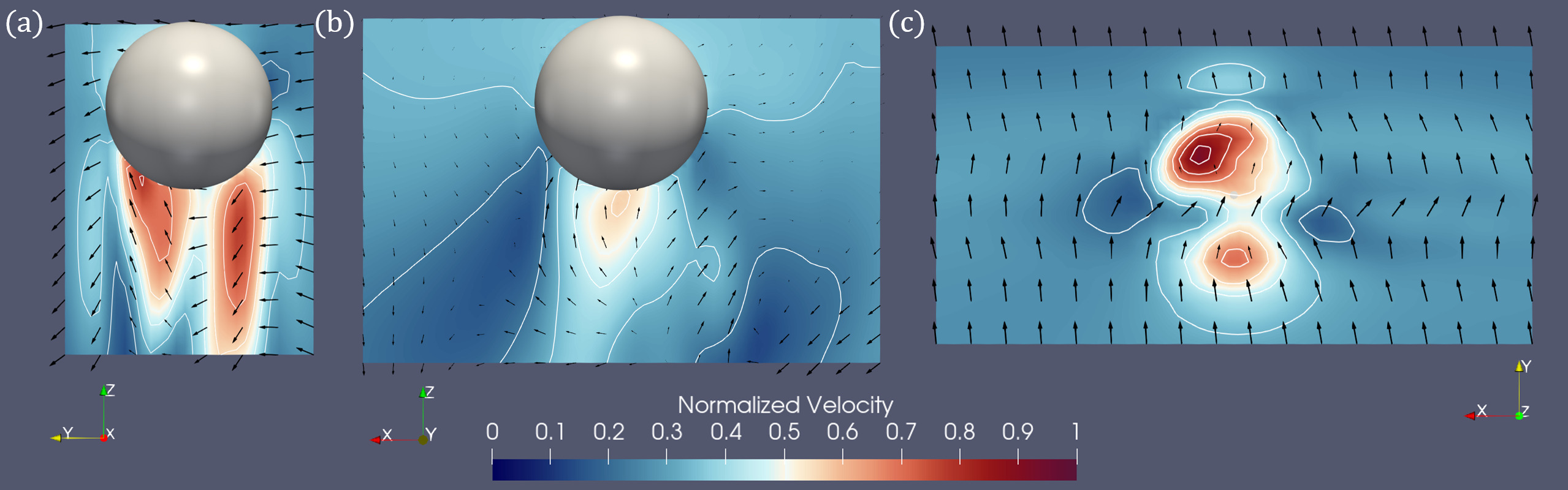}
    \caption{The 7-th bubble velocity field mode for $30~sccm$ with applied MF: (a) \textit{YZ} plane, (b) \textit{XZ} plane and (c) \textit{XY} plane.}
    \label{fig:u-bubble-30-on-m7}
\end{figure}

\begin{figure}[H]
    \centering
    \includegraphics[width=1\textwidth]{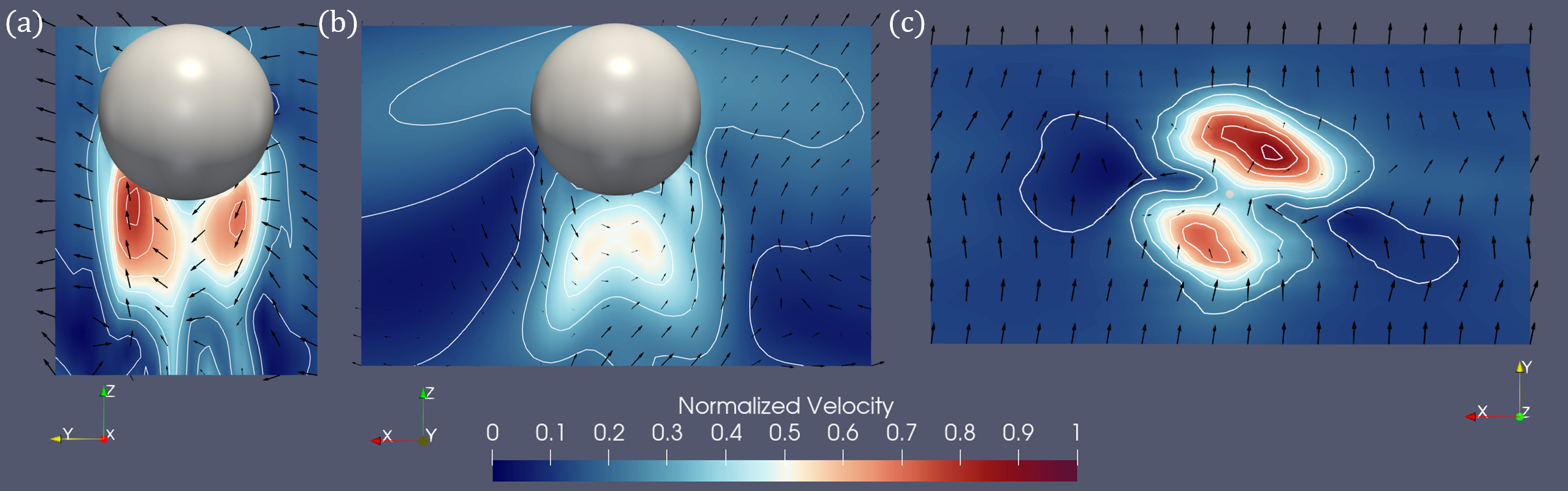}
    \caption{The 13-th bubble velocity field mode for $30~sccm$ with applied MF: (a) \textit{YZ} plane, (b) \textit{XZ} plane and (c) \textit{XY} plane.}
    \label{fig:u-bubble-30-on-m13}
\end{figure}

\begin{figure}[H]
    \centering
    \includegraphics[width=1\textwidth]{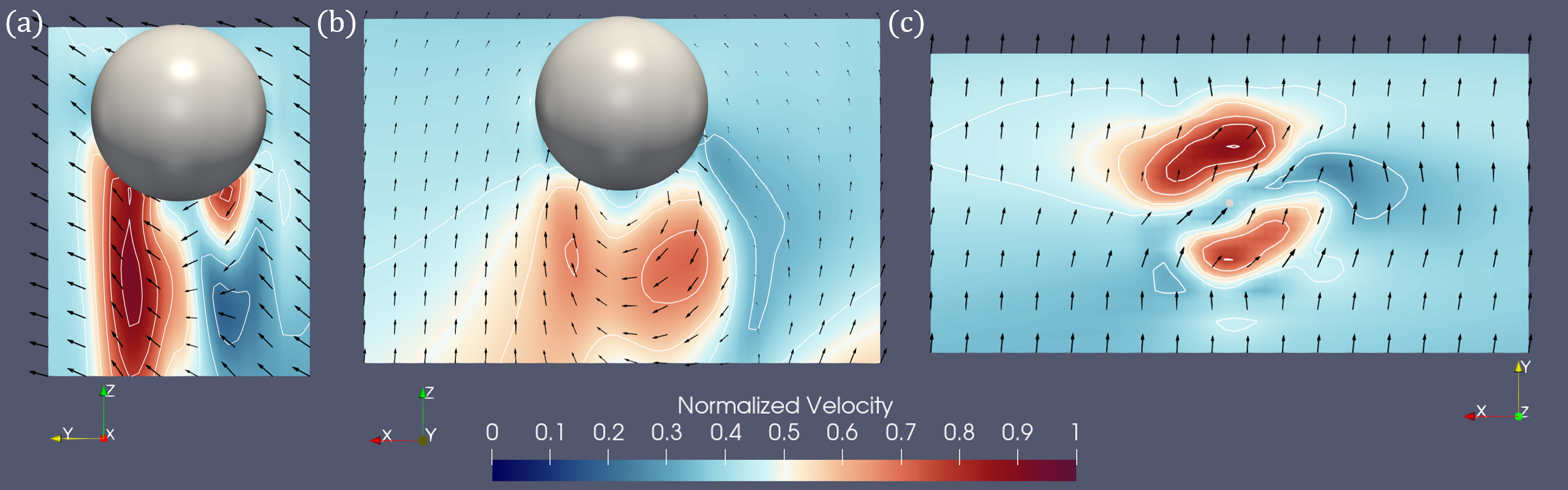}
    \caption{The 8-th bubble velocity field mode for $30~sccm$ with applied MF: (a) \textit{YZ} plane, (b) \textit{XZ} plane and (c) \textit{XY} plane.}
    \label{fig:u-bubble-30-on-m8}
\end{figure}

\begin{figure}[H]
    \centering
    \includegraphics[width=1\textwidth]{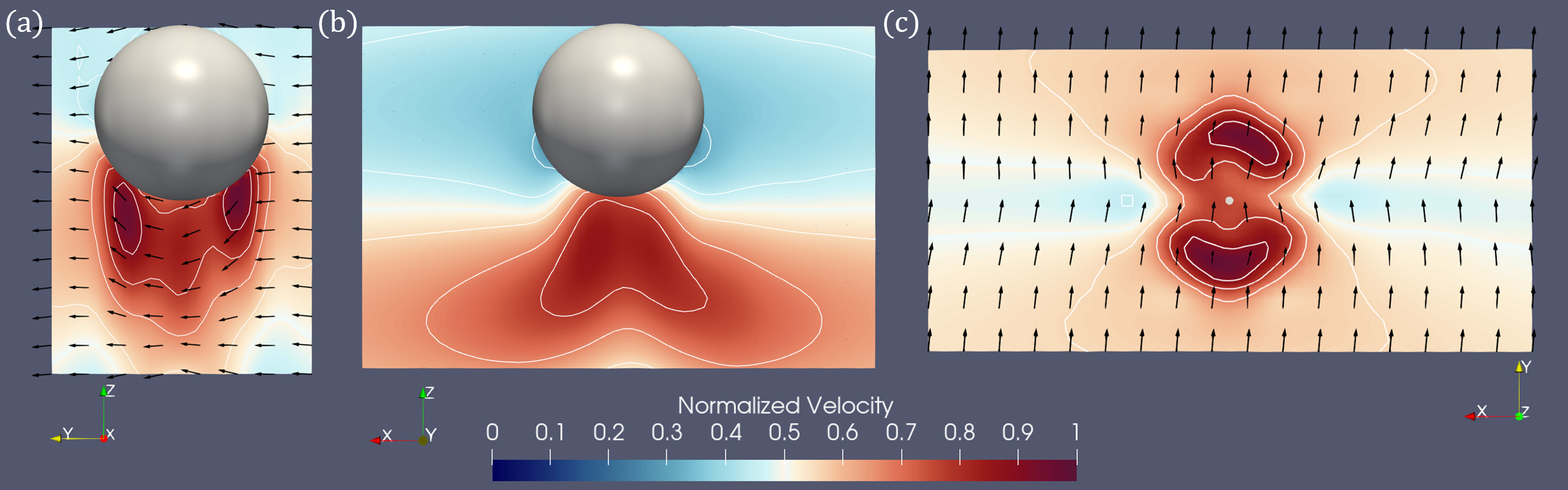}
    \caption{The 11-th bubble velocity field mode for $30~sccm$ with applied MF: (a) \textit{YZ} plane, (b) \textit{XZ} plane and (c) \textit{XY} plane.}
    \label{fig:u-bubble-30-on-m11}
\end{figure}

To understand the effects that this has on the overall velocity field, one must examine the higher order mode flow patterns -- these are shown in Figures \ref{fig:u-bubble-30-on-m12}-\ref{fig:u-bubble-30-on-m11}, in order of descending mean initial amplitude. This order of presentation is kept throughout this section. Modes 1 to 4 are not shown here, as modes 1 and 2 are slightly distorted versions of the zeroth mode (notice in Figure \ref{fig:u-bubble-mode-correlations-30-on} that modes 1 and 2 are strongly correlated to it), while modes 3 and 4 are much weaker versions of 12 and 7, respectively. Combined, modes 1 to 4 initially constitute at most $< 8 \%$ of the zeroth mode amplitude and considerably decline over trajectory time (Figure \ref{fig:u-bubble-mode-growth-30-on}).

Mode 12 ($\omega_{12} \sim 9.22 ~ Hz$, $a_{12} \sim -14 ~ s^{-1}$), as seen in Figure \ref{fig:u-bubble-30-on-m12}, mostly has a very pronounced \textit{X} component with a lesser \textit{Y} direction contribution meaning that this mode represents velocity field oscillations about the bubble mostly in the \textit{XZ} plane with less pronounced \textit{
YZ} oscillations. This could be interpreted as the flow pattern responsible for bubble trajectory oscillations in the \textit{XZ} plane in the initial stages of trajectories, as seen in Figure \ref{fig:trajectories-30-sccm-field-on} -- note the green lines representing trajectory projections onto the \textit{XZ} plane. The exponential decay of mode 12 which decreases in amplitude $\sim$ threefold over the first third of the trajectory time interval and its frequency are consistent with the rapidly damped oscillations observed for the \textit{XZ} projections. Figure \ref{fig:u-bubble-mode-rms-amps-30-on} suggests this is the case for most of the trajectories.

Modes 7 ($\omega_{7} \sim 5.2 ~ Hz$, $a_{7} \sim -8.3 ~ s^{-1}$) and 13 ($\omega_{13} \sim 11 ~ Hz$, $a_{13} \sim -15 ~ s^{-1}$), on the other hand, contain velocity field oscillations in the \textit{YZ} plane, \textit{Y} direction with minor out-of-plane perturbations (Figures \ref{fig:u-bubble-30-on-m7} and \ref{fig:u-bubble-30-on-m13}). While these modes decay rather quickly, modes 8 ($\omega_{8} \sim 5.7 ~ Hz$, $a_{8} \sim -0.8 ~ s^{-1}$) and 11 ($\omega_{11} \sim 8.6 ~ Hz$, $a_{11} \sim -0.5 ~ s^{-1}$) which are similar have very small growth rates and thus persist over the mean trajectory time. Mode 9 is not shown here because it is much weaker than and remarkably similar to mode 12.

It could be speculated that these modes determine the trajectory oscillations in the \textit{YZ} plane, especially considering that these oscillations have increased amplitudes for later trajectories (Figure \ref{fig:trajectories-30-sccm-field-on}), which is consistent with somewhat greater RMS amplitudes observed for modes 7, 8 and 11 for later trajectories (Figure \ref{fig:u-bubble-mode-rms-amps-30-on}). Importantly, the mode periods are on the order of or less than the mean trajectory time. Note that this behaviour is not necessarily evident from inspections of vessel modes for $30~sccm$ with applied MF (e.g. Figure \ref{fig:vessel-mode-30-sccm-on-m0}), indicating the potential usefulness of applying DMD to bubble wakes separately. However, longer DMD sampling times are required to be certain. 

It should also be noted that considerable correlation is seen between the above modes: $7 \leftrightarrow (8,12)$, $8 \leftrightarrow 12$, $11 \leftrightarrow 13$ (Figure \ref{fig:u-bubble-mode-correlations-30-on}). Note also that the showcased modes exhibit a great deal of symmetry and none of them contain small scale spatial structures, indicating wake flow laminarization by applied MF, which is, again, consistent with bubble wakes observed in Figures \ref{fig:bubbles-lic-30-sccm-on} and \ref{fig:bubbles-q-30-sccm-on}.

\begin{figure}[H]
    \centering
    \includegraphics[width=0.4\textwidth]{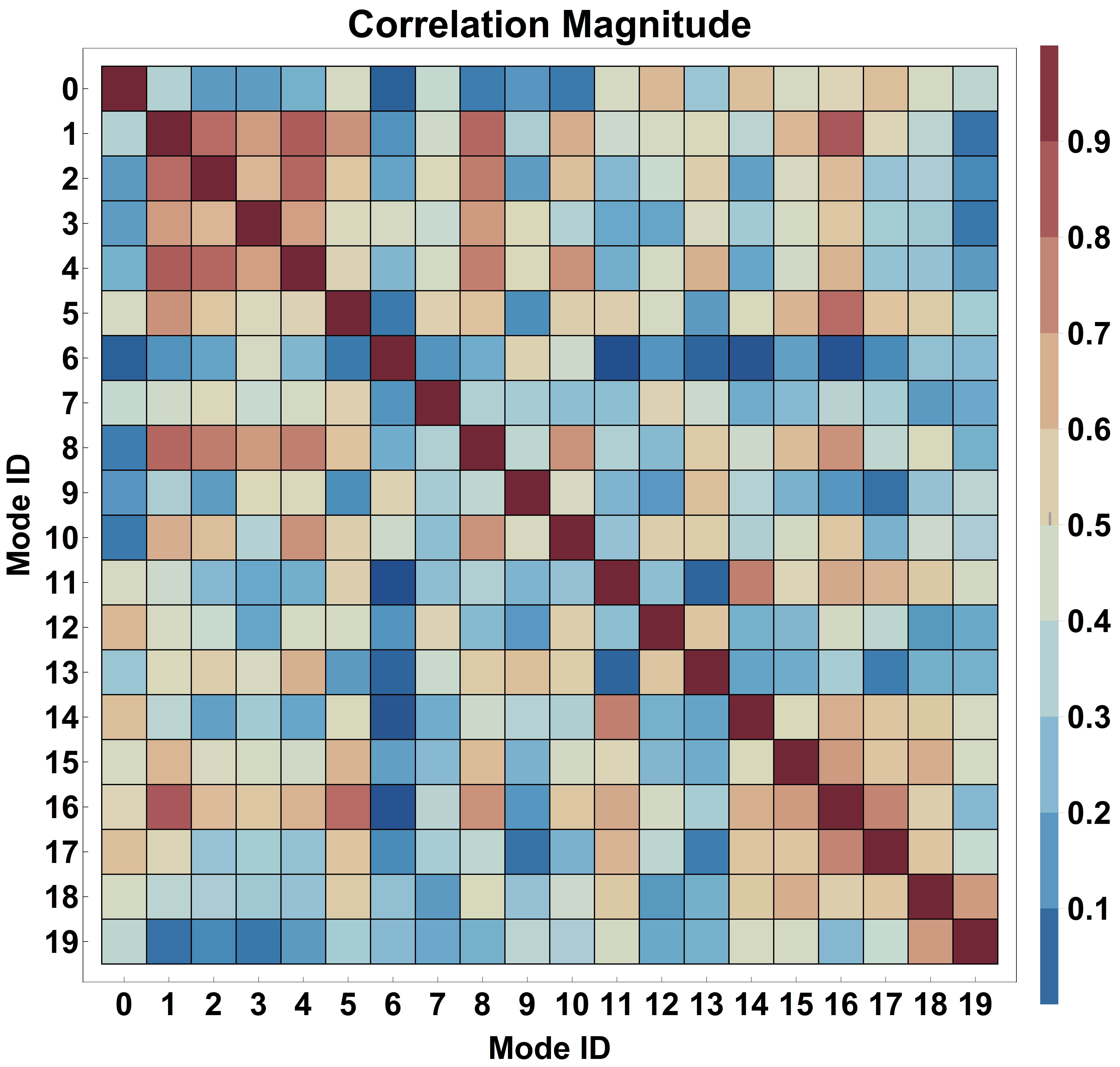}
    \caption{The bubble DMD mode correlation magnitude matrix for $100~sccm$ with applied MF.}
    \label{fig:u-bubble-mode-correlations-100-on}
\end{figure}

\begin{figure}[H]
    \centering
    \includegraphics[width=1\textwidth]{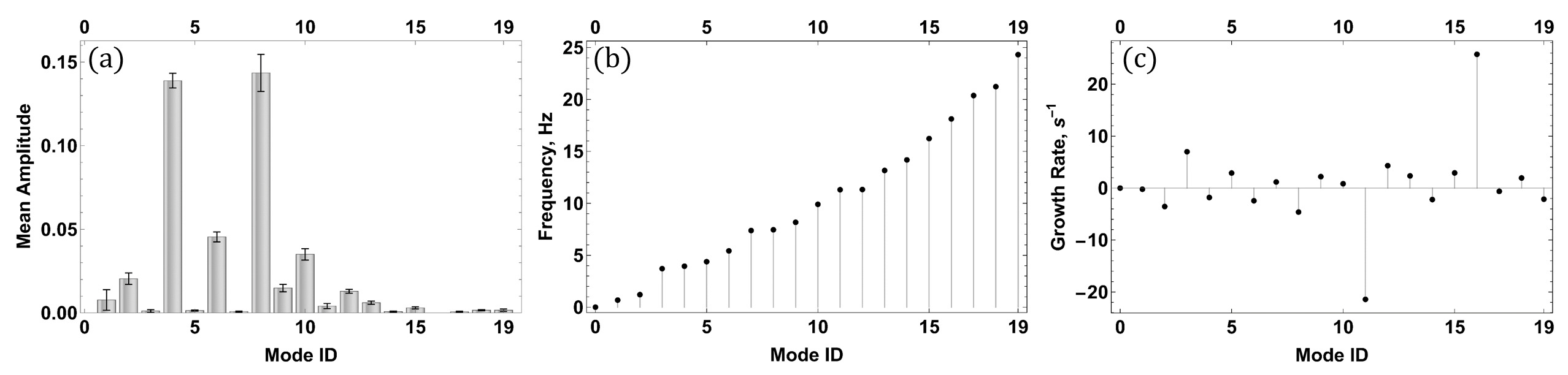}
    \caption{$100~sccm$ with applied MF: bubble velocity field mode (a) normalized initial amplitudes, (b) frequencies and (c) growth rates.}
    \label{fig:u-bubble-mode-stats-100-on}
\end{figure}

\begin{figure}[H]
    \centering
    \includegraphics[width=1\textwidth]{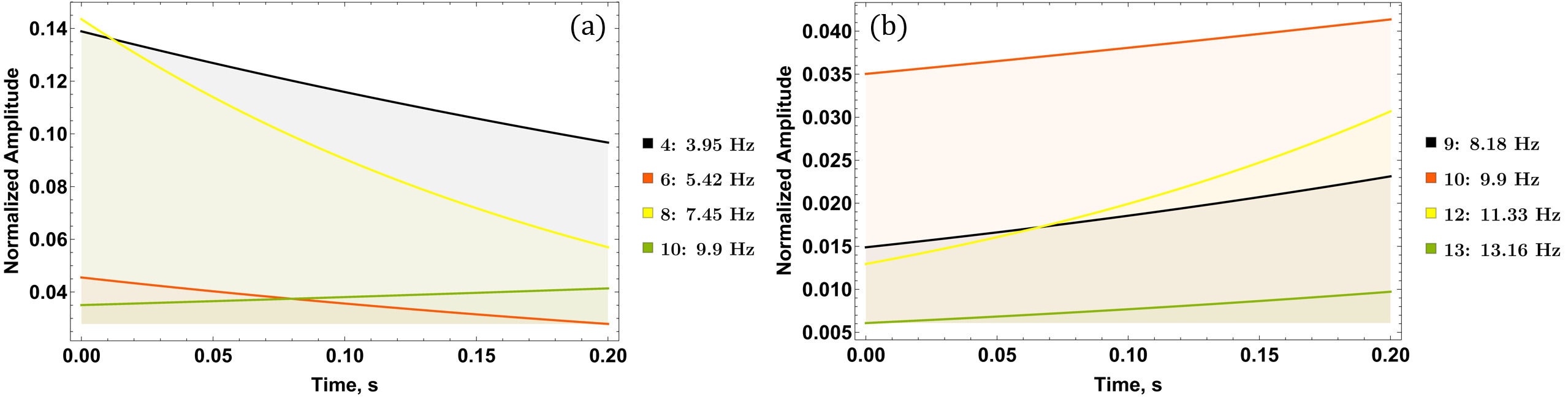}
    \caption{$100~sccm$ with applied MF: amplitude dynamics over the mean trajectory time for significant modes.}
    \label{fig:u-bubble-mode-growth-100-on}
\end{figure}

\begin{figure}[H]
    \centering
    \includegraphics[width=0.8\textwidth]{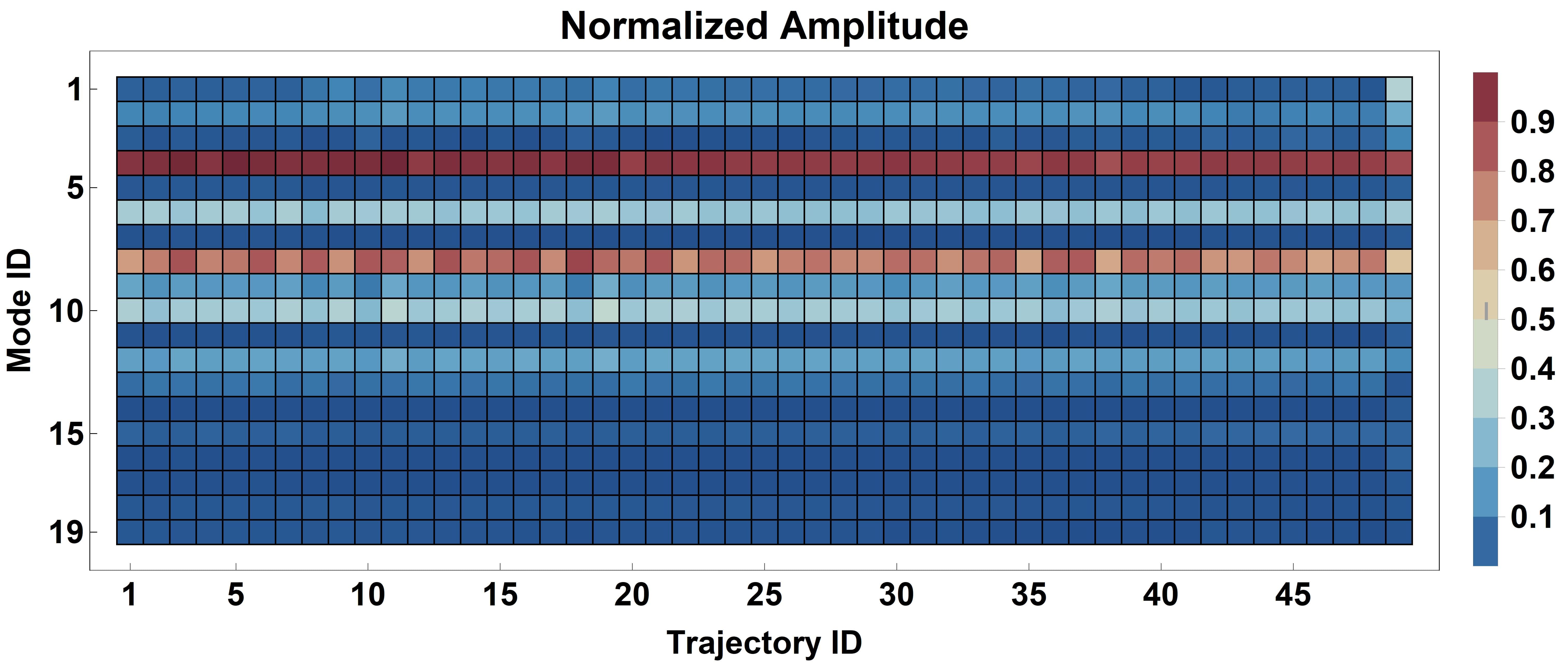}
    \caption{$100~sccm$ with applied MF: normalized RMS amplitudes for modes over all processed trajectories.}
    \label{fig:u-bubble-mode-rms-amps-100-on}
\end{figure}

Mode analysis results for $100~sccm$ with applied MF are shown in Figures \ref{fig:u-bubble-mode-correlations-100-on}-\ref{fig:u-bubble-mode-rms-amps-100-on}. As in the $30~sccm$ case, many of the DMD modes are significantly correlated (Figure \ref{fig:u-bubble-mode-correlations-100-on}). Figure \ref{fig:u-bubble-mode-stats-100-on}a indicates that there are only two modes with relatively high initial amplitudes but, unlike the $30~sccm$ case, here one can see in Figure \ref{fig:u-bubble-mode-stats-100-on}c that there are quite a few modes with positive growth rates, even though the rate magnitudes are relatively small with the exception of mode 16 -- this mode, however, has an extremely small initial amplitude and despite the highest growth rate has an insignificant RMS amplitude, as seen in Figure \ref{fig:u-bubble-mode-rms-amps-100-on}. It is also interesting to note that the mode growth damping from $30~sccm$ to $100~sccm$ with applied MF observed in Figures \ref{fig:u-bubble-mode-stats-30-on}c and \ref{fig:u-bubble-mode-stats-100-on}c is also obverved for vessel DMD modes (Figures \ref{fig:stats-vessel-modes-30-on}c and \ref{fig:stats-vessel-modes-100-on}c) with the distinction that in the vessel mode case one has mostly negative growth rates for $100~sccm$, not $30~sccm$ as it is with the bubble modes. Figures \ref{fig:u-bubble-mode-growth-100-on} and \ref{fig:u-bubble-mode-rms-amps-100-on} indicate that it makes sense to take a closer look at modes 4, 6, 8-10 and 12 as these are either initially dominant by a considerable margin or persist at or grow to a significant amplitude.

\begin{figure}[H]
    \centering
    \includegraphics[width=1\textwidth]{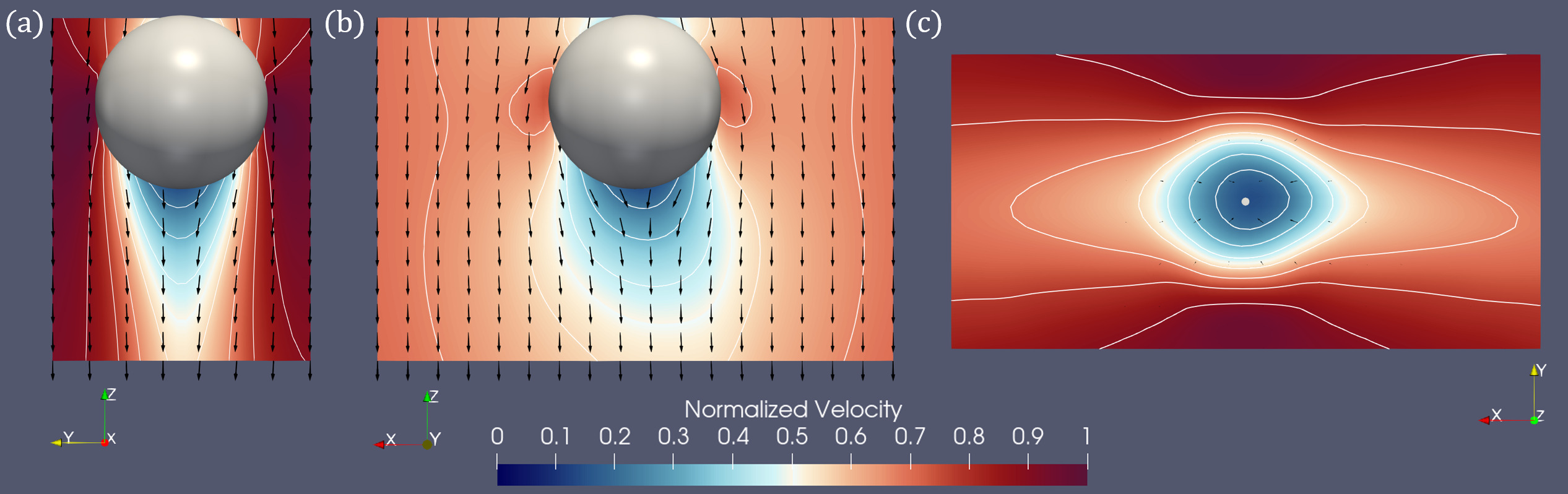}
    \caption{The zeroth bubble velocity field mode for $100~sccm$ with applied MF: (a) \textit{YZ} plane, (b) \textit{XZ} plane and (c) \textit{XY} plane.}
    \label{fig:u-bubble-100-on-m0}
\end{figure}

\begin{figure}[H]
    \centering
    \includegraphics[width=1\textwidth]{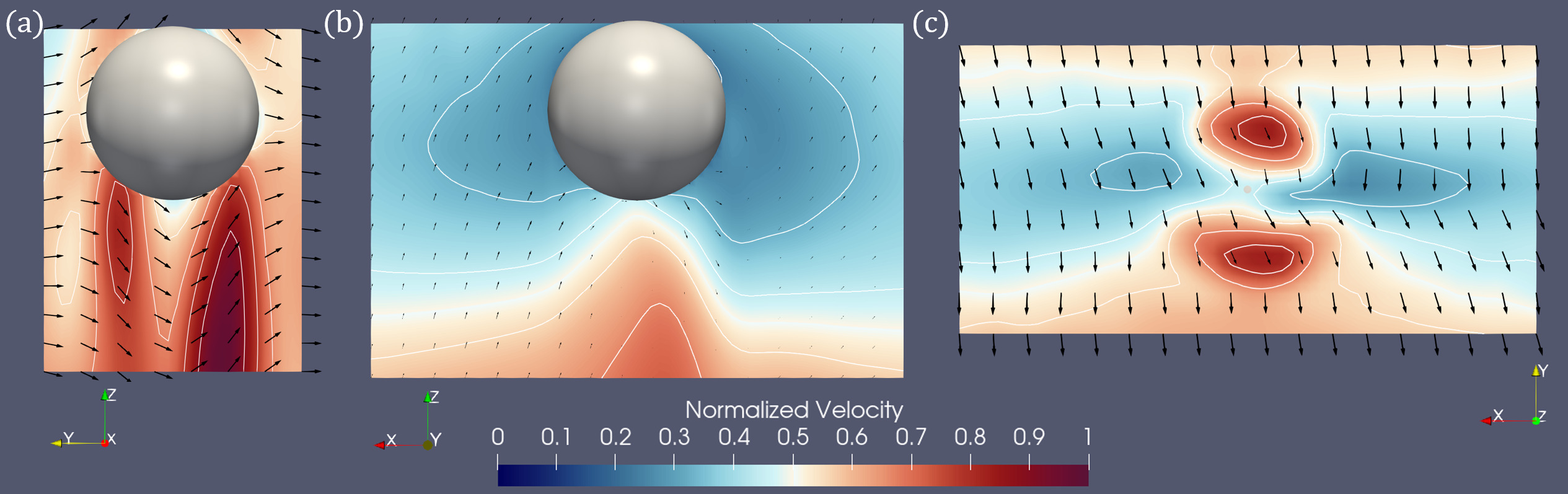}
    \caption{The 4-th bubble velocity field mode for $100~sccm$ with applied MF: (a) \textit{YZ} plane, (b) \textit{XZ} plane and (c) \textit{XY} plane.}
    \label{fig:u-bubble-100-on-m4}
\end{figure}

\begin{figure}[H]
    \centering
    \includegraphics[width=1\textwidth]{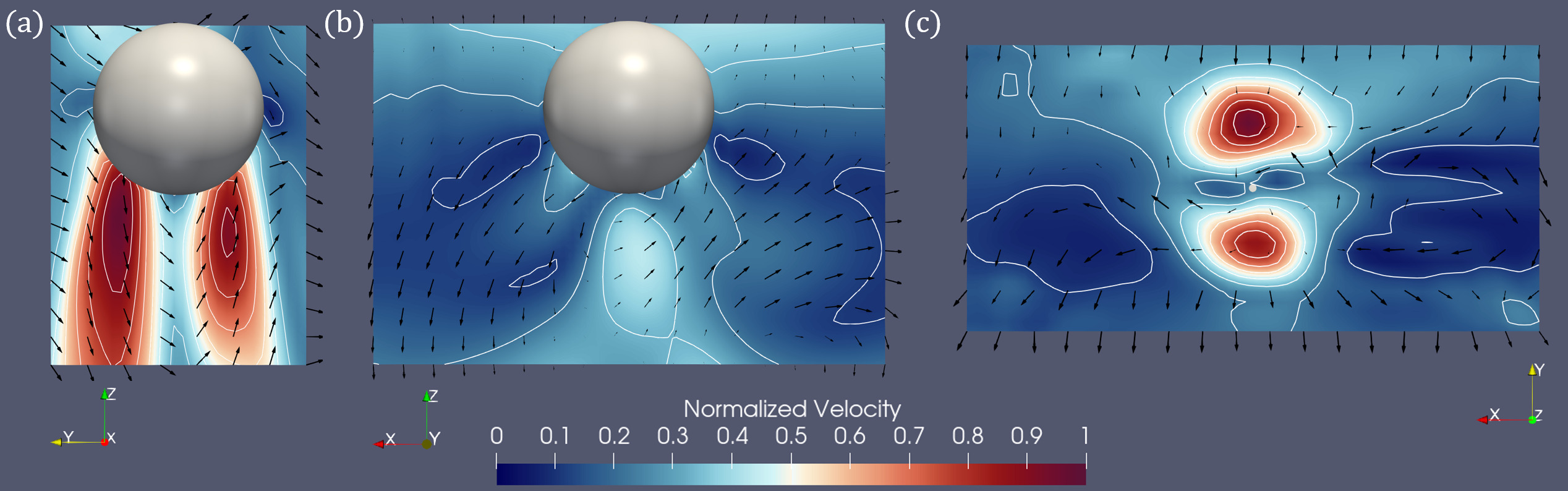}
    \caption{The 8-th bubble velocity field mode for $100~sccm$ with applied MF: (a) \textit{YZ} plane, (b) \textit{XZ} plane and (c) \textit{XY} plane.}
    \label{fig:u-bubble-100-on-m8}
\end{figure}

\begin{figure}[H]
    \centering
    \includegraphics[width=1\textwidth]{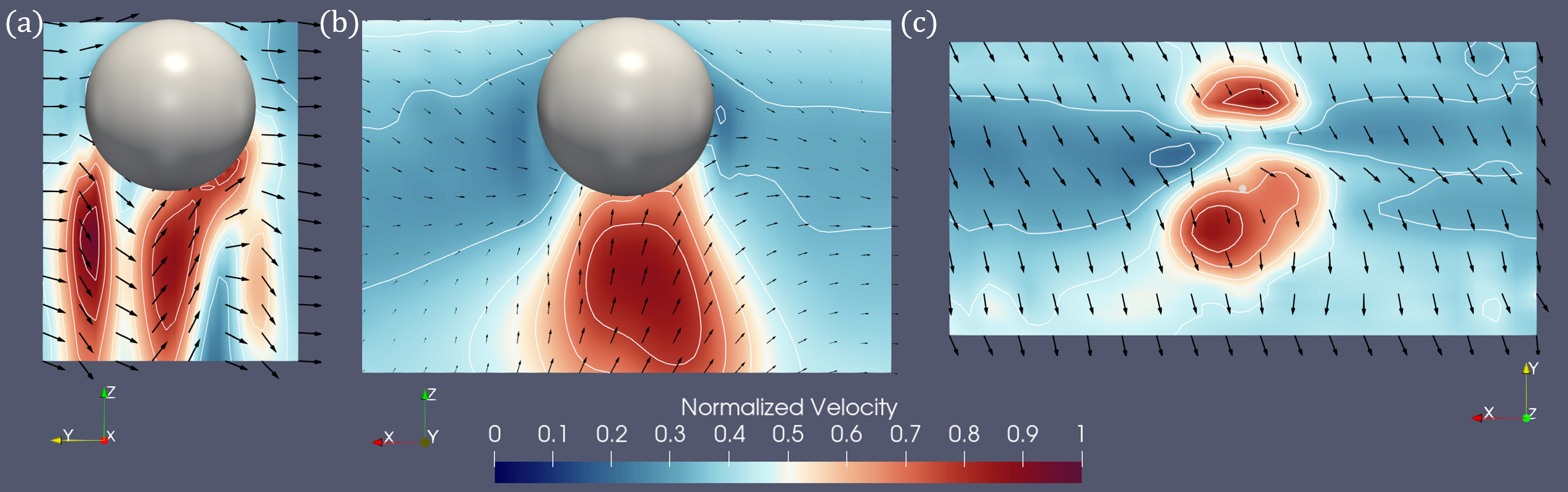}
    \caption{The 10-th bubble velocity field mode for $100~sccm$ with applied MF: (a) \textit{YZ} plane, (b) \textit{XZ} plane and (c) \textit{XY} plane.}
    \label{fig:u-bubble-100-on-m10}
\end{figure}

\begin{figure}[H]
    \centering
    \includegraphics[width=1\textwidth]{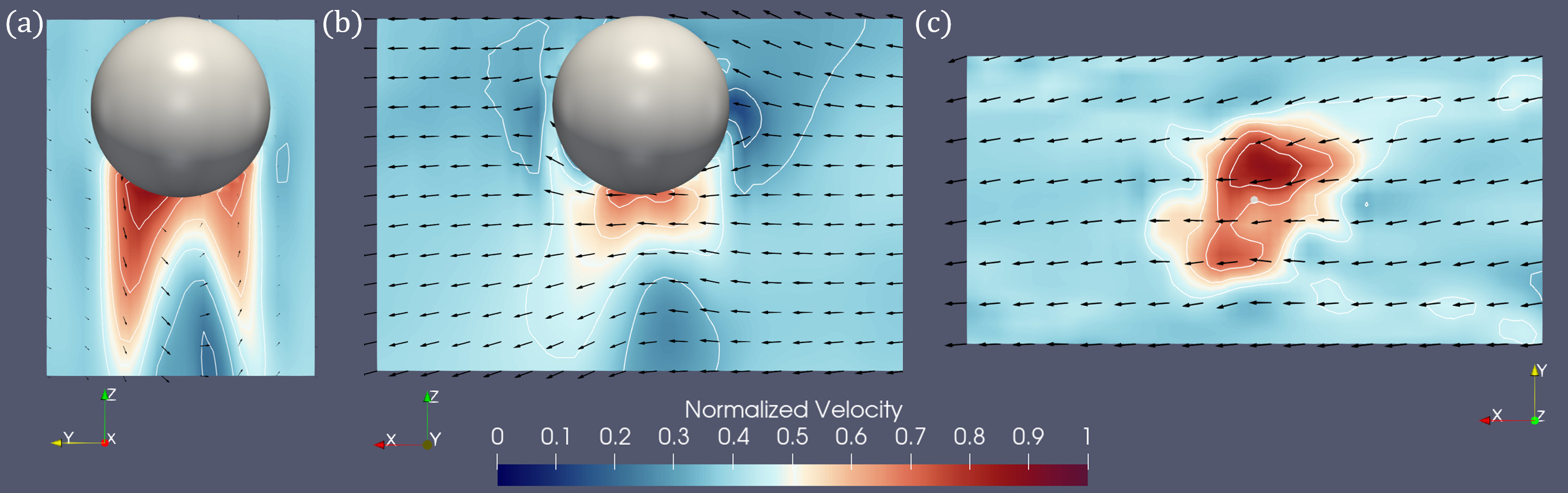}
    \caption{The 6-th bubble velocity field mode for $100~sccm$ with applied MF: (a) \textit{YZ} plane, (b) \textit{XZ} plane and (c) \textit{XY} plane.}
    \label{fig:u-bubble-100-on-m6}
\end{figure}

\begin{figure}[H]
    \centering
    \includegraphics[width=1\textwidth]{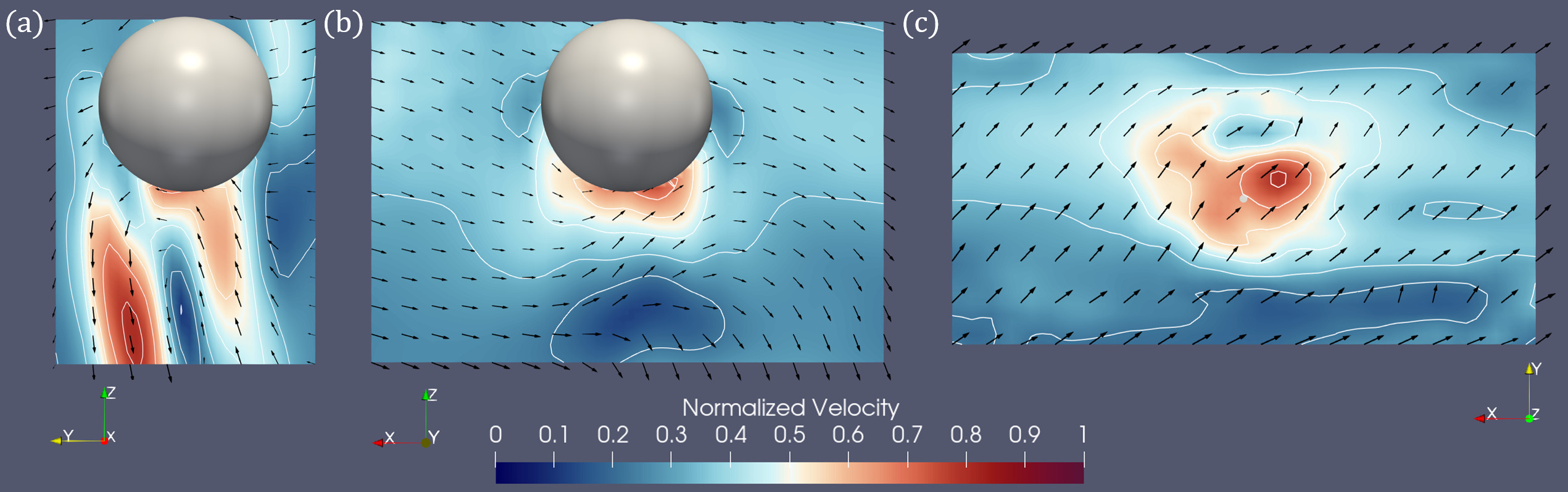}
    \caption{The 9-th bubble velocity field mode for $100~sccm$ with applied MF: (a) \textit{YZ} plane, (b) \textit{XZ} plane and (c) \textit{XY} plane.}
    \label{fig:u-bubble-100-on-m9}
\end{figure}

\begin{figure}[H]
    \centering
    \includegraphics[width=1\textwidth]{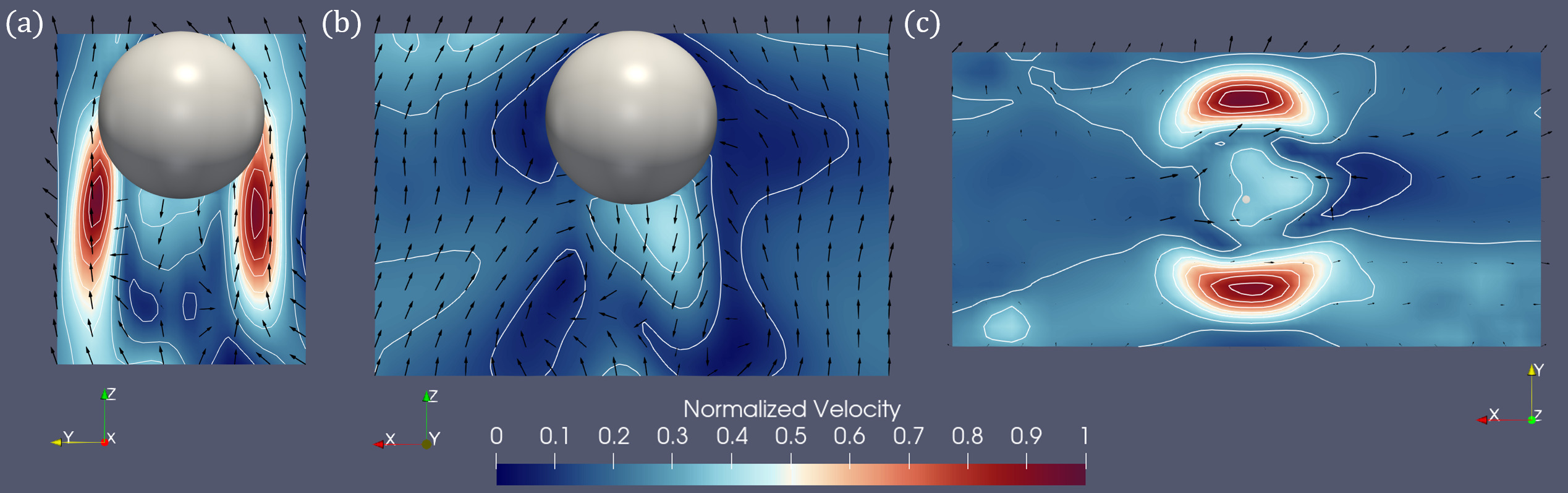}
    \caption{The 12-th bubble velocity field mode for $100~sccm$ with applied MF: (a) \textit{YZ} plane, (b) \textit{XZ} plane and (c) \textit{XY} plane.}
    \label{fig:u-bubble-100-on-m12}
\end{figure}

Significant bubble modes for $100~sccm$ with applied MF are presented in Figures \ref{fig:u-bubble-100-on-m0}-\ref{fig:u-bubble-100-on-m12}. It is immediately evident that the zeroth mode ($\omega_0 \sim 2.3~ mHz$, $a_0 \sim - 4.6 \cdot 10^{-3} ~ s^{-1}$, amplitude $\sim 1$ for all trajectories, with a $0.66\%$ deviation) in Figure \ref{fig:u-bubble-100-on-m0} is very similar to its $30~sccm$ counterpart (Figure \ref{fig:u-bubble-30-on-m0}). The difference lies in the more pronounced larger stagnation zone below the bubble. An important distinction between the modes in this case versus $30~sccm$ is that for $100~sccm$ there are high amplitude modes with much stronger \textit{Z} components within the bubble wake, i.e. modes 8 ($\omega_8 \sim 7.5 ~ Hz$, $a_8 \sim -4.6 ~ s^{-1}$, Figure \ref{fig:u-bubble-100-on-m8}), 10 ($\omega_{10} \sim 9.9 ~ Hz$, $a_{10} \sim 0.83 ~ s^{-1}$, Figure \ref{fig:u-bubble-100-on-m10}), 9 ($\omega_9 \sim 8.2 ~ Hz$, $a_9 \sim 2.2 ~ s^{-1}$, Figure \ref{fig:u-bubble-100-on-m9}) and especially 12 ($\omega_{12} \sim 11 ~ Hz$, $a_{12} \sim 4.3 ~ s^{-1}$, Figure \ref{fig:u-bubble-100-on-m12}) versus what is seen in Figures \ref{fig:u-bubble-30-on-m7}a, \ref{fig:u-bubble-30-on-m13}a and \ref{fig:u-bubble-30-on-m8}a.

One may notice a difference between Figures \ref{fig:trajectories-30-sccm-field-on} and \ref{fig:trajectories-100-sccm-field-on} -- in the $100~sccm$ case there is no pronounced initial bubble displacement in the \textit{XZ} plane unlike for $30~sccm$. Looking at the modes with significant \textit{X} components, note that for $30~sccm$ there is mode 12 (Figure \ref{fig:u-bubble-30-on-m12}) that is initially the one with the greatest magnitude after the zeroth mode; for $100~sccm$, on the other hand, mode 6 ($\omega_6 \sim 5.4 ~ Hz$, $a_6 \sim -2.4 ~ s^{-1}$, Figure \ref{fig:u-bubble-100-on-m6}) has $\sim 3$ times lower initial amplitude compared to modes 4 ($\omega_4 \sim 3.9 ~ Hz$, $a_4 \sim -1.8 ~ s^{-1}$, Figure \ref{fig:u-bubble-100-on-m4}) and 8, and mode 9 has a less pronounced \textit{X} component and consistently an even lower amplitude throughout trajectory time. The relatively small \textit{X} component contribution from these modes likely explains the differences in the \textit{XZ} projections for trajectories.

Consider also the \textit{YZ} projections of trajectories for $30~sccm$ (Figure \ref{fig:trajectories-30-sccm-field-on}) and $100~sccm$ (\ref{fig:trajectories-100-sccm-field-on}) -- notice that the deviations in the \textit{YZ} plane are overall greater in the $100~sccm$ case, even for initial trajectories (Figure \ref{fig:trajectories-100-sccm-field-on}a). Modes 4 ($\omega_4 \sim 3.9 ~ Hz$, $a_4 \sim -1.8 ~ s^{-1}$, Figure \ref{fig:u-bubble-100-on-m4}), 8, 10 and 9 all contribute to the bubble wake velocity field oscillations in the \textit{XY} plane, \textit{Z} direction, and remain dominant over trajectory time. Moreover, their combined relative contribution is overall significantly greater than that of similar modes in the $30~sccm$ case. While this might explain the trajectory oscillations in the \textit{YZ} plane -- mode frequencies suggest this is plausible -- it is not evident from Figure \ref{fig:u-bubble-mode-rms-amps-100-on} that the above mentioned modes should cause, over flow time, the transition from a still roughly rectilinear trajectory as seen in Figure \ref{fig:trajectories-100-sccm-field-on}a to trajectories with zig-zag patterns in the upper half of the metal container like in Figures \ref{fig:trajectories-100-sccm-field-on}b and \ref{fig:trajectories-100-sccm-field-on}c.

\begin{figure}[H]
    \centering
    \includegraphics[width=1\textwidth]{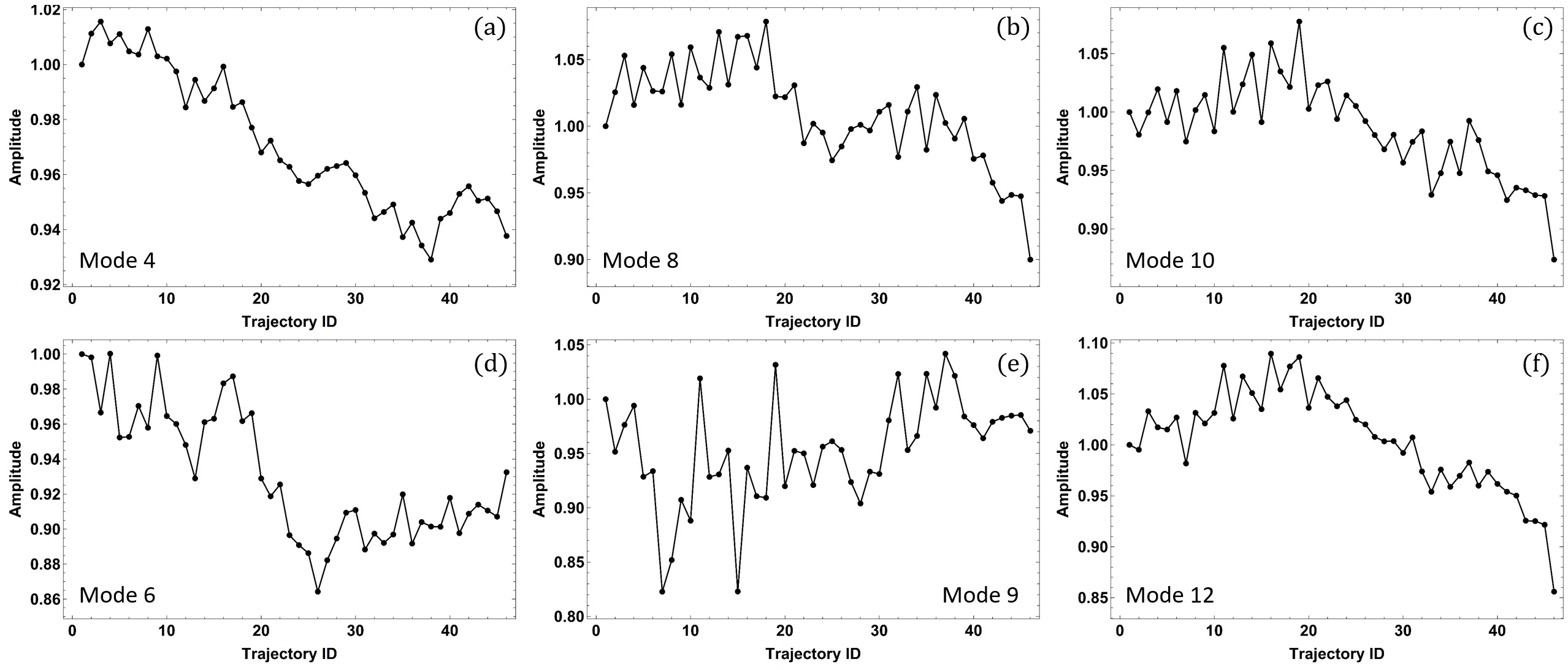}
    \caption{$100~sccm$ with applied MF: averaged RMS amplitude history over flow time (trajectory IDs) for significant bubble velocity field modes. The averaging window width is 5 trajectories. The amplitude normalization is post-averaging and separate for each mode.}
    \label{fig:u-bubble-amplitude-history-100-on}
\end{figure}

As mentioned before, this largely stems from the dominant zeroth vessel reference frame velocity field mode (Figure \ref{fig:vessel-mode-100-sccm-on-m0}) which gradually becomes even more prominent than other quickly decaying modes (Figure \ref{fig:stats-vessel-modes-100-on}c). Taking a closer look at the history of RMS amplitudes of the significant bubble modes (Figure \ref{fig:u-bubble-amplitude-history-100-on}), one can see that, if anything, modes 4, 8, 10 exhibit a slight negative trend. The fact (as far as the presented results indicate) that the trajectory forms and transitions thereof over time in this case are explained by the vessel DMD modes rather than the bubble modes would suggest that larger scale flow structures are responsible, not the bubble wake flow structures. However, it is reasonable to assume that the initial flow asymmetry/perturbations -- and therefore the corresponding bubble modes -- in the \textit{YZ} plane of the metal vessel are responsible for the onset of the larger scale patterns. The \textit{YZ} plane perturbations themselves might originate from asymmetric bubble detachment from the nozzle -- it would make sense that this asymmetry is amplified with increased gas flow rate. It is then of interest to investigate how the \textit{YZ} symmetry is broken as the flow rate varies from $30~sccm$ to $100~sccm$ and up, and to see how that is reflected in the bubble DMD modes.

Among the considered modes the following are significantly correlated (Figure \ref{fig:u-bubble-mode-correlations-100-on}): $8 \leftrightarrow (6,9)$, $10 \leftrightarrow (4,12)$.

\begin{figure}[H]
    \centering
    \includegraphics[width=0.4\textwidth]{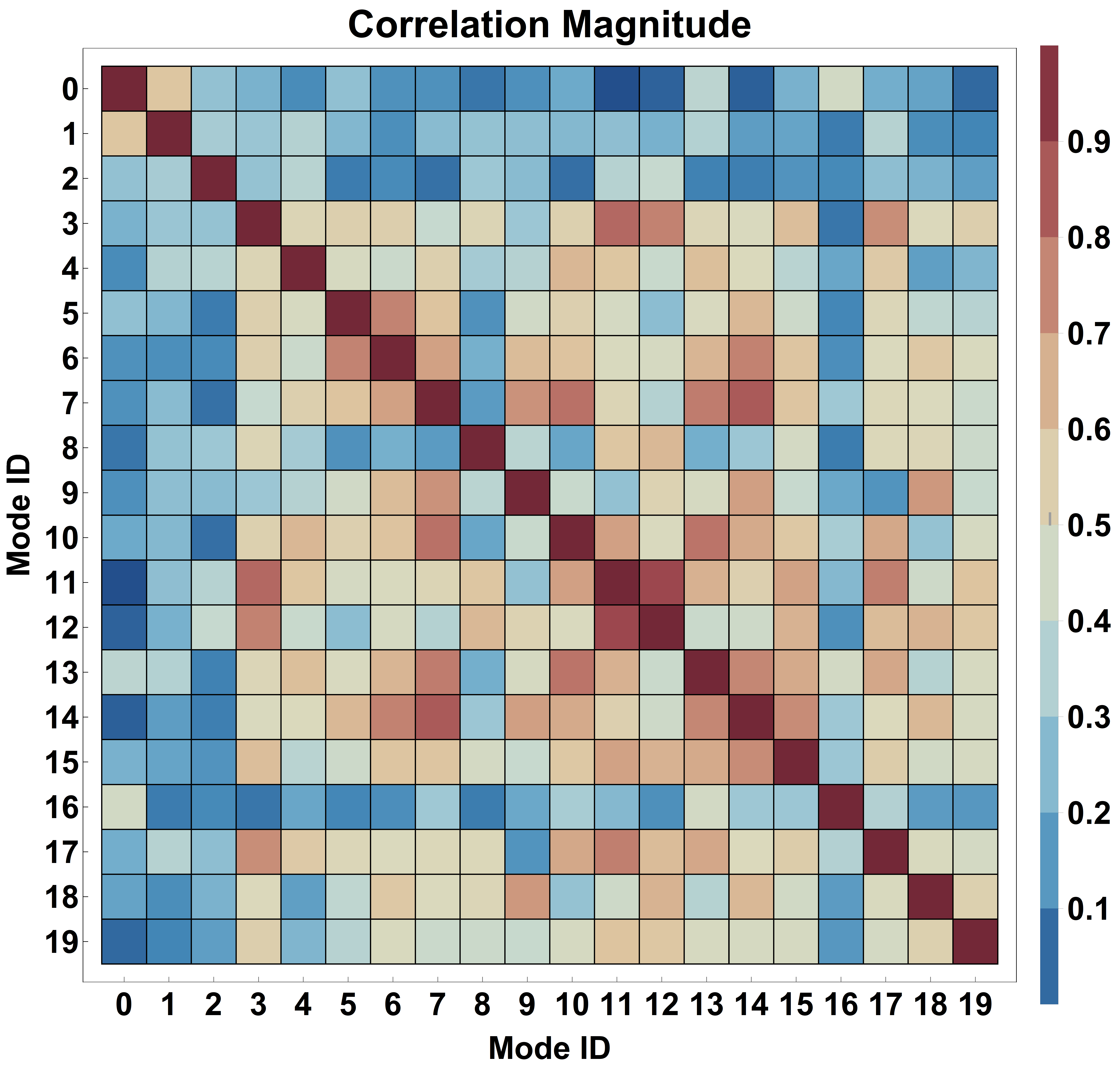}
    \caption{The bubble DMD mode correlation magnitude matrix for $30~sccm$ without applied MF.}
    \label{fig:u-bubble-mode-correlations-30-off}
\end{figure}

\begin{figure}[H]
    \centering
    \includegraphics[width=1\textwidth]{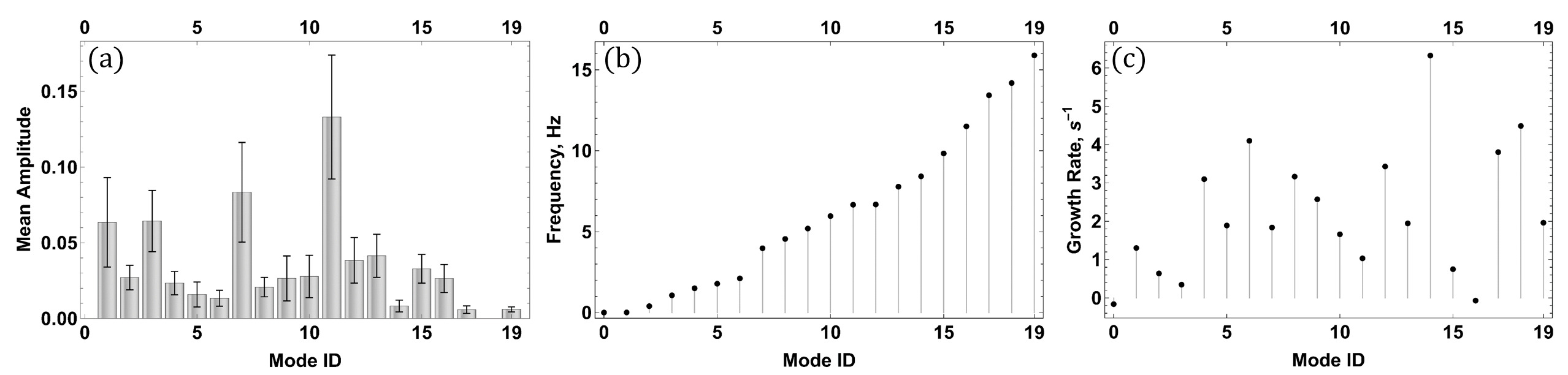}
    \caption{$30~sccm$ without applied MF: bubble velocity field mode (a) normalized initial amplitudes, (b) frequencies and (c) growth rates.}
    \label{fig:u-bubble-mode-stats-30-off}
\end{figure}

\begin{figure}[H]
    \centering
    \includegraphics[width=1\textwidth]{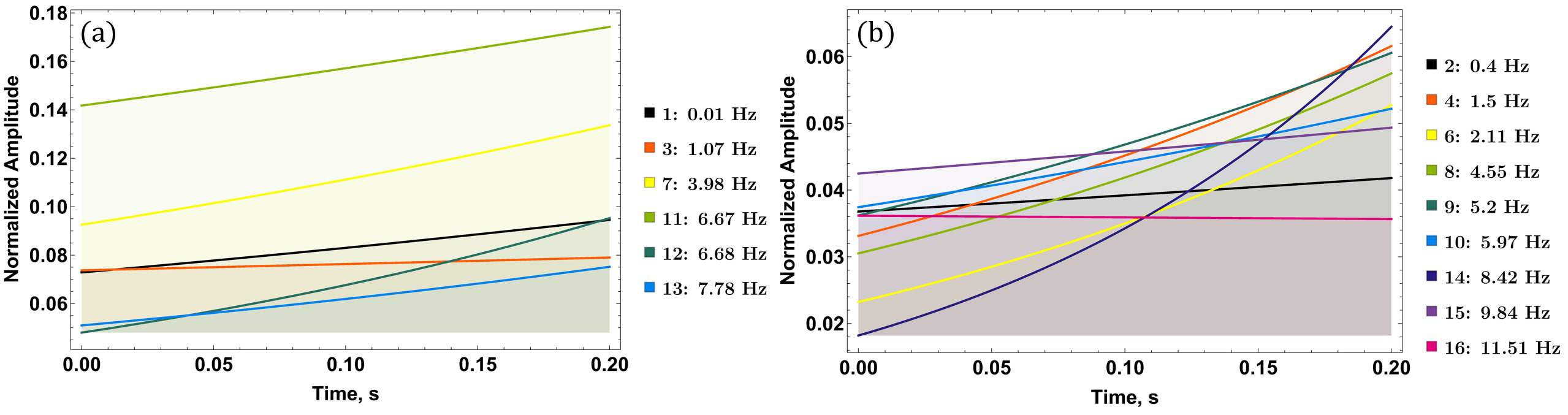}
    \caption{$30~sccm$ without applied MF: amplitude dynamics over the mean trajectory time for significant modes.}
    \label{fig:u-bubble-mode-growth-30-off}
\end{figure}

\begin{figure}[H]
    \centering
    \includegraphics[width=0.8\textwidth]{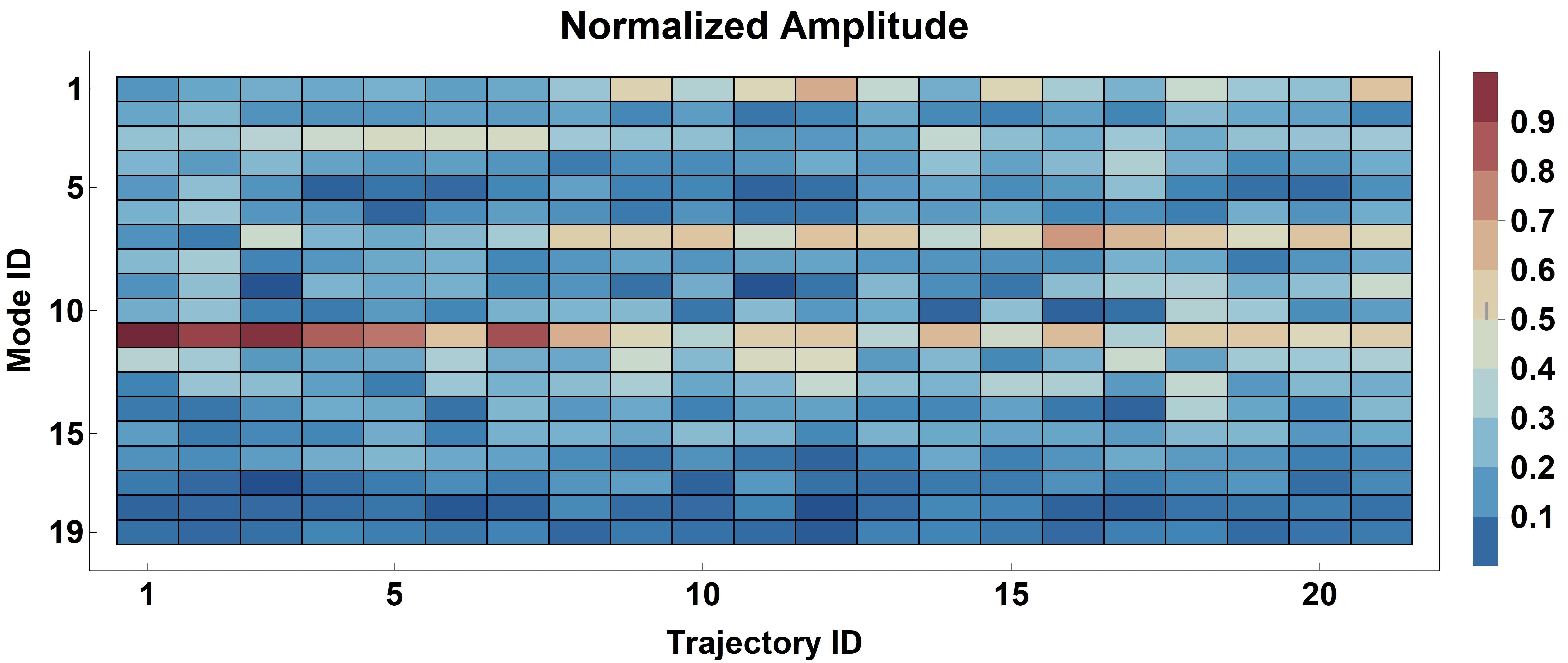}
    \caption{$30~sccm$ without applied MF: normalized root mean square (RMS) amplitudes for modes over all processed trajectories.}
    \label{fig:u-bubble-mode-rms-amps-30-off}
\end{figure}

Transitioning to the cases without MF, consider first the $30~sccm$ case: DMD mode analysis results are shown in Figures \ref{fig:u-bubble-mode-correlations-30-off}-\ref{fig:u-bubble-mode-rms-amps-30-off}.

One may notice that the mode frequencies in Figure \ref{fig:u-bubble-mode-stats-30-off}b are overall lower than in the two cases with applied MF (Figures \ref{fig:u-bubble-mode-stats-30-on}b and \ref{fig:u-bubble-mode-stats-100-on}b) -- this tendency was also observed above for the vessel velocity field modes (Figures \ref{fig:stats-vessel-modes-30-off}b-\ref{fig:stats-vessel-modes-100-on}b). Another difference clearly seen in Figure \ref{fig:u-bubble-mode-growth-30-off} is that both the first few significant modes that are dominant (a) and the lesser modes (b) constitute a very significant fraction of the zeroth mode's amplitude: $\sim 0.48$ and $\sim 0.29$, respectively.

\begin{figure}[H]
    \centering
    \includegraphics[width=1\textwidth]{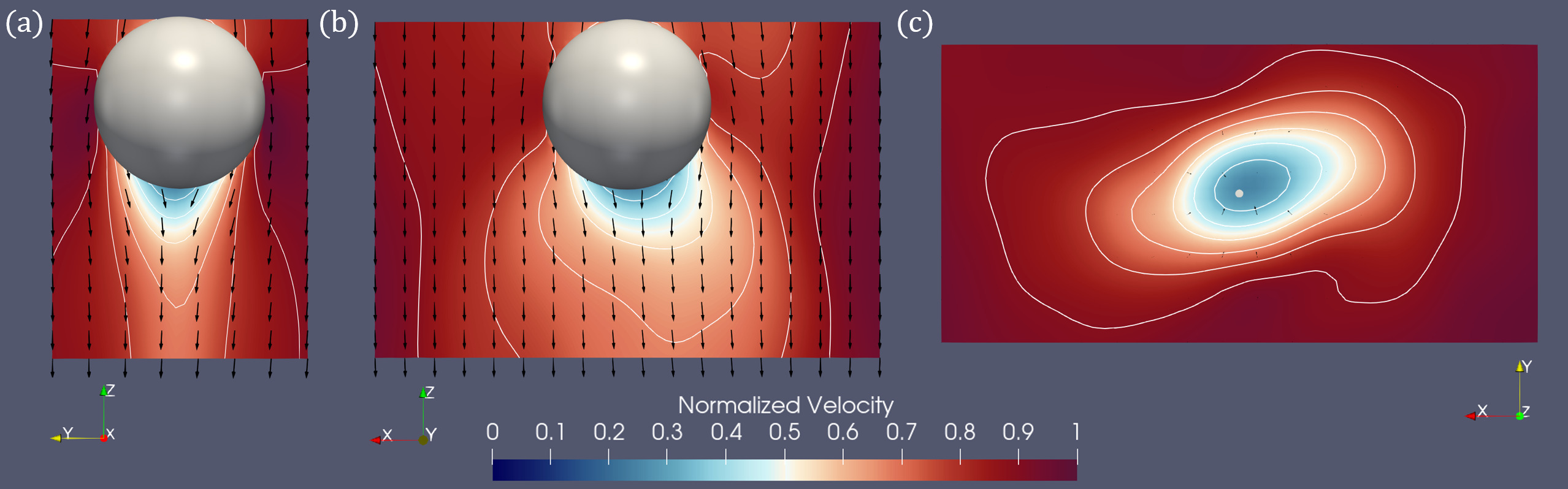}
    \caption{The zeroth bubble velocity field mode for $30~sccm$ without applied MF: (a) \textit{YZ} plane, (b) \textit{XZ} plane and (c) \textit{XY} plane.}
    \label{fig:u-bubble-30-off-m0}
\end{figure}

\begin{figure}[H]
    \centering
    \includegraphics[width=1\textwidth]{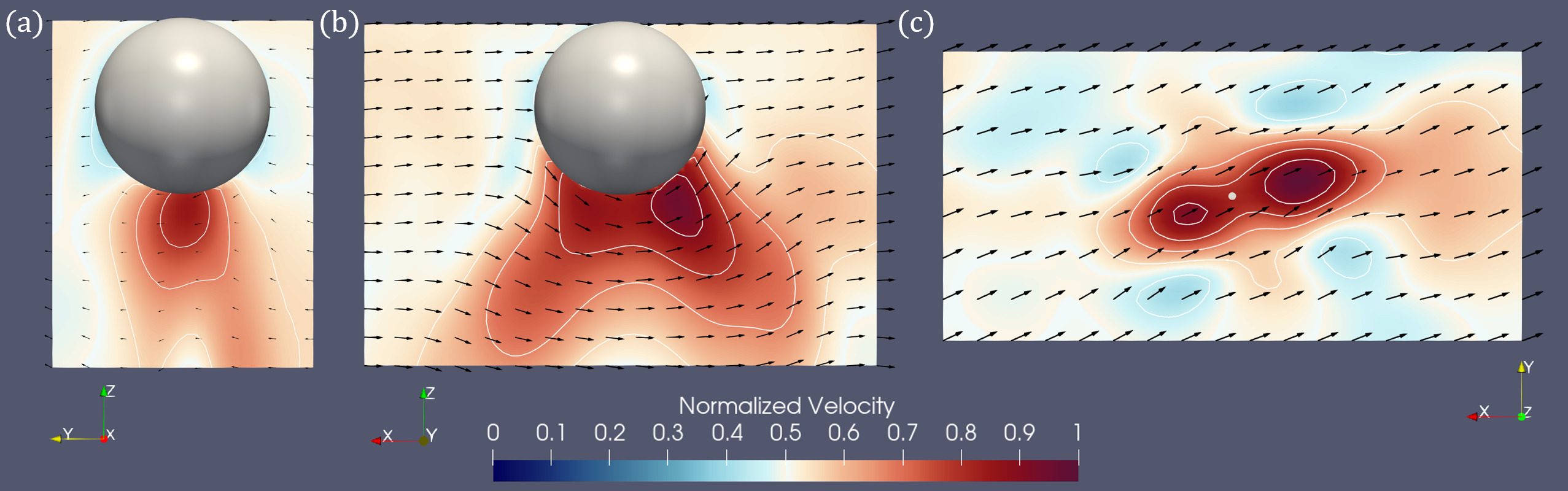}
    \caption{The 11-th bubble velocity field mode for $30~sccm$ without applied MF: (a) \textit{YZ} plane, (b) \textit{XZ} plane and (c) \textit{XY} plane.}
    \label{fig:u-bubble-30-off-m11}
\end{figure}

\begin{figure}[H]
    \centering
    \includegraphics[width=1\textwidth]{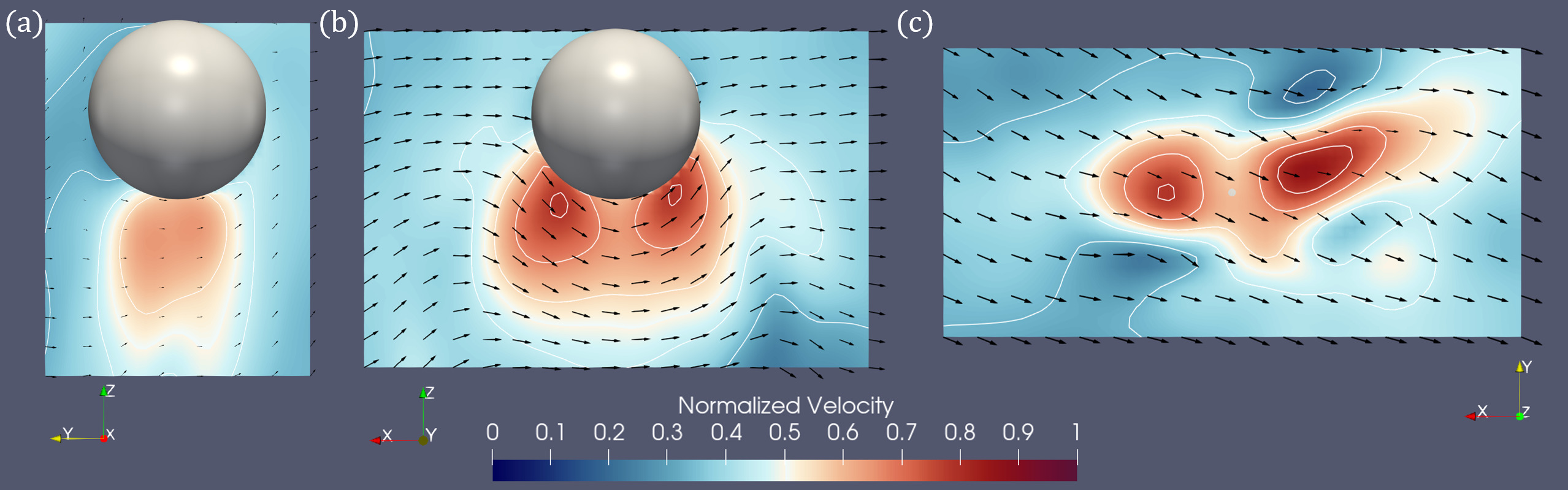}
    \caption{The 7-th bubble velocity field mode for $30~sccm$ without applied MF: (a) \textit{YZ} plane, (b) \textit{XZ} plane and (c) \textit{XY} plane.}
    \label{fig:u-bubble-30-off-m7}
\end{figure}

\begin{figure}[H]
    \centering
    \includegraphics[width=1\textwidth]{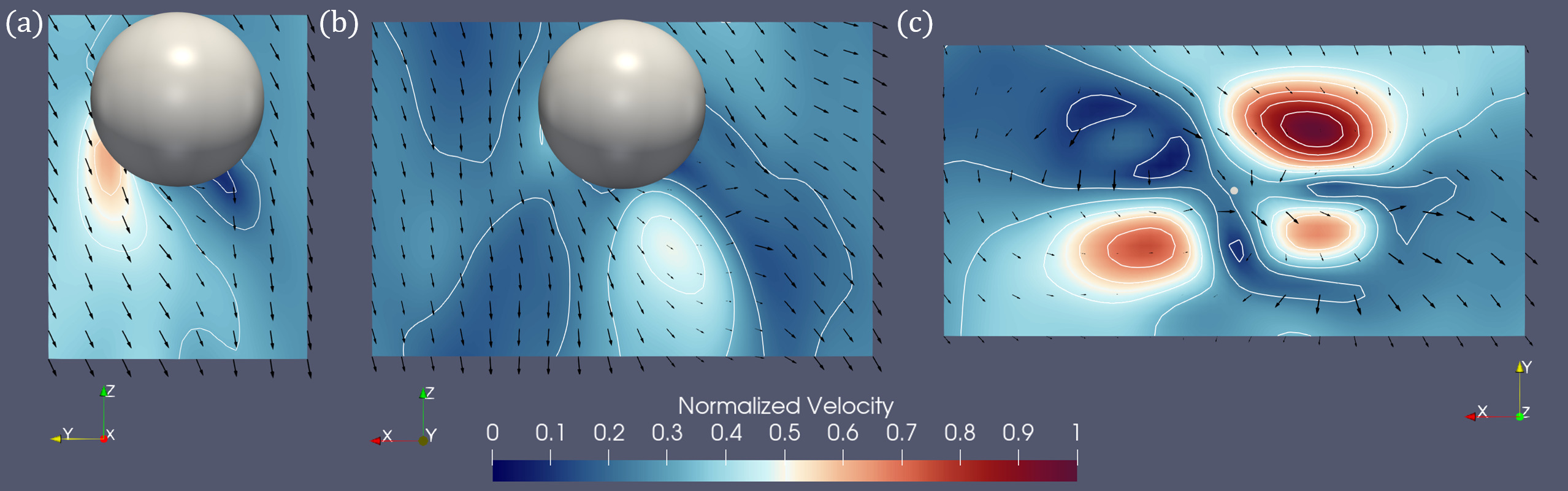}
    \caption{The 1-st bubble velocity field mode for $30~sccm$ without applied MF: (a) \textit{YZ} plane, (b) \textit{XZ} plane and (c) \textit{XY} plane.}
    \label{fig:u-bubble-30-off-m1}
\end{figure}

\begin{figure}[H]
    \centering
    \includegraphics[width=1\textwidth]{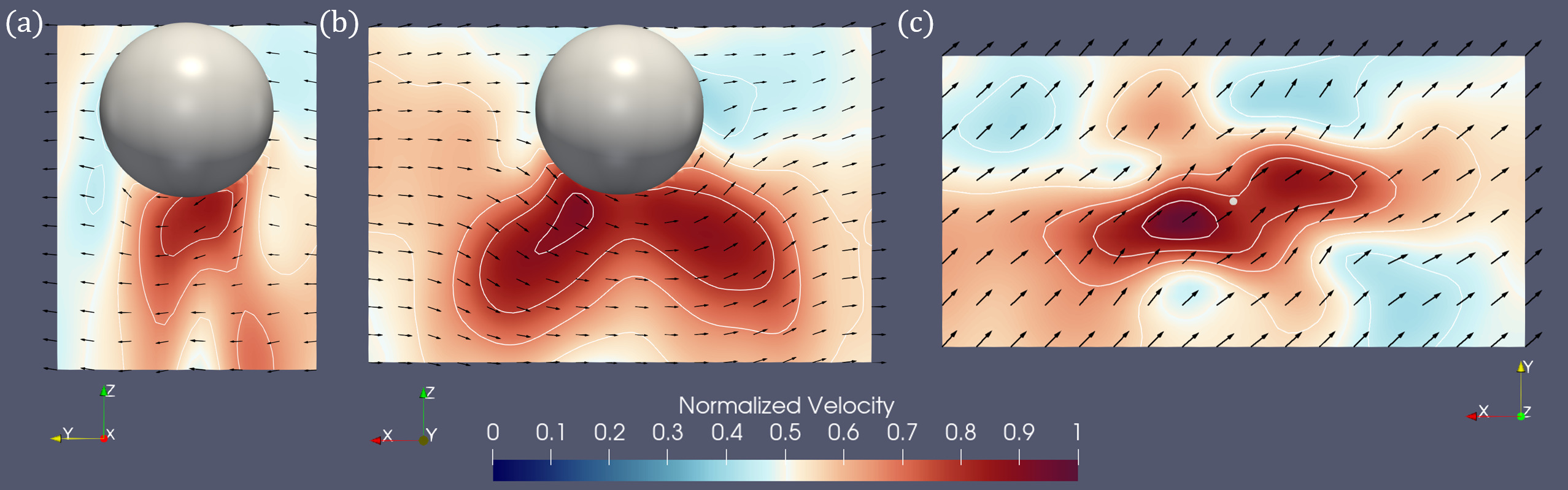}
    \caption{The 12-th bubble velocity field mode for $30~sccm$ without applied MF: (a) \textit{YZ} plane, (b) \textit{XZ} plane and (c) \textit{XY} plane.}
    \label{fig:u-bubble-30-off-m12}
\end{figure}

\begin{figure}[H]
    \centering
    \includegraphics[width=1\textwidth]{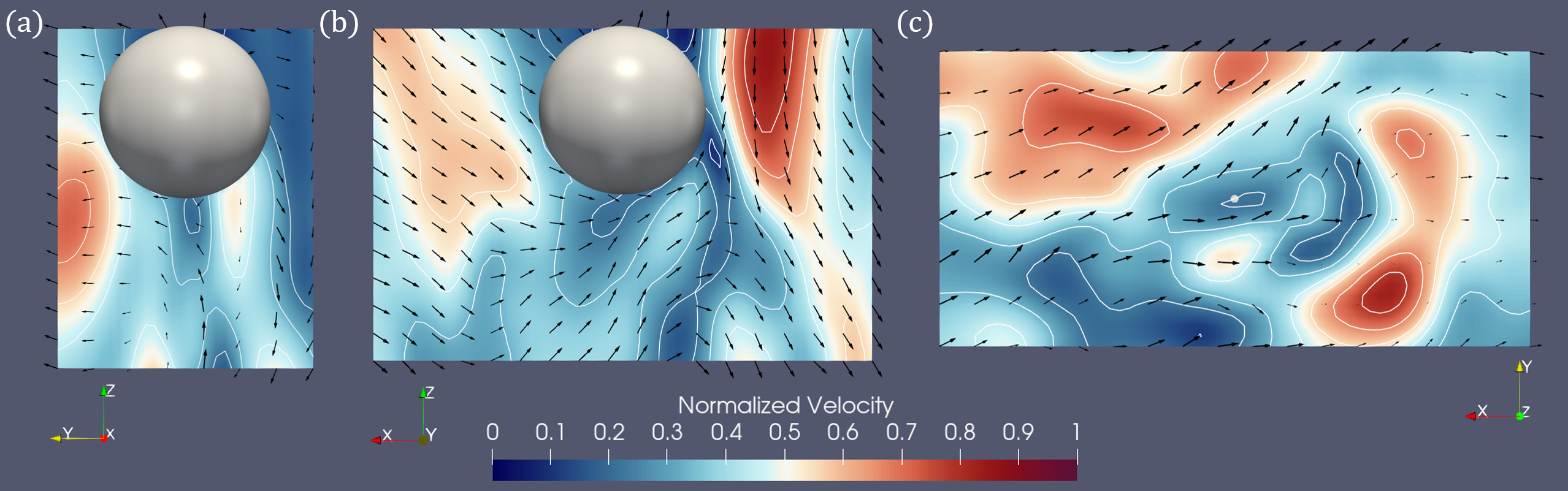}
    \caption{The 3-rd bubble velocity field mode for $30~sccm$ without applied MF: (a) \textit{YZ} plane, (b) \textit{XZ} plane and (c) \textit{XY} plane.}
    \label{fig:u-bubble-30-off-m3}
\end{figure}

\begin{figure}[H]
    \centering
    \includegraphics[width=1\textwidth]{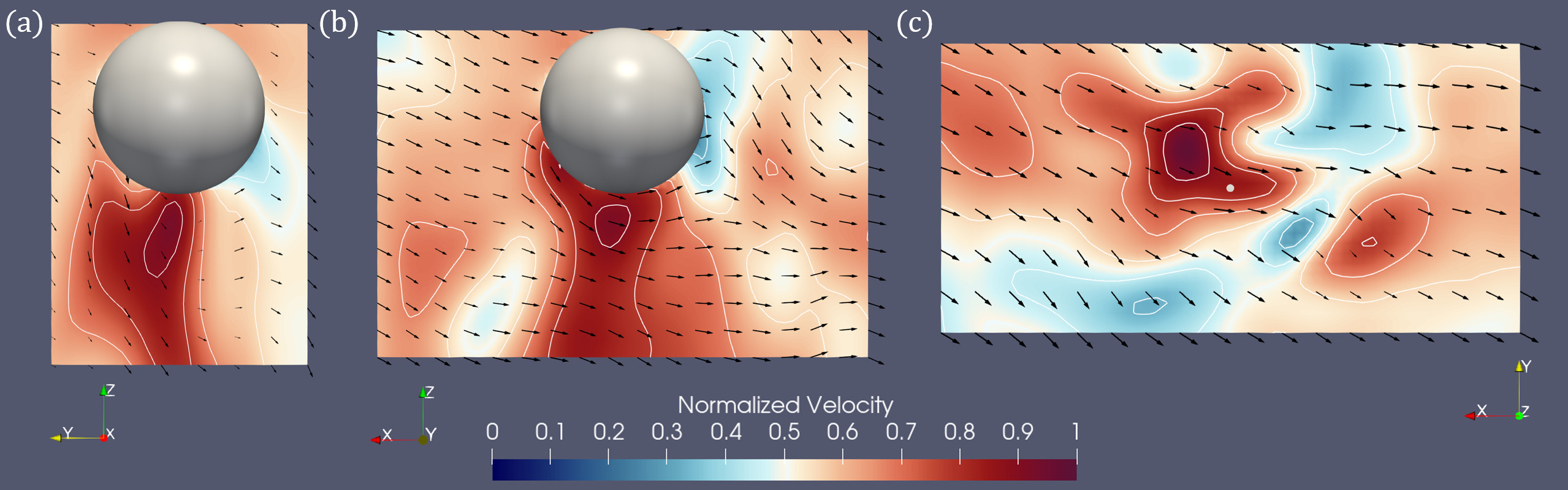}
    \caption{The 13-th bubble velocity field mode for $30~sccm$ without applied MF: (a) \textit{YZ} plane, (b) \textit{XZ} plane and (c) \textit{XY} plane.}
    \label{fig:u-bubble-30-off-m13}
\end{figure}

As seen in Figure \ref{fig:u-bubble-mode-stats-30-off}c, only two modes, the zeroth and 16-th, have slight negative growth rates while every other mode has a positive growth rate leading to an overall increase in the relative importance of non-zeroth modes over the mean trajectory time: amplitude sums for the outlined mode groups become $\sim 0.65$ and $\sim 0.48$, respectively. On this note, it is interesting that the lower frequency modes, 1 to 3, exhibit very small relative amplitude growth. Moreover, Figure \ref{fig:u-bubble-mode-rms-amps-30-off} indicates that the RMS amplitude of mode 7 (second strongest mode after mode 0) increases overall over the flow time, while for mode 11 (the dominant mode second to mode 0) the RMS amplitude shows an overall decline. The most significant modes are shown in Figures \ref{fig:u-bubble-30-off-m0}-\ref{fig:u-bubble-30-off-m13}.

The zeroth mode ($\omega_0 \sim 5.4 ~ mHz$, $a_0 \sim -0.16~ s^{-1}$, Figure \ref{fig:u-bubble-30-off-m0}) is again with the highest amplitude that is consistently $\sim 1$ for all trajectories with a $\sim 4.6\%$ deviation. Although its growth rate is higher then in the cases with applied MF roughly by an order of magnitude , its amplitude still only decays to $0.96$ on average. It is clear, especially from Figure \ref{fig:u-bubble-30-off-m0}c, that in this case the zeroth mode is very asymmetric in the \textit{XY} plane while mostly retaining symmetry in the \textit{YZ} plane (a). It is interesting to note that the next two dominant modes, 11 ($\omega_{11} \sim 6.7~ Hz$, $a_{11} \sim 1.0~ s^{-1}$, Figure \ref{fig:u-bubble-30-off-m11}) and 7 ($\omega_7 \sim 4.0~Hz$, $a_7 \sim 1.8~ s^{-1}$, Figure \ref{fig:u-bubble-30-off-m7}), are rather symmetric as well, especially in the \textit{XZ} plane. Note also the very pronounced wake zone beneath the bubble in both cases, as well as that both modes indicate dominant flow in the \textit{X} direction with a smaller contribution in the \textit{Y} direction. The velocity field patterns seen in Figures \ref{fig:u-bubble-30-off-m11}b and \ref{fig:u-bubble-30-off-m7}b and the mode frequencies consistent with zig-zag wavelengths given the mean trajectory time suggest these modes might be responsible for vortex shedding and trajectory zig-zagging in the \textit{XZ} plane (Figure \ref{fig:trajectories-30-sccm-field-off}), since velocity field oscillations in the \textit{X} direction are the strongest in areas where wake vortices form (Figure \ref{fig:bubbles-lic-30-sccm-off}). Mode 12 ($\omega_{12} \sim 6.7~Hz$, $a_{12} \sim 3.4~ s^{-1}$, Figure \ref{fig:u-bubble-30-off-m12}) is rather similar to 11 and 7, but is less symmetric and has a greater \textit{Y} component. Modes 11 and 12 are also very strongly correlated (Figure \ref{fig:u-bubble-mode-correlations-30-off}).

The 1-st mode ($\omega_1 \sim 12 ~ mHz$, $a_1 \sim 1.3~ s^{-1}$, Figure \ref{fig:u-bubble-30-off-m1}) seems to be related to the zeroth mode in that it exhibits intense flow in the \textit{Z} direction and the flow pattern in Figure \ref{fig:u-bubble-30-off-m1}c suggests it might be a higher order spatial/temporal harmonic of mode 0 (frequencies differ by a factor of $\sim 2.2$). Note that according to Figure \ref{fig:u-bubble-mode-correlations-30-off} modes 0 and 1 are significantly correlated.

The other two of the dominant modes are 3 ($\omega_3 \sim 1.1~ Hz$, $a_3 \sim 0.35~ s^{-1}$, Figure \ref{fig:u-bubble-30-off-m3}) and 13 ($\omega_{13} \sim 7.8~ Hz$, $a_{13} \sim 1.9~ s^{-1}$, Figure \ref{fig:u-bubble-30-off-m13}) -- these exhibit rather complex flow patterns and have no obvious symmetries or clear general velocity field directions. Modes that are grouped in Figure \ref{fig:u-bubble-mode-growth-30-off}b generally contribute velocity fields that are largely distorted versions of modes 11, 7, 1 or exhibit no clear symmetry as in modes 3 and 13.

Modes 15 ($\omega_{15} \sim 9.8~ Hz$, $a_{15} \sim 0.75~ s^{-1}$) and 16 ($\omega_{16} \sim 12~ Hz$, $a_3 \sim -6.9 \cdot 10^{-3}~ s^{-1}$) are interesting in that they are essentially perturbed variations of mode 12 seen for $100~sccm$ with applied MF (Figure \ref{fig:u-bubble-100-on-m12}) with flow velocity magnitude maxima rotated by $\pi/2$ in the \textit{XY} plane.

\begin{figure}[H]
    \centering
    \includegraphics[width=1\textwidth]{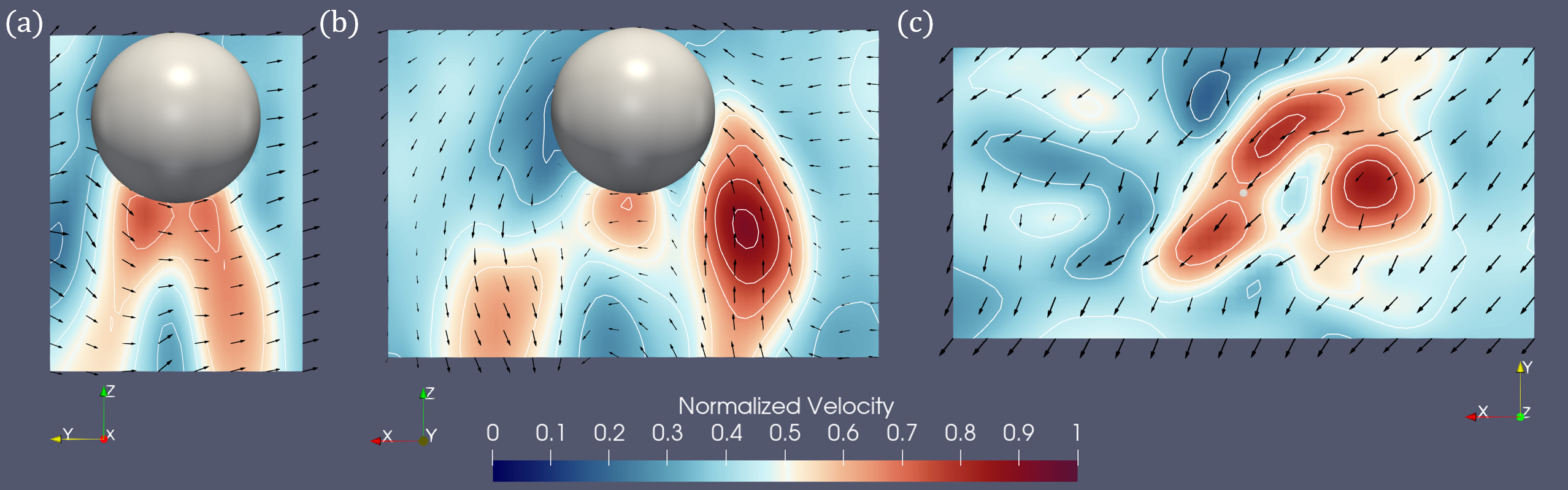}
    \caption{The 8-th bubble velocity field mode for $30~sccm$ without applied MF: (a) \textit{YZ} plane, (b) \textit{XZ} plane and (c) \textit{XY} plane.}
    \label{fig:u-bubble-30-off-m8}
\end{figure}

Mode 8 ($\omega_8 \sim 4.6~ Hz$, $a_8 \sim 3.2~ s^{-1}$), shown in Figure \ref{fig:u-bubble-30-off-m8}, deserves special attention despite its relatively low amplitude. The velocity field pattern seen in Figure \ref{fig:u-bubble-30-off-m8}b suggests that this mode is likely connected to vortex shedding -- not directly like modes 11 and 7, but rather it represents a common wake pattern right after vortex detachment where there is an upward flow about the bubble at the side (in the \textit{XZ} plane) wherefrom detachment occurred. In this case, or rather looking at the given mode phase, detachment took place to the right of the bubble and a vortex with clockwise rotation induced the upwards flow region. Flow field aside, two more factors suggest this might be the right interpretation. First, the frequency is roughly adequate, as it corresponds to a period of about the mean trajectory time. Looking at trajectories (Figure \ref{fig:trajectories-30-sccm-field-off}) one finds that the zig-zagging motion in the \textit{XZ} plane fits $\sim 1$-$1.5$ wavelengths into the, which is slightly off the mode frequency, but is close enough to arouse suspicion. Second, Figure \ref{fig:u-bubble-mode-growth-30-off}b indicates that the 8-th mode's amplitude increases significantly (versus its initial value) over trajectory time which makes sense given that, as the bubble accelerates during ascension, the detached vortices also exhibit greater velocity and vorticity magnitudes. Another minor argument is that mode 8 is rather strongly correlated with modes 11 and 12 (Figure \ref{fig:u-bubble-mode-correlations-30-off}) that are also in all likelihood, as noted above, linked to vortex shedding.

One may note that trajectory \textit{XY} projections (Figure \ref{fig:trajectories-30-sccm-field-off}) in the for of distorted spirals and tightly packed self-intersecting lines are due to the relatively much more disordered and asymmetric mode flow patterns seen in this case as opposed to the cases with applied MF. Finally, to reiterate on the correlation between the modes, one has the following dependencies (Figure \ref{fig:u-bubble-mode-correlations-30-off}): $0 \leftrightarrow 1$, $13 \leftrightarrow (7,11)$; correlated triplets $(3,11,12)$ and $(8,11,12)$.

\begin{figure}[H]
    \centering
    \includegraphics[width=0.4\textwidth]{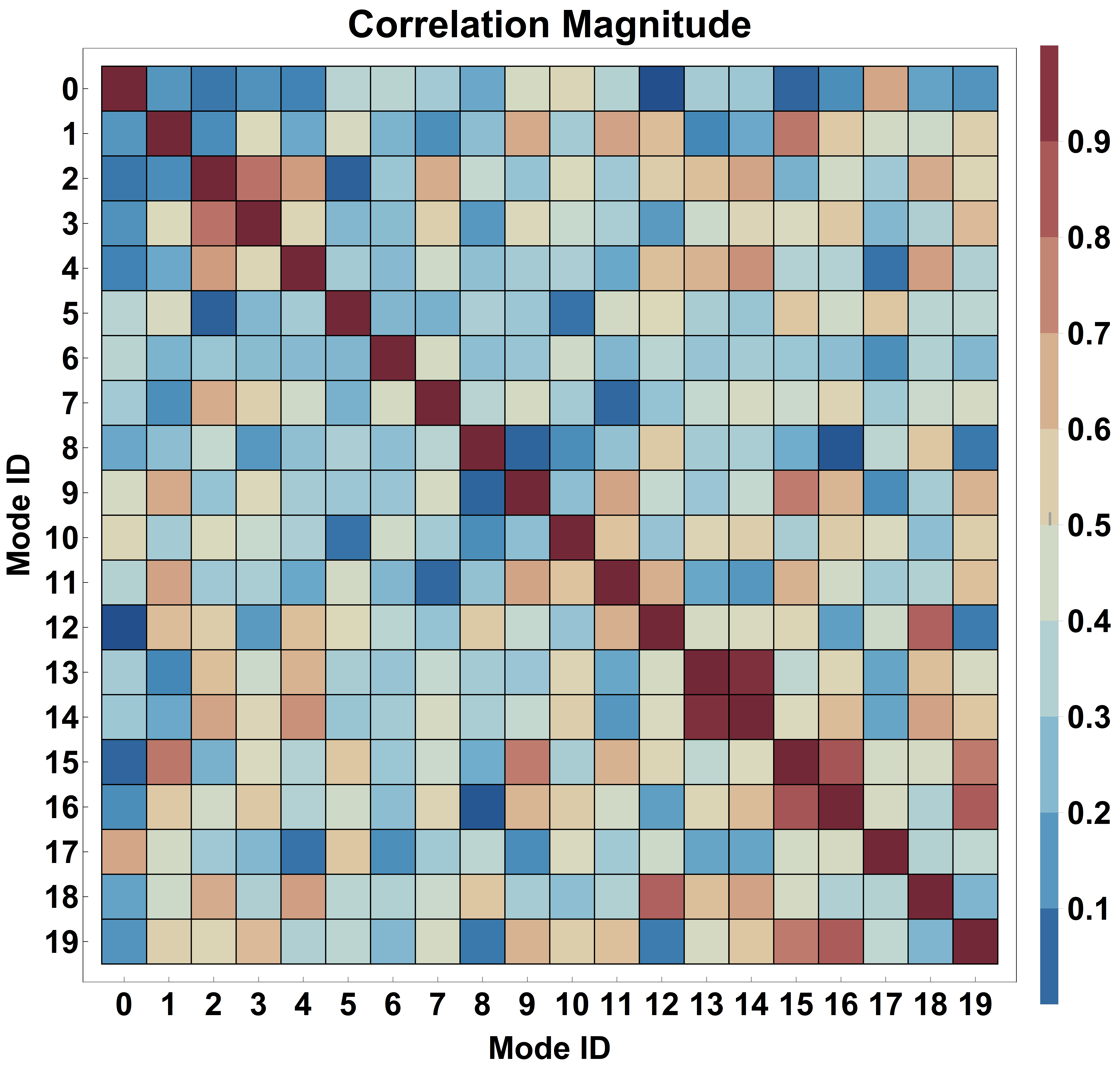}
    \caption{The bubble DMD mode correlation magnitude matrix for $100~sccm$ without applied MF.}
    \label{fig:u-bubble-mode-correlations-100-off}
\end{figure}

\begin{figure}[H]
    \centering
    \includegraphics[width=1\textwidth]{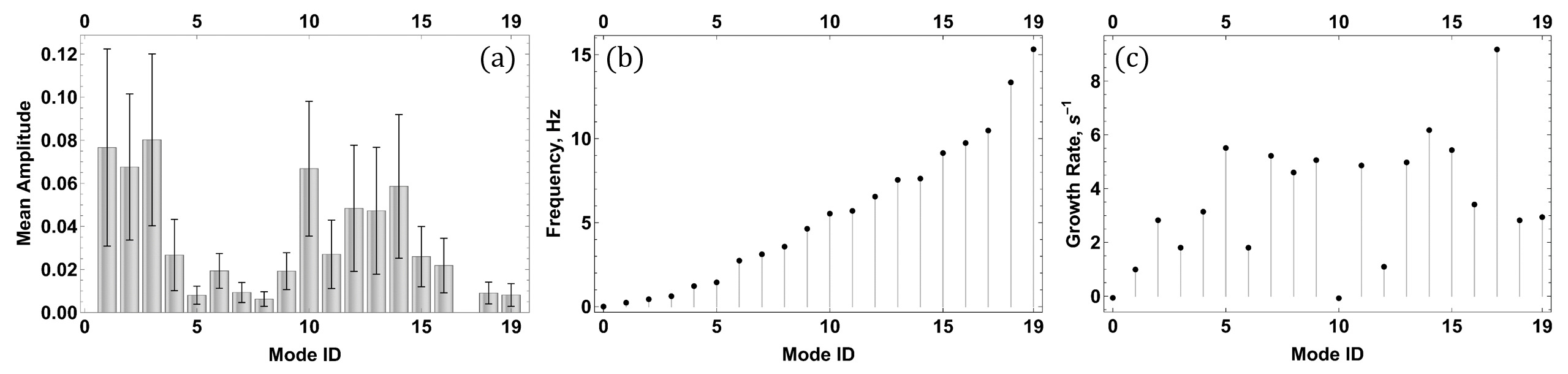}
    \caption{$100~sccm$ without applied MF: bubble velocity field mode (a) normalized initial amplitudes, (b) frequencies and (c) growth rates.}
    \label{fig:u-bubble-mode-stats-100-off}
\end{figure}

\begin{figure}[H]
    \centering
    \includegraphics[width=1\textwidth]{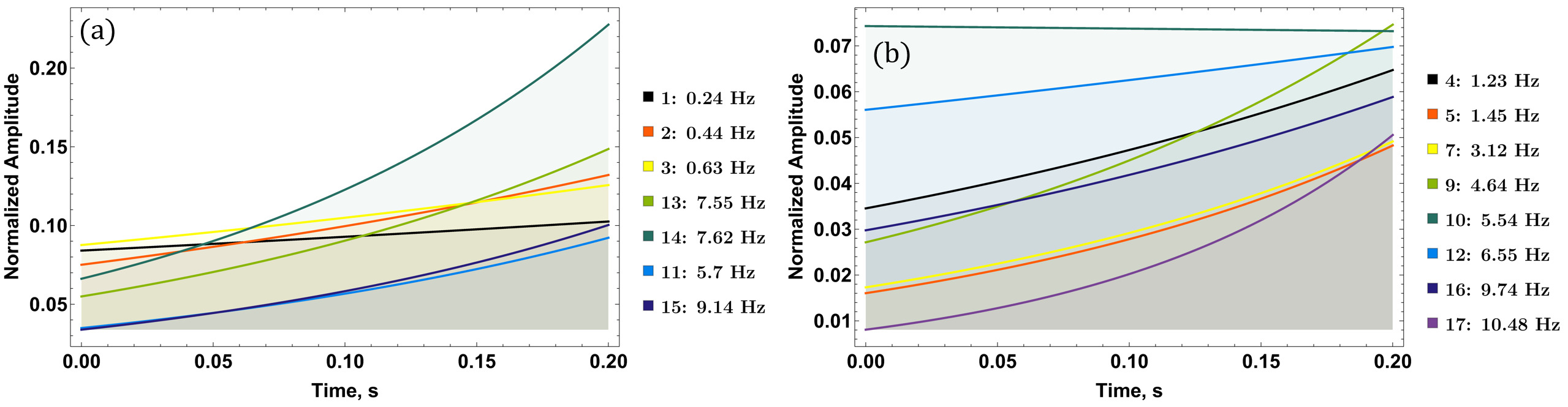}
    \caption{$100~sccm$ without applied MF: amplitude dynamics over the mean trajectory time for significant modes.}
    \label{fig:u-bubble-mode-growth-100-off}
\end{figure}

\begin{figure}[H]
    \centering
    \includegraphics[width=0.8\textwidth]{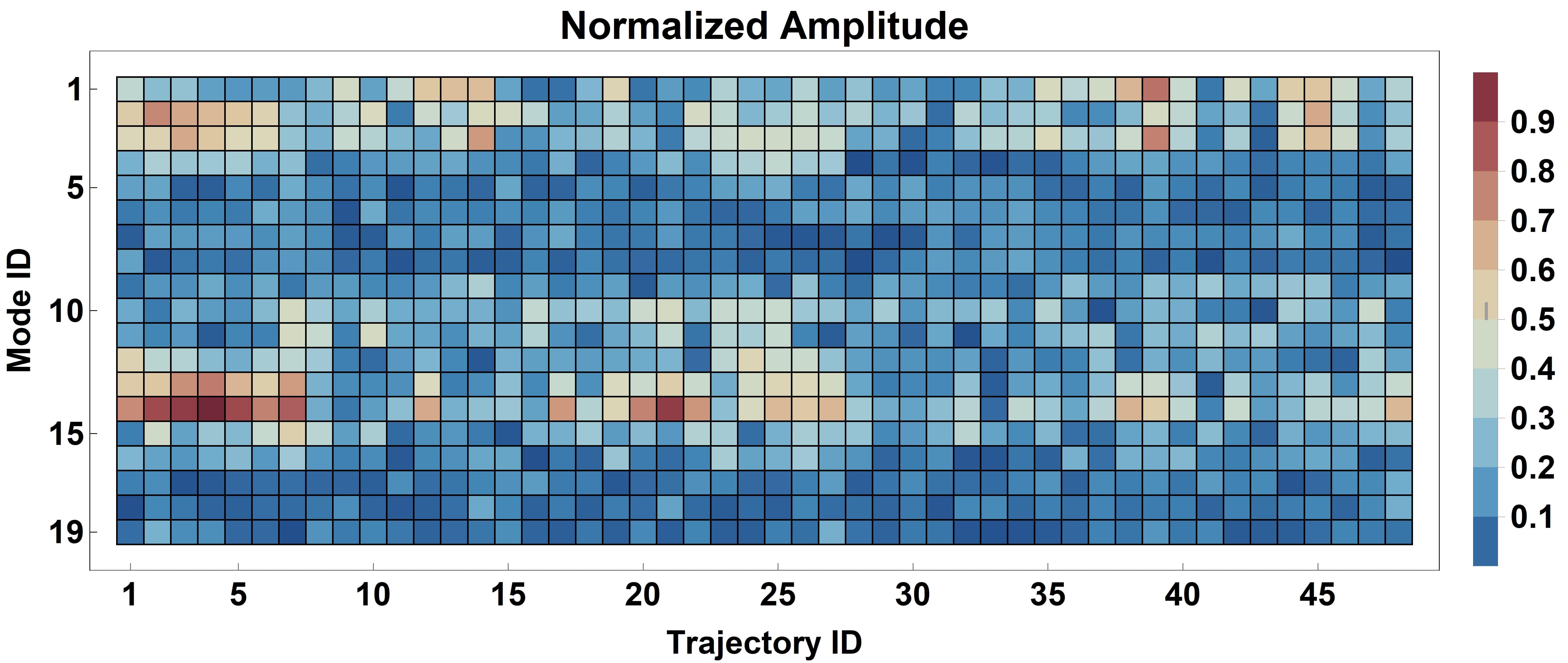}
    \caption{$100~sccm$ without applied MF: normalized root mean square (RMS) amplitudes for modes over all processed trajectories.}
    \label{fig:u-bubble-mode-rms-amps-100-off}
\end{figure}

Given the DMD results for $30~sccm$ above, it is important to see how they are different from $100~sccm$ without applied MF. DMD mode analysis results are shown in Figures \ref{fig:u-bubble-mode-correlations-100-off}-\ref{fig:u-bubble-mode-rms-amps-100-off}. As with $30~sccm$ without MF, the frequencies for $100~sccm$ are consistently lower than in the cases with applied MF (Figure \ref{fig:u-bubble-mode-stats-100-off}b). Note also that the growth rates for $100~sccm$ (Figure \ref{fig:u-bubble-mode-stats-100-off}c) are, like for $30~sccm$, positive except for modes 0 and 10 (0 and 16 for $30~sccm$, as seen in Figure \ref{fig:u-bubble-mode-stats-30-off}c). The growth rate magnitudes are greater overall, though. Another difference is that in this case higher frequency modes have greater initial amplitudes than with $30~sccm$ (Figure \ref{fig:u-bubble-mode-stats-100-off}a versus Figure \ref{fig:u-bubble-mode-stats-30-off}a). Figures \ref{fig:u-bubble-mode-growth-100-off}a and \ref{fig:u-bubble-mode-growth-100-off}b show the dominant modes and the ones with lesser overall amplitudes, respectively -- note that the two overall strongest modes in \ref{fig:u-bubble-mode-growth-100-off}a, modes 14 and 13, have higher frequencies that the two strongest dominant modes in Figure \ref{fig:u-bubble-mode-growth-30-off}a. Since grouping modes into dominant/secondary groups is difficult to do strictly, in this case the 10-th mode (Figure \ref{fig:u-bubble-mode-growth-100-off}b) was used as a nominal boundary since it changes very little over the mean trajectory time and its and the 9-th mode's final amplitudes are on average just below that of mode 11 (Figure \ref{fig:u-bubble-mode-growth-30-off}a).

Another major difference is that the dominant modes for $100~sccm$ have lower relative initial amplitudes than in the $30~sccm$ case and are have a less sparse value distribution (Figure \ref{fig:u-bubble-mode-growth-100-off}a). In addition the differences in amplitudes at the start and end of mean trajectory time are greater for most modes at $100~sccm$. Specifically, amplitudes of dominant dynamic modes initially constitute $\sim 0.44$ and this sum increases to $\sim 0.93$ over the mean trajectory time (Figure \ref{fig:u-bubble-mode-growth-100-off}a), while lesser amplitudes initially amount to $\sim 0.26$ which becomes $\sim 0.48$ over trajectory time. The zeroth mode which stands for the mean flow field has an amplitude of $\sim 1.08$ on average ($1$ initially) with a $5.2\%$ deviation, meaning that the dominant non-stationary modes practically overshadow the mean flow field with secondary modes also contributing significant perturbations. This leads one to expect much more disturbed wake patterns than in the $30~sccm$ case, as it should be.

\begin{figure}[H]
    \centering
    \includegraphics[width=1\textwidth]{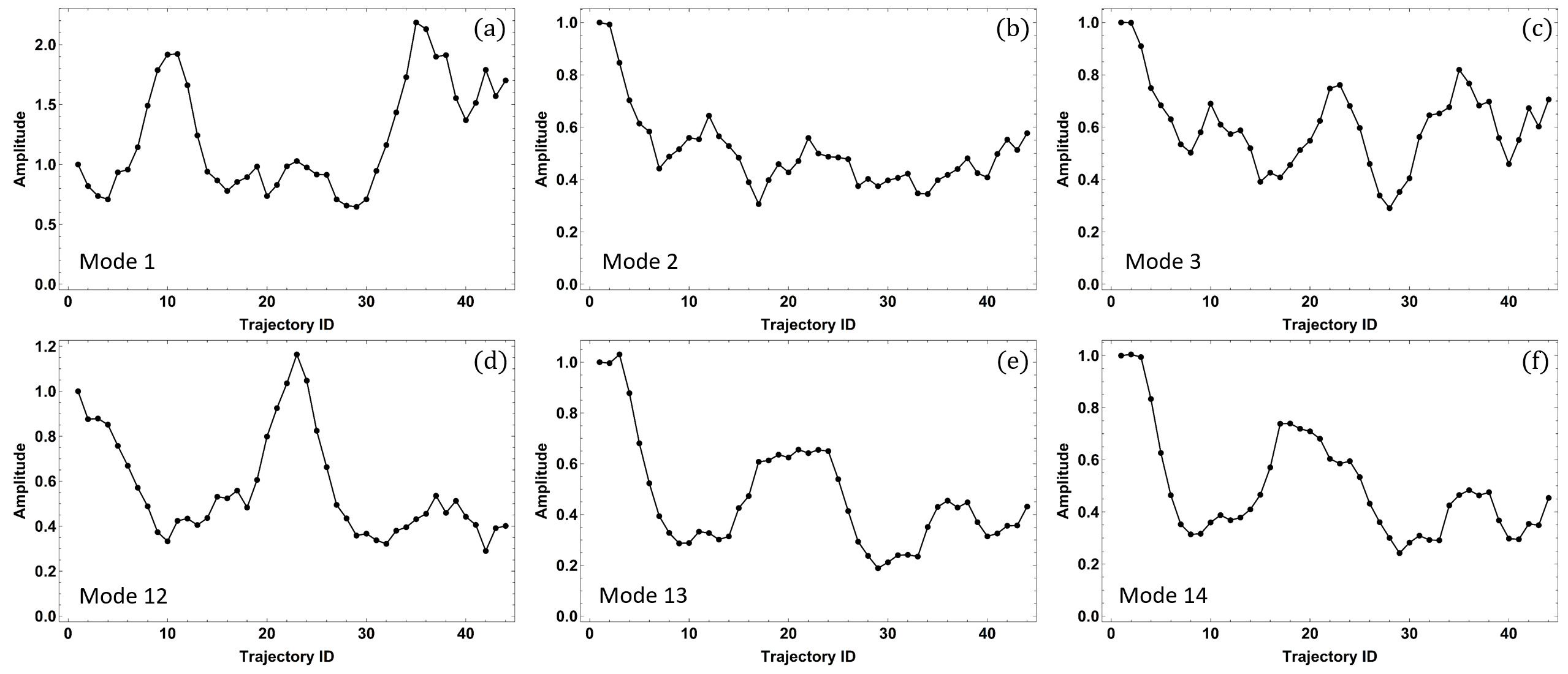}
    \caption{$100~sccm$ without applied MF: averaged RMS amplitude history over flow time (trajectory IDs) for significant bubble velocity field modes.The averaging window width is 5 trajectories.}
    \label{fig:u-bubble-amplitude-history-100-off}
\end{figure}

It is also interesting to see that mode RMS amplitude values over consecutive trajectories (Figure \ref{fig:u-bubble-mode-rms-amps-100-off}) seem to exhibit oscillatory trends for five of the dominant modes as well as mode 12 -- close inspection indicates that it is indeed so, as seen in Figure \ref{fig:u-bubble-amplitude-history-100-off}. This is interesting as it raises the question of how these trends are connected to the vessel reference frame velocity field modes. Given that the trajectories processed by the DMD algorithm cover $\sim 5~s$ of flow time and the frequencies of significant extrema seen in Figure \ref{fig:u-bubble-amplitude-history-100-off}, it seems likely that the considered bubble wake modes are modulated by vessel flow modes with roughly $< 1~ Hz$ frequencies -- this would correspond to modes 1 to 4 which are the strongest after the zeroth vessel flow mode (Figure \ref{fig:stats-vessel-modes-100-off}).

\begin{figure}[H]
    \centering
    \includegraphics[width=1\textwidth]{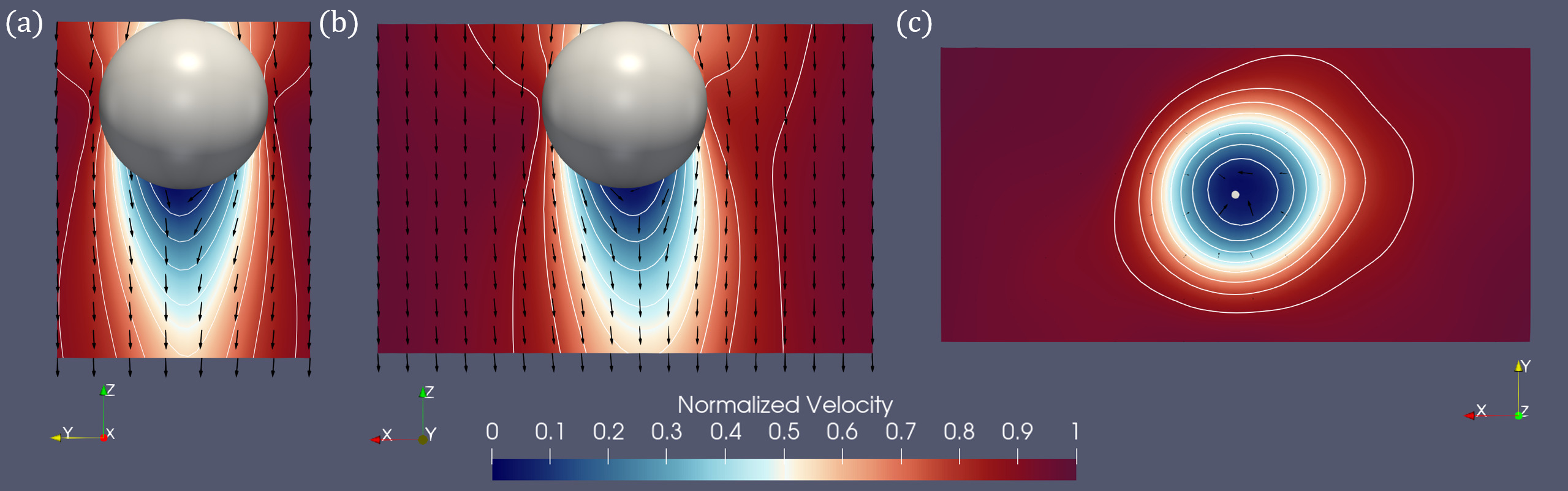}
    \caption{The zeroth bubble velocity field mode for $100~sccm$ without applied MF: (a) \textit{YZ} plane, (b) \textit{XZ} plane and (c) \textit{XY} plane.}
    \label{fig:u-bubble-100-off-m0}
\end{figure}

\begin{figure}[H]
    \centering
    \includegraphics[width=1\textwidth]{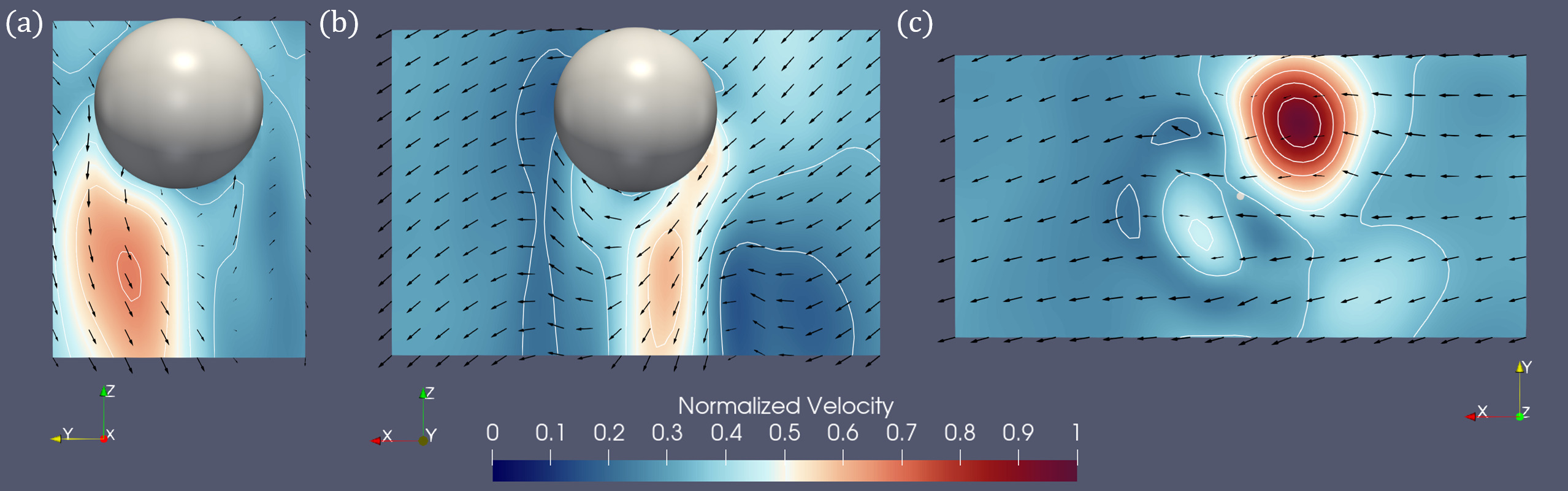}
    \caption{The 14-th bubble velocity field mode for $100~sccm$ without applied MF: (a) \textit{YZ} plane, (b) \textit{XZ} plane and (c) \textit{XY} plane.}
    \label{fig:u-bubble-100-off-m14}
\end{figure}

\begin{figure}[H]
    \centering
    \includegraphics[width=1\textwidth]{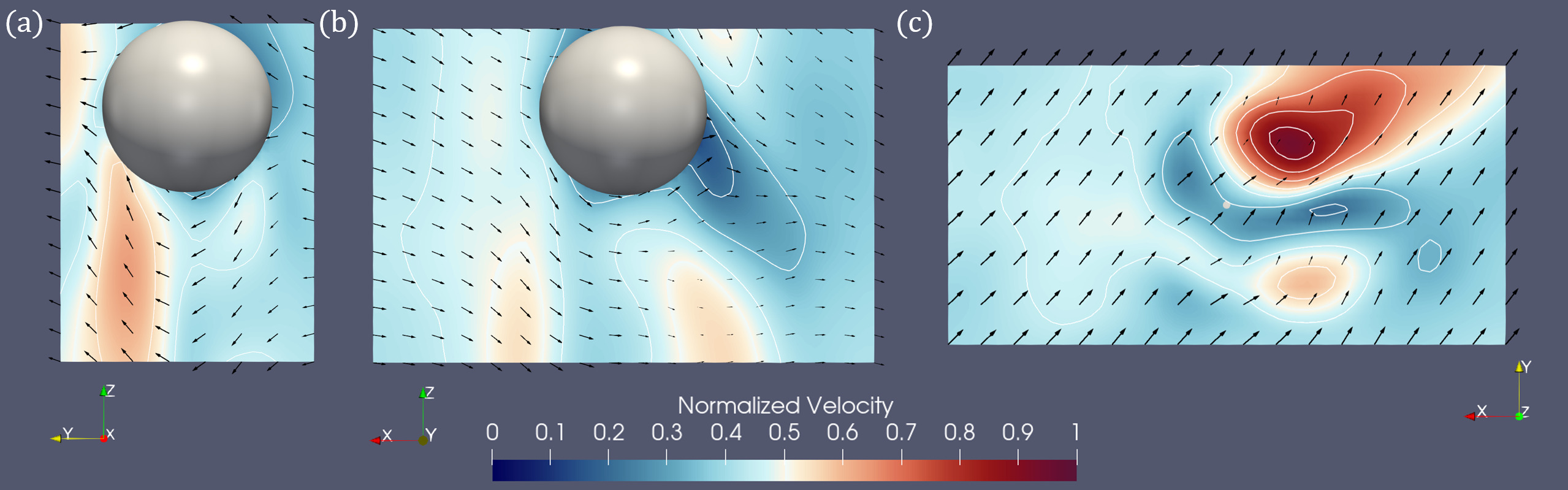}
    \caption{The 2-nd bubble velocity field mode for $100~sccm$ without applied MF: (a) \textit{YZ} plane, (b) \textit{XZ} plane and (c) \textit{XY} plane.}
    \label{fig:u-bubble-100-off-m2}
\end{figure}

\begin{figure}[H]
    \centering
    \includegraphics[width=1\textwidth]{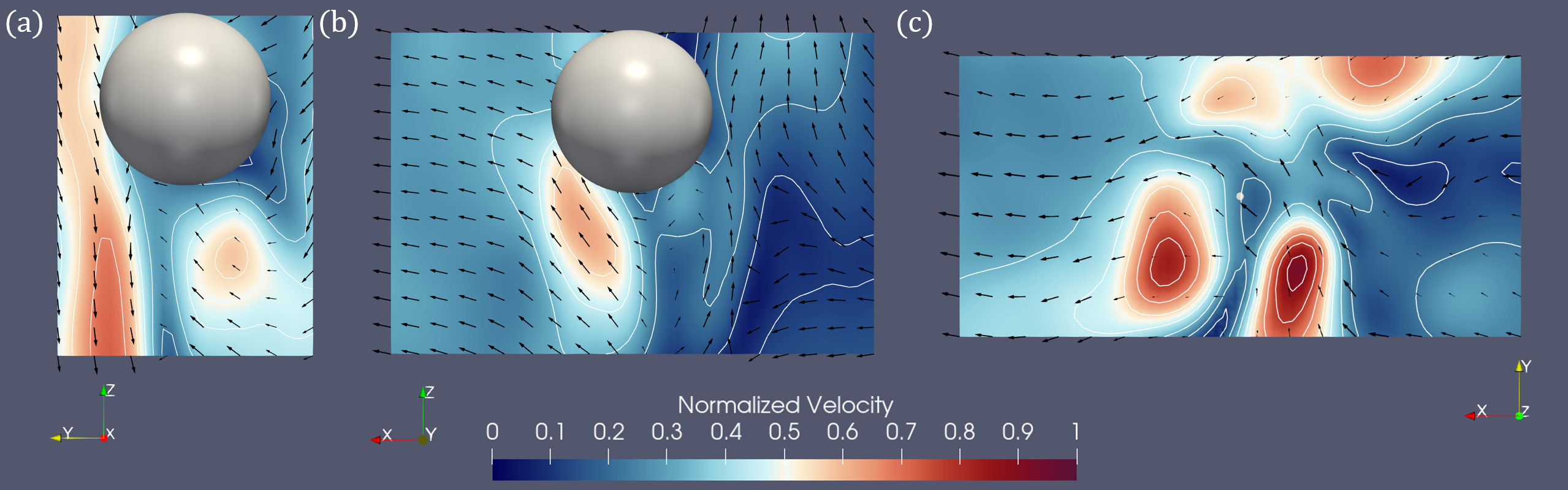}
    \caption{The 3-rd bubble velocity field mode for $100~sccm$ without applied MF: (a) \textit{YZ} plane, (b) \textit{XZ} plane and (c) \textit{XY} plane.}
    \label{fig:u-bubble-100-off-m3}
\end{figure}

\begin{figure}[H]
    \centering
    \includegraphics[width=1\textwidth]{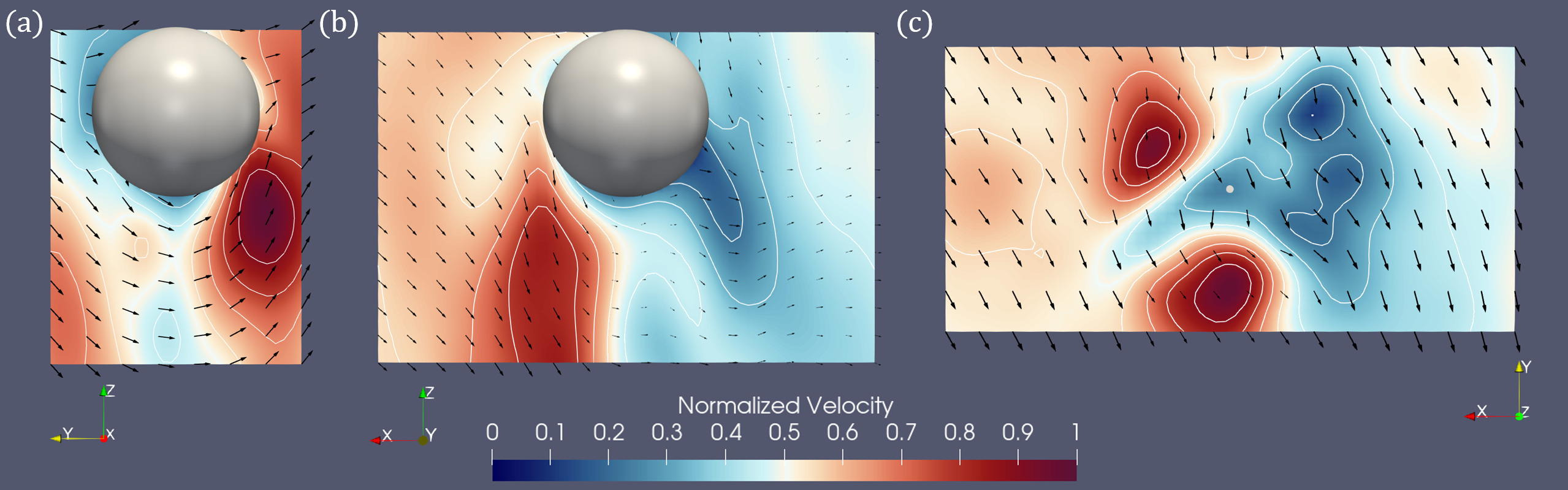}
    \caption{The 1-st bubble velocity field mode for $100~sccm$ without applied MF: (a) \textit{YZ} plane, (b) \textit{XZ} plane and (c) \textit{XY} plane.}
    \label{fig:u-bubble-100-off-m1}
\end{figure}

\begin{figure}[H]
    \centering
    \includegraphics[width=1\textwidth]{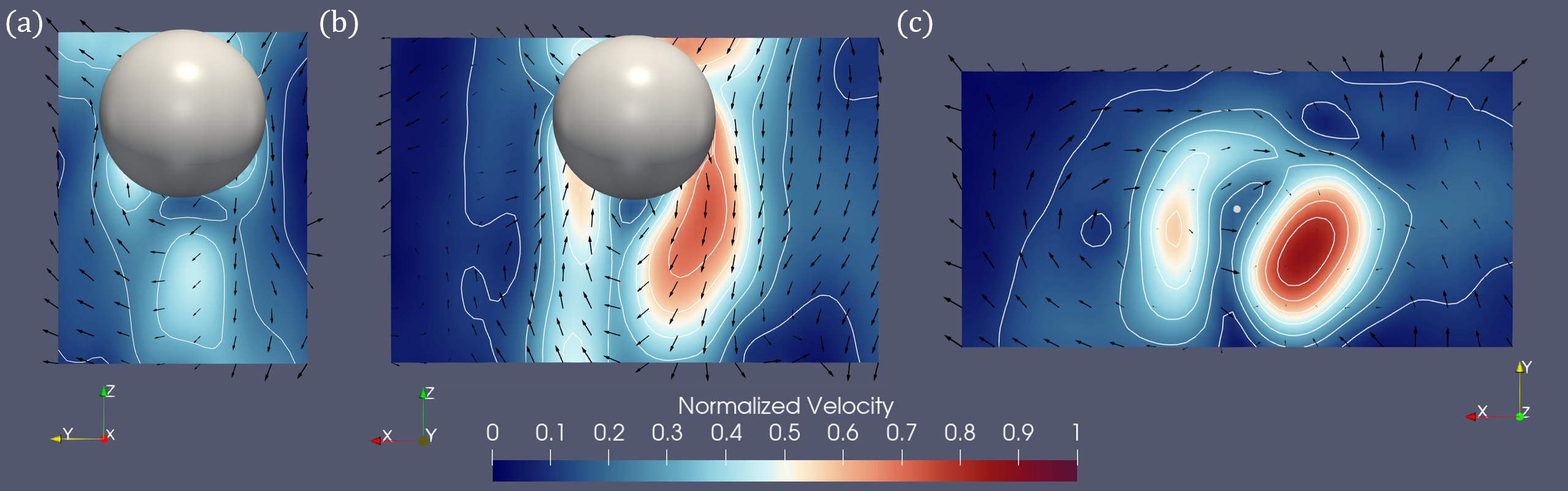}
    \caption{The 15-th bubble velocity field mode for $100~sccm$ without applied MF: (a) \textit{YZ} plane, (b) \textit{XZ} plane and (c) \textit{XY} plane.}
    \label{fig:u-bubble-100-off-m15}
\end{figure}

\begin{figure}[H]
    \centering
    \includegraphics[width=1\textwidth]{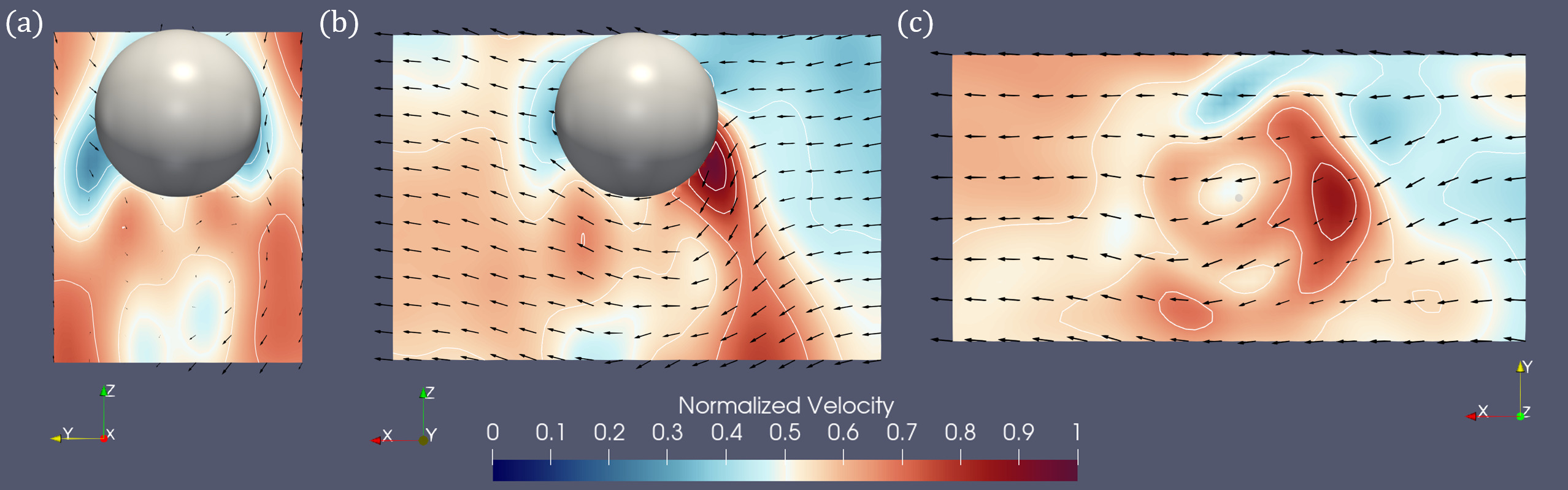}
    \caption{The 11-th bubble velocity field mode for $100~sccm$ without applied MF: (a) \textit{YZ} plane, (b) \textit{XZ} plane and (c) \textit{XY} plane.}
    \label{fig:u-bubble-100-off-m11}
\end{figure}

The significant bubble wake modes for $100~sccm$ with applied MF are shown in Figures \ref{fig:u-bubble-100-off-m0}-\ref{fig:u-bubble-100-off-m11}. The zeroth mode is as in other cases stationary ($\omega_0 \sim 6.6~ mHz$, $a_0 \sim -6.2 \cdot 10^{-2}~ s^{-1}$, Figure \ref{fig:u-bubble-100-off-m0}) and, rather curiously, is not quite as asymmetric as its counterpart for $30~sccm$ (Figure \ref{fig:u-bubble-30-off-m0}). It also expectantly has a much more developed bubble wake.

Mode 14 ($\omega_{14} \sim 7.6~ Hz$, $a_{14} \sim 6.2~ s^{-1}$, Figure \ref{fig:u-bubble-100-off-m14}) exhibits a velocity field pattern that is consistent with bubble wake structure during bubble tilting in the \textit{XZ} plane as seen in Figure \ref{fig:bubbles-lic-100-sccm-off}, e.g. (b), (f), where a stagnation zone forms on one side of the wake while higher velocity is observed on the other side (\textit{XZ} plane). Another argument for this conjecture is that the mode frequency is a good match for the mean observed trajectory oscillation frequency (Figure \ref{fig:trajectories-100-sccm-field-off}). Mode 13 ($\omega_{13} \sim 7.5~ Hz$, $a_{13} \sim 5.0~ s^{-1}$) has the next highest RMS amplitude overall, but is not shown here since its flow pattern is remarkably similar to that of mode 14. In fact, Figure \ref{fig:u-bubble-mode-correlations-100-off} indicates that modes 13 and 14 are extremely strongly correlated ($\sim 0.97$ correlation value) and suggests that mode 13 might simply be a slightly phase shifted version of mode 14, especially given that $\omega_{14} \approx \omega_{13}$.

Mode 2 ($\omega_2 \sim 0.45~ Hz$, $a_2 \sim 2.8~ s^{-1}$, Figure \ref{fig:u-bubble-100-off-m2}) does not seem to be directly linked to a certain momentum transfer mechanism within the wake but rather exhibits flow oscillations mostly in the \textit{XY} plane with a very low frequency. Mode 3 ($\omega_3 \sim 0.63~ Hz$, $a_3 \sim 1.8~ s^{-1}$) is likewise difficult to interpret in terms of clear processes that one would expect in a bubble wake, but rather it shows disordered low frequency oscillations mostly in the \textit{Z} direction in the form of four "jets" seen in Figure \ref{fig:u-bubble-100-off-m3}c as four velocity field maxima with with almost zero \textit{XY} component -- two of these are partially captured in Figures \ref{fig:u-bubble-100-off-m3}a and \ref{fig:u-bubble-100-off-m3}b. It must be noted that modes 2 and 3 are significantly correlated (Figure \ref{fig:u-bubble-mode-correlations-100-off}).

Mode 1 ($\omega_1 \sim 0.23~ Hz$, $a_2 \sim 1.0~ s^{-1}$, Figure \ref{fig:u-bubble-100-off-m1}) contributes velocity field oscillations in the \textit{Y} direction with a relatively weak \textit{Z} component and the flow pattern suggests that the mode may be linked to wake oscillations in the \textit{YZ} plane. It is, however, unclear how given the low frequency.

Mode 15 ($\omega_{15} \sim 9.1~ Hz$, $a_{15} \sim 5.4~ s^{-1}$, Figure \ref{fig:u-bubble-100-off-m15}) on the other hand has both the time scale and the flow field that fit wake oscillations as the bubble tilts periodically. This mode is also rather symmetric in the \textit{YZ} plane in terms of velocity magnitude (Figure \ref{fig:u-bubble-100-off-m15}a) and exhibits a swirl flow pattern in the \textit{XY} plane (Figure \ref{fig:u-bubble-100-off-m15}c). Mode 11 ($\omega_{11} \sim 5.7~ Hz$, $a_{11} \sim 4.9~ s^{-1}$, Figure \ref{fig:u-bubble-100-off-m11}) is interesting in that is exhibits flow field oscillations almost entirely in the \textit{X} direction with an asymmetric velocity maximum located just above (Figure \ref{fig:u-bubble-100-off-m11}b) the zone where the wake vortex should be expected to form (Figure \ref{fig:bubbles-lic-100-sccm-off}). This mode's frequency is within the range where the mode could be responsible for affecting vortex shedding via velocity pulsations in the \textit{X} direction. Mode 11 is also quite symmetric in the \textit{YZ} plane.

Other notable modes include mode 12 ($\omega_{12} \sim 5.7~ Hz$, $a_{12} \sim 4.9~ s^{-1}$) which is somewhat similar in the flow field structure to mode 8 for $100~sccm$ with applied MF (Figure \ref{fig:u-bubble-100-on-m8}), but with a more developed bubble wake zone. Mode 9 ($\omega_9 \sim 4.6~ Hz$, $a_9 \sim 5.1~ s^{-1}$) has the velocity field resembling mode 7 for $30~sccm$ without applied MF (Figure \ref{fig:u-bubble-30-off-m7}), but with much more pronounced velocity maximum zones beneath the bubble that are elongated in the \textit{Z} direction and an overall greater \textit{Y} component. Mode 10 ($\omega_{10} \sim 5.5~ Hz$, $a_{10} \sim -7.5 \cdot 10^{-3} ~ s^{-1}$) in turn closely resembles modes 2, 13 and 14 for this case and is somewhat correlated to all three (Figure \ref{fig:u-bubble-mode-correlations-100-off}). Mode 16 ($\omega_{16} \sim 9.7~ Hz$, $a_{16} \sim 3.4 ~ s^{-1}$) is essentially mode 15, but with a smaller RMS amplitude and a slightly greater frequency ($\omega_{15} \sim 9.1~ Hz$). Note that modes 15 and 16 are very strongly correlated (Figure \ref{fig:u-bubble-mode-correlations-100-off}).

It should be said that, while some of the modes for $100~sccm$ without MF are physically interpretable, at least hypothetically, it is very difficult to attribute them to the various features of trajectories (Figure \ref{fig:trajectories-100-sccm-field-off}). While one could argue that the initial few trajectories as in Figure \ref{fig:trajectories-100-sccm-field-off}a can be explained by the initial dominance of modes 10, 13 and 14 that would fit the oscillations seen in the \textit{XY} projections of bubble trajectories, later trajectories are highly disordered, some without clearly discernible patterns, making it very difficult to attribute specific modes to the observed behaviour given the present analysis. While some of the seemingly random trajectory perturbations (Figure \ref{fig:trajectories-100-sccm-field-off}) could be due to the oscillations in the RMS amplitudes of some of the key modes (Figure \ref{fig:u-bubble-amplitude-history-100-off}), it is currently unclear to what extent.

To summarize the mode correlations, one has the following dependencies (Figure \ref{fig:u-bubble-mode-correlations-100-off}): $14 \leftrightarrow (2,13)$, $2 \leftrightarrow (3,14)$, $11 \leftrightarrow (1,9)$, $15 \leftrightarrow (1,9,16)$.

\clearpage

\section{Conclusions \& outlook}

To summarize the results presented in the paper: we have developed a custom memory efficient implementation of noise resilient DMD combining higher order DMD and streaming SVD (MOSES SVD) approaches, as well as real-to-complex domain mapping for DMD input. Since the code was developed to perform DMD for data from MHD bubble flow simulations, we have also implemented a methodology for computing the DMD of bubble wake flow in the bubble reference frame, combining the \textit{VTK} and \textit{Python} libraries with our own recently developed \textit{MHT-X} object tracer.

The DMD code is available at \textit{GitHub} \cite{martinKlevsPyDMD2021}. The newly implemented classes names start with \textit{mosesdmd}. They all use MOSES-SVD implemented in \cite{martinKlevsMOSESSVD2021} (\textit{GitHub}).

These tools were applied to the output of MHD bubble flow simulations of a model downscaled liquid metal system to assess the potential for an in-depth analysis of the underlying physical processes. Relatively coarse resolution yet representative and physically meaningful simulations were performed for bubble chain flow with a quasi-single bubble flow regime at a lower flow rate ($30~sccm$) as well as for a higher flow rate ($100~sccm$) with strong bubble collective dynamics. Both cases were simulated without and with applied static horizontal MF.

The results of DMD of the velocity field in both the liquid metal vessel reference frame and the bubble wake flow in the bubble reference frame indicate that DMD is indeed a viable and very useful tool for MHD bubble flow analysis. This is indicated by the fact that even for the currently coarse spatial resolution and limited simulated flow time the obtained DMD modes largely explain, in some cases at least hypothetically, the observed velocity and Q-factor field dynamics, as well as bubble trajectories. Moreover, DMD mode statistical and flow field analysis identified several noteworthy effects and observations that warrant further investigation:

\begin{itemize}
    \item DMD modes in the vessel reference frame show that for the cases without applied MF the zeroth (mean/stationary) velocity field modes exhibit symmetry mostly with respect to the \textit{YZ} mid-plane, whereas when horizontal MF is applied one observes very clear symmetry in the \textit{XZ} mid-plane as well the \textit{YZ} mid-plane for the lower flow rate. With MF the symmetries are clearer.
    
    \item For the lower flow rate the \textit{XZ} symmetry manifests in the form of three parallel metal flow sheets over the \textit{Y} dimension: the central sheet about the bubble chain core with flow in the positive \textit{Z} direction and two sheets on its either side with counter flow in the negative \textit{Z} direction. For higher flow rate the \textit{XZ} symmetry is partially broken in the upper part of the vessel and the three counter flow layers assume zig-zag shapes (projected) in the \textit{YZ} plane.
    
    \item Standing velocity field waves with flow oscillations in the \textit{Y} direction develop in the upper part of the vessel over time rather quickly and persist throughout the rest of the flow time.
\end{itemize}

It is important to devote further research to elucidate exactly how the transition between symmetry states and the formation of counter flow sheets occur. It is planned to perform simulations with varying MF strength and orientation, as well as flow rate, and find out if the transition is gradual or if there exists a threshold parameter combination in terms of the Reynolds ($Re$), Stuart ($N$)/Hartmann ($Ha$) and Eötvös ($Eo$) numbers. It is also of interest to see how increasing flow rate disrupts flow structures and symmetry while causing the onset of standing waves in the vessel velocity field.

Nonzero order vessel reference frame modes also exhibited physically interesting traits that should be investigated further:

\begin{itemize}
    \item Modes associated with velocity field oscillations within the bubble chain indicate that the latter can be assigned a coherence length which can be measured given a criterion (one used herein or a better alternative) -- here one finds that bubble chain coherence lengths for all cases except for $100~sccm$ without MF exceed the vertical dimension of the liquid metal vessel, whereas in the exception case the chain coherence length is $\sim 3/4$ of the vessel height.
    
    \item A finite coherence length indicates the onset of bubble chain instability beyond a certain elevation threshold -- this destabilization could be related to other vessel reference frame modes.
    
    \item Higher order modes for the cases without applied MF contain fine flow structures due to turbulent pulsations that develop within the vessel over time -- DMD modes seem to separate these into different spatial/temporal scale groups.
\end{itemize}

Measurements of chain coherence lengths and characteristic length scales of turbulent flow structures revealed by DMD should enable one to localize the spatial origins of bubble chain instability onsets and investigate how MF strength and orientation affects the formation of turbulent flow structures -- all of this can be quantified. The latter, however, would require simulations with a finer spatial resolution.

DMD analysis of bubble wake flow in the bubble reference frame yielded several insights:

\begin{itemize}
    \item In the higher flow rate cases it would appear that the bubble wake modes are considerably modulated by processes outside of the bubble chain volume (vessel modes) -- while this makes sense qualitatively, it is important that DMD enables to quantify this.
    
    \item Bubble wake mode modulation by the vessel modes is more pronounced in the $100~sccm$ case without MF.
    
    \item Results for $100~sccm$ with applied MF may suggest that the origins of the onset of flow asymmetrization in the \textit{YZ} plane should be sought by analyzing the bubble wake mode dynamics.
\end{itemize}

This means that DMD of the output of simulations with longer flow time and greater spatial resolution could reveal how flow structures within the vessel and about the bubbles develop over time and affect one another. Quantifying this via a more in-depth analysis of modes than was performed in this paper (not feasible with the current mesh) at different points in the $Re/N(Ha)/Eo$ space could reveal important details regarding the feedback between near field flow about the bubbles and far field flow within the vessel.

Beyond these points the authors would refrain from further analysis of the present data -- this is beyond the scope of this proof-of-concept endeavour due to the low resolution. However, the results obtained herein indicate that the developed methodology could and should be applied to other flow fields of interest as well:

\begin{itemize}
    \item By analyzing the vorticity field one could observe flow structures and effects otherwise overlooked in the velocity field analysis.
    \item Performing DMD for the volume fraction field in the bubble reference frame should enable one to quantify the effects of applied MF and gas flow rate on bubble shape oscillations by measuring dominant surface wave length scales and temporal frequencies.
\end{itemize}

Furthermore, the latter can also be done for bubble shape projections obtained via, for example, dynamic neutron/X-ray imaging of model bubble flow systems like the ones studied in \cite{birjukovsPhaseBoundaryDynamics2020, birjukovsArgonBubbleFlow2020}. Applying DMD to bubble wake flow captured by direct numerical simulations with/without applied MF is also of great interest.

Recognising the need to work with large, high resolution simulation datasets in the future, the authors also plan to augment the presented DMD solution with specialized methods for memory efficient QR decomposition of high aspect ratio matrices encountered during DMD analysis.

Finally, the authors must point out a very clear limitation of the proposed approach for DMD analysis in the bubble reference frame -- one cannot treat cases where bubbles travel too close to one another or collide, coalesce and/or split. This is because masking within the DMD sampling volume must be consistent over time. This limitation extends to similar applications outside of the physical system investigated in this paper, i.e. near field flow analysis for particles travelling within fluids.

\section{Acknowledgements}

This research is a part of the ERDF project ”Development of numerical modelling approaches to study complex multiphysical interactions in electromagnetic liquid metal technologies” (No. 1.1.1.1/18/A/108).

\printbibliography

\end{document}